\documentclass[aps,prd,preprint,superscriptaddress,nofootinbib,showpacs,eqsecnum]{revtex4-2}

\usepackage{amsmath}    
\usepackage{graphicx}   
\usepackage{amssymb}    
\usepackage{enumitem}
\usepackage{color}
\usepackage{mathrsfs}
\usepackage{bm}
\usepackage{bbm}
\usepackage{multirow}
\usepackage{easyReview}
\usepackage{tikz}
\usepackage{array}

\usetikzlibrary{decorations.markings,decorations.pathmorphing}
\tikzset{
	photon/.style={decorate, draw=black,
	decoration={coil,aspect=0,amplitude=1.5pt, segment length=6pt}}
}

\usepackage{hyperref}
\usepackage{cleveref}
\crefname{equation}{Eq.}{Eqs.}
\crefname{figure}{Fig.}{Figs.}
\crefname{section}{Sect.}{Sects.}
\crefname{appendix}{Appendix}{Appendices}
\crefname{table}{Table}{Tables}

\def\sect#1{Sect.~\ref{#1}}

\def\eqn#1{Eq.~(\ref{#1})}

\def\nn{\nonumber}
\def\phib{\bar\phi}
\def\cc{\text{c.c}}
\def\S{\mathbb{S}}
\def\lc{\varepsilon}
\def\pol{\epsilon}
\def\polM{\mathcal{E}}
\def\polMb{\bar{\mathcal{E}}}
\def\polm{\varepsilon}
\def\polmb{\bar{\varepsilon}}
\def\kk{q}
\def\LEM{\mathcal{L}_{\text{EM}}}
\def\Lmin{\mathcal{L}_{\text{min}}}
\def\Lnonmin{\mathcal{L}_{\text{non-min}}}
\def\Lspins{\mathcal{L}^{s}_{\text{min}}}

\def\Ltwospin{\mathcal{L}^{s,s-1}_{\text{min}}}
\def\Lnonmints{\mathcal{L}_{\text{non-min}}^{s,s-1}}

\def\Mtree{\mathcal{M}_{4,\text{cl.}}^{\text{tree}}}
\def\Moneloop{\mathcal{M}_{4,\text{cl.}}^{(1)}}
\def\Mtri{\mathcal{M}_{\Uptri + \Downtri}}

\def\mb{\bar{m}}
\def\pb{\bar{p}}
\def\ub{\bar{u}}
\def\imUnit{i}
\def\hamO{\mathbb{O}}

\def\ie{i0}
\def\dd{\text{d}}
\def\lm{\ell}
\def\gS{\mathsf{S}}
\def\identity{1\kern-0.25em\text{l}}
\def\id{1\kern-0.25em\text{l}}

\def\opcS{\hat{S}}
\def\opS{{\hat{\textbf{S}}}}
\def\clS{\textbf{S}}
\def\opH{\mathcal{H}}

\def\opp{\bm p}
\def\clp{\bm p}
\def\clpb{\bar{\bm p}}
\def\clq{\bm q}
\def\opr{\bm r}
\def\clr{\bm r}
\def\opK{{\hat{\textbf{K}}}}
\def\opcK{\hat{K}}
\def\clK{\textbf{K}}
\def\opL{\bm L}
\def\clL{\bm L}
\def\momL{\bm L_q}

\def\hamOperatorId{ }
\def\hamOperatorS{\frac{\opL  \cdot \opS_1}{\opr^2}}
\def\hamOperatorK{\frac{\opr \cdot \opK_1}{\opr^2}}

\def\ampOperatorS{\momL \cdot \clS_1}
\def\ampOperatorK{\imUnit \bm q \cdot \clK_1}
\def\spinH{\opH_{1}}
\def\boostH{\opH_{2}}
\def\coupling{\alpha}

\def\ampEFT{\mathbb{M}}
\newcommand{\PL}[1]{#1\text{PL}}
\def\pder{\mathcal{D}}

\def\com{CoM } 

\usepackage{tikz}
\usetikzlibrary{decorations}
\usetikzlibrary{trees}
\usetikzlibrary{decorations.pathmorphing}
\usetikzlibrary{decorations.markings}
\usetikzlibrary{external}
\usetikzlibrary{intersections}
\usetikzlibrary{shapes,arrows}
\usetikzlibrary{arrows.meta}
\usetikzlibrary{calc}
\usetikzlibrary{shapes.misc}
\usetikzlibrary{decorations.text}
\usetikzlibrary{backgrounds}
\usetikzlibrary{fadings}
\usetikzlibrary{tikzmark,calc,arrows,shapes,decorations.pathreplacing}

\tikzset{
	graviton/.style={decorate,line width=0.15mm, 
	decoration={snake,amplitude=.6mm, segment length=1.5mm}
	},
	scalar/.style={postaction={decorate},
	},
	massive/.style={postaction={decorate},
		line width=0.75mm,
	},
	spinSm1/.style={postaction={decorate},
        line width=0.75mm,draw=black,
	},
	spinSm2/.style={postaction={decorate},
		line width=0.75mm,draw=blue,dashed,
	},
	massless/.style={postaction={decorate},
	},
	masslessWithDot/.style={postaction={decorate},
		decoration={
			markings,
			mark=at position 0.5 with {\fill circle (2pt);}}
	},
	massiveWithDot/.style={postaction={decorate},
		line width=0.5mm,
		decoration={
			markings,
			mark=at position 0.5 with {\fill circle (2pt);}}
	},
	massiveWithArrow/.style={postaction={decorate},
		line width=0.75mm,
		decoration={
			markings,
			mark=at position 0.5 with {\arrow{latex}}}
	},
	massiveWithArrowB/.style={postaction={decorate},
		line width=0.75mm,
		decoration={
			markings,
			mark=at position 0.95 with {\arrow{latex}}}
	},
	massiveLin/.style={postaction={decorate},
		double,
		thick,
		fill=white
	},
	massivePhi/.style={postaction={decorate},
		line width=0.75mm,
		dashed
	},
	masslessPhi/.style={postaction={decorate},
		dashed
	},
	unitaryCut/.style={postaction={draw,densely dashed,blue,thin},
		line width = 0.2cm,white
	},
	gluon/.style={decorate, draw=magenta,
		decoration={coil,amplitude=4pt, segment length=5pt}},
	partial ellipse/.style args={#1:#2:#3}{
		insert path={+ (#1:#3) arc (#1:#2:#3)}
	},
	cross/.style={cross out, draw=black, minimum size=2*(#1-\pgflinewidth), inner sep=0pt, outer sep=0pt},
	branchCut/.style={postaction={decorate},
		snake=zigzag,
		decoration = {snake=zigzag,segment length = 2mm, amplitude = 2mm}	
	}
	cross/.default={1pt}
}

\colorlet{mred}{black!30!red}
\colorlet{mgreen}{black!30!green}
\colorlet{mblue}{black!30!blue}
\colorlet{morange}{blue!70!red}

\newcommand{\TreeSymbol}{
	\begin{tikzpicture}[scale=0.1]
		\draw (-1,-1) -- (1,-1)  (-1,1) -- (1,1) (0,-1) -- (0,1);
	\end{tikzpicture}
}

\newcommand{\BoxSymbol}{
\begin{tikzpicture}[scale=0.1]
	\draw (-1,-1) -- (-1,1) -- (1,1) -- (1,-1) -- cycle;
\end{tikzpicture}
}
\newcommand{\CboxSymbol}{
	\begin{tikzpicture}[scale=0.1]
		\draw (-1,-1) -- (1,1) -- (-1,1) -- (1,-1) -- cycle;
	\end{tikzpicture}
}

\newcommand{\Uptri}{
	\begin{tikzpicture}[scale=0.2]
		\draw (-0.5,0) -- (0.5,0) -- ++(120:1) -- cycle;
	\end{tikzpicture}
}
\newcommand{\Downtri}{
	\begin{tikzpicture}[scale=0.2]
		\draw (0,0) -- ++(60:1) -- ++(-1,0) -- cycle;
	\end{tikzpicture}
}

\newcommand{\FigLowerSpinExchangeInCompton}{
\begin{tikzpicture}
        \coordinate (e1) at (-1, 0);  
		\coordinate (e2) at (0, 1);
		\coordinate (e3) at (1, 1);
		\coordinate (e4) at (2,0);
		
		\coordinate (v1) at (0,0);  
		\coordinate (v2) at (1,0);  
		
		\draw[massless] (e1) -- (v1);
		\draw[graviton] (e2) -- (v1);
		\draw[massive] (v1) -- (v2);
		\draw[graviton] (e3) -- (v2);
		\draw[massless] (e4) -- (v2);
		
		\node[below left] at (e1) {$1, p_1$};
        \node[above]  at (e2) {$2, q_2$};
        \node[above] at (e3) {$3, q_3$};
        \node[below right] at (e4) {$4, p_4$};
		
        \draw[fill=lightgray, opacity=1] (v1) ellipse (0.1 and 0.1);
        \draw[fill=lightgray, opacity=1] (v2) ellipse (0.1 and 0.1);        
       
\end{tikzpicture}}

\newcommand{\FigLowerSpinExchangeInComptonX}{
\begin{tikzpicture}
        \coordinate (e1) at (-1, 0);  
		\coordinate (e2) at (0, 1);
		\coordinate (e3) at (1, 1);
		\coordinate (e4) at (2,0);
		
		\coordinate (v1) at (0,0);  
		\coordinate (v2) at (1,0);  
		
		\draw[massless] (e1) -- (v1);
		\draw[graviton] (e2) -- (v1);
		\draw[massive] (v1) -- (v2);
		\draw[graviton] (e3) -- (v2);
		\draw[massless] (e4) -- (v2);
		
		\node[below left] at (e1) {$1, p_1$};
        \node[above]  at (e2) {$3, q_3$};
        \node[above] at (e3) {$2, q_2$};
        \node[below right] at (e4) {$4, p_4$};
		
        \draw[fill=lightgray, opacity=1] (v1) ellipse (0.1 and 0.1);
        \draw[fill=lightgray, opacity=1] (v2) ellipse (0.1 and 0.1);        
       
\end{tikzpicture}}

\newcommand{\FigLowerSpinThreePTsn}{
\begin{tikzpicture}
        \coordinate (e1) at (-1, 0);  
		\coordinate (e2) at (0, 1);
		\coordinate (e3) at (1, 0);
		
		\coordinate (v1) at (0,0);  
		
		\draw[massless] (e1) -- (v1);
		\draw[graviton] (e2) -- (v1);
		\draw[spinSm1] (v1) -- (e3);
		
		\node[below left] at (e1) {$\varepsilon_1^{(s)},\,p_1$};
        \node[above]  at (e2) {$\varepsilon,\,q$};
        \node[below right] at (e3) {$\varepsilon_2^{(s-n)},\,p_2$};
		
        \draw[fill=lightgray, opacity=1] (v1) ellipse (0.1 and 0.1);      
\end{tikzpicture}}

\begin{document}

\title{\Large Quantum Field Theory, Worldline Theory, \\and Spin Magnitude Change in Orbital Evolution}

\author{Zvi~Bern}
\affiliation{
	Mani L. Bhaumik Institute for Theoretical Physics,
	University of California at Los Angeles,
	Los Angeles, CA 90095, USA}
\author{Dimitrios~Kosmopoulos}
\affiliation{
         D\'epartement de Physique Th\'eorique, 
         Universit\'e de Gen\`eve, 
         CH-1211 Geneva, 
         Switzerland
}		
\author{Andres~Luna}
\affiliation{
	Niels Bohr International Academy,
	Niels Bohr Institute, University of Copenhagen,
	Blegdamsvej 17, DK-2100, Copenhagen \O , Denmark}
\author{Radu~Roiban}
\affiliation{Institute for Gravitation and the Cosmos,
	Pennsylvania State University,
	University Park, PA 16802, USA}
\author{Trevor~Scheopner}	
\affiliation{
	Mani L. Bhaumik Institute for Theoretical Physics,
	University of California at Los Angeles,
	Los Angeles, CA 90095, USA}
\author{Fei~Teng}
\affiliation{Institute for Gravitation and the Cosmos,
	Pennsylvania State University,
	University Park, PA 16802, USA}
\author{Justin~Vines}
\affiliation{
	Mani L. Bhaumik Institute for Theoretical Physics,
	University of California at Los Angeles,
	Los Angeles, CA 90095, USA}
	
\begin{abstract}
{
A previous paper~\cite{Bern:2022kto} identified a puzzle stemming from the amplitudes-based approach to spinning bodies in general relativity: additional Wilson coefficients appear compared to current worldline approaches to conservative dynamics of generic astrophysical objects, including neutron stars. In this paper we clarify the nature of analogous Wilson coefficients in the simpler theory of electrodynamics. We analyze the original field-theory construction, identifying definite-spin states some of which have negative norms, and relating the additional Wilson coefficients in the classical theory to transitions between different quantum spin states. We produce a new version of the theory which also has additional Wilson coefficients, but no negative-norm states. We match, through $\mathcal O(\alpha^2)$ and $\mathcal O(S^2)$, the Compton amplitudes of these field theories with those of a modified worldline theory with extra degrees of freedom introduced by releasing the spin supplementary condition. We build an effective two-body Hamiltonian that matches the impulse and spin kick of the modified field theory and of the worldline theory, displaying additional Wilson coefficients compared to standard worldline approaches. The results are then compactly expressed in terms of an eikonal formula. Our key conclusion is that, contrary to standard approaches, while the magnitude of the spin tensor is still conserved, the magnitude of the spin vector can change under conserved Hamiltonian dynamics and this change is governed by the additional Wilson coefficients.  For specific values of Wilson coefficients the results are equivalent to those from a definite spin obeying the spin supplementary condition, but for generic values they are physically inequivalent. These results warrant detailed studies of the corresponding issues in general relativity.
}

\end{abstract} 

\maketitle

\newpage
	
\tableofcontents

\newpage

\section{Introduction}
\label{sec:intro} 

\subsection{General Overview}

The landmark detection of gravitational waves by the LIGO/Virgo
collaboration~\cite{LIGOScientific:2016aoc,  LIGOScientific:2017vwq} opened a new  era in astronomy, cosmology and perhaps even particle physics.
As gravitational-wave detectors become more
sensitive~\cite{Punturo:2010zz,  LISA:2017pwj,  Reitze:2019iox}, the
spin of objects such as black holes and neutron stars will
play an increasingly important role in identifying and interpreting
signals.  
Spin also  leads to much richer three-dimensional dynamics because of the
exchange of angular momentum between bodies and their orbital motion.
Its precise definition leads to interesting and subtle
theoretical questions, some of which we address here.

The study of the dynamics of spinning objects in general relativity~\cite{Mathisson:1937zz,   Papapetrou:1951pa,   Pirani:1956tn,   Tulczyjew} has a long history, in both the post-Newtonian
(PN) framework~\cite{Barker:1970zr, Barker:1975ae, Kidder:1992fr, Kidder:1995zr, Blanchet:1998vx, Tagoshi:2000zg,
Porto:2005ac, Faye:2006gx, Blanchet:2006gy, Damour:2007nc, Steinhoff:2007mb, Levi:2008nh,
Steinhoff:2008zr, Steinhoff:2008ji, Marsat:2012fn,Hergt:2010pa,
 Porto:2010tr,  Levi:2010zu,  Porto:2010zg, Levi:2011eq,
 Porto:2012as,  Hergt:2012zx,  Bohe:2012mr,  Hartung:2013dza,  Marsat:2013wwa,  Levi:2014gsa,
Vaidya:2014kza, Bohe:2015ana,  Bini:2017pee,  Siemonsen:2017yux,
 Porto:2006bt,  Porto:2007tt,  Porto:2008tb,  Porto:2008jj,  Levi:2014sba,
Levi:2015msa, 
Levi:2015uxa,  Levi:2015ixa,  Levi:2016ofk,   Levi:2019kgk,  Levi:2020lfn,
 Levi:2020kvb,  Levi:2020uwu,  Kim:2021rfj,  Maia:2017gxn,  Maia:2017yok,  Cho:2021mqw,  Cho:2022syn,
 Kim:2022pou,  Mandal:2022nty, Kim:2022bwv, Mandal:2022ufb, Levi:2022dqm,  Levi:2022rrq},
where observables are simultaneously expanded in Newton's constant $G$ and in the velocity $v$, and the post-Minkowskian (PM) framework~\cite{Bini:2017xzy, Bini:2018ywr, 
Maybee:2019jus,   Guevara:2019fsj,
  Chung:2020rrz,   Guevara:2017csg,   Vines:2018gqi,  Damgaard:2019lfh,
Aoude:2020onz,
Vines:2017hyw,
Guevara:2018wpp,   Chung:2018kqs,
Chung:2019duq,
Bern:2020buy,
Kosmopoulos:2021zoq,
Liu:2021zxr,
Aoude:2021oqj,
Jakobsen:2021lvp,
Jakobsen:2021zvh,
Chen:2021kxt,
Chen:2022clh,
Cristofoli:2021jas,
Chiodaroli:2021eug,
Cangemi:2022abk,
Cangemi:2022bew,
Haddad:2021znf,
Aoude:2022trd,
Menezes:2022tcs,
Bern:2022kto,
Chen:2022clh,
Alessio:2022kwv,
Alessio:2023kgf,
Bjerrum-Bohr:2023jau,
Damgaard:2022jem,
Haddad:2023ylx,
Aoude:2023vdk,
Jakobsen:2023ndj,
Jakobsen:2023hig,
Heissenberg:2023uvo,
Bianchi:2023lrg},
where observables are expanded only in Newton's constant with exact velocity dependence.
In both approaches the interaction of spinning objects with the  gravitational field is described in terms of a set of higher-dimension operators whose Wilson coefficients encode the detailed properties of the objects. 
For the interesting case of black holes, the values of these coefficients at $\mathcal O(G)$ are known~\cite{Vines:2017hyw}, with proposals for the additional coefficients at $\mathcal O(G^2)$ recently given based on a shift symmetry~\cite{Aoude:2022trd, Bern:2022kto, Haddad:2023ylx,  Aoude:2023vdk} already present at~$\mathcal O(G)$.\footnote{\baselineskip=14pt Through $O(S^4)$ 
Refs.~\cite{Bautista:2022wjf,Bautista:2023szu} find that the Compton amplitude derived by solving the Teukolsky equation agrees with with these previous results. However,
the predictions based on shift symmetry at ${\cal O}(S^5)$ are in tension with results 
from the Teukolsky equation, though the latter involve a subtle analytic continuation between the black-hole and naked-singularity regimes.}    The electromagnetic case is similar in structure~\cite{Westpfahl:1979gu, Damour:1990jh, Buonanno:2000qq, Kosower:2018adc, Saketh:2021sri, Bern:2021xze, Bern:2023ccb} (see also Refs.~\cite{delaCruz:2020bbn, delaCruz:2021gjp} for non-abelian generalisations), with the post-Coulombian (PC) and post-Lorentzian (PL) expansions being the respective analogs of the gravitational PN and PM expansions.

A primary purpose of this paper is to explore puzzles identified in Ref.~\cite{Bern:2022kto} regarding the description of spinning bodies in general relativity.
In that paper, results for the conservative two-body scattering angle were obtained through fifth power in the spin using a scattering-amplitudes-based method.
A rather striking outcome, which follows from the fact that the field-theory Lagrangian is not directly expressed in terms of particles' spin tensor, is that the field-theory approach of Ref.~\cite{Bern:2020buy} has a larger number of independent Wilson coefficients for a given power of spin than standard (worldline) methods.  
While at 1PM (tree) level the number of independent Wilson coefficients is identical in the two approaches, matching of physical observables starting at 2PM (one loop) and third power of the spin
can only be attained by setting some of the field-theory Wilson coefficients to definite numerical values, so that they are no longer independent.  
This implies that the field theory contains a larger number of physically-relevant independent Wilson coefficients. 
For the special case of Kerr black holes it appears that the additional Wilson coefficients present in the field theory are not needed~\cite{Vines:2017hyw, Chen:2021kxt,Chen:2022clh, Aoude:2022trd, Bern:2022kto}.  In electrodynamics we find a similar situation for the root-Kerr solution~\cite{Arkani-Hamed:2019ymq}, related to the Kerr solution via the double copy~\cite{Kawai:1985xq,Bern:2008qj,Bern:2010ue,Monteiro:2014cda,Luna:2016hge, Bern:2019prr}.

The connection between scattering amplitudes and effective two-body interactions has been known for some time~\cite{Iwasaki:1971vb, Iwasaki:1971iy,  Gupta:1979br, Donoghue:1994dn,  Bjerrum-Bohr:2002gqz,  Holstein:2008sx,  Neill:2013wsa,  Bjerrum-Bohr:2013bxa, Damour:2016gwp,   Damour:2017zjx,   Bjerrum-Bohr:2018xdl}.  
Recent years have seen the construction of new systematic methods for
extracting potentials and physical observables at high orders from
scattering amplitudes~\cite{Cheung:2018wkq, Kosower:2018adc,
Bern:2019nnu,   Bern:2019crd,   Cristofoli:2019neg,   Bjerrum-Bohr:2019kec,   Brandhuber:2021eyq}, which leverage modern methods for calculating scattering amplitudes, including generalized unitarity~\cite{Bern:1994zx,   Bern:1994cg,   Bern:1995db,   Bern:1997sc,   Britto:2004nc,   Bern:2007ct}, the double copy~\cite{Kawai:1985xq,   Bern:2008qj,   Bern:2010ue,   Bern:2019prr} and advanced integration techniques~\cite{Chetyrkin:1981qh,   Kotikov:1990kg,   Bern:1993kr,   Remiddi:1997ny,   Remiddi:1997ny, Laporta:2000dsw, Parra-Martinez:2020dzs}.   
The extraction of classical physics from quantum scattering is greatly simplified by concepts from effective field theories (EFTs), systematized for the gravitational-wave problem in Ref.~\cite{Goldberger:2004jt} and applied to the PM framework in Ref.~\cite{Cheung:2018wkq}.  
By manifestly maintaining Lorentz invariance, the amplitudes approach fits naturally in the PM or PL frameworks,
and produced the first conservative spinless two-body Hamiltonian at $\mathcal O(G^3)$ and $\mathcal O(G^4)$~\cite{Bern:2019nnu,  Bern:2019crd,  Bern:2021dqo,  Bern:2021yeh} (see also Refs.~\cite{Cheung:2020gyp,  Kalin:2020fhe,  Dlapa:2021npj,  Dlapa:2021vgp,   Manohar:2022dea, Dlapa:2022lmu, Bjerrum-Bohr:2022ows, Jakobsen:2023ndj}). 
Such methods also led to new perspectives on the gravitational interactions of spinning particles~\cite{Cachazo:2017jef,Chung:2018kqs,Guevara:2018wpp,Guevara:2019fsj,Chung:2019duq, Vines:2018gqi,Damgaard:2019lfh, Bern:2020buy, Aoude:2020onz,Chen:2021kxt, Bern:2022kto} and on tidal effects~\cite{Bini:2020flp,   Kalin:2020mvi,   Cheung:2020sdj,   Haddad:2020que,   Kalin:2020lmz,   Bern:2020uwk,   Cheung:2020gbf,   Aoude:2020ygw}.                               

Here we use both the amplitudes-based method and the more standard 
worldline approach~\cite{Goldberger:2004jt,Porto:2006bt,
 Porto:2016pyg, Levi:2015msa,JanSteinhoff:2015ist,Levi:2018nxp} to study the interactions of spinning particles.
Since they describe the same physics, one may expect that there is a (usually nontrivial) correspondence between the operators (as well as between their Wilson coefficients) describing these interactions in the two approaches.
Each type of object, whether a Kerr black hole or neutron star, is described by particular values for the Wilson coefficients, which are determined by an appropriate matching calculation.  
In the worldline approach  one imposes a spin supplementary condition (SSC)~\cite{Fleming} that identifies the three physical spin degrees of freedom.
This condition has been interpreted in terms of a spin-gauge symmetry which formalizes the freedom to shift the worldline in the ambient space~\cite{Steinhoff:2015ksa,Levi:2015msa, Vines:2016unv} without changing the physics.
An important aspect of an SSC is that it reduces the number of possible independent operators---and consequently the number of Wilson coefficients---by equating operators whose difference is proportional to the SSC. 
Here we use the dynamical mass function formalism of Ref.~\cite{JanSteinhoff:2015ist} to explore the consequences of relaxing the SSC and to help interpret the additional degrees of freedom. 

An interesting subtlety in the amplitudes approach is whether the complete description of a spinning compact body is provided by a single quantum spin $s\gg 1$ or by a suitable combination of multiple quantum spins, with possible transitions between them.
For the sake of simplicity, the field theory of Ref.~\cite{Bern:2020buy}---meant to be valid only in the classical limit---is based on the matter states forming an irreducible representation of the Lorentz group but a reducible representation of the rotation group; some of its components have negative norm. 
One might worry these negative-norm states might lead to some difficulties in the classical limit~\cite{Kim:2023drc}.  In addition, projecting onto the physical states of a quantum spin $s$~\cite{Kim:2023drc, Haddad:2023ylx} appears to effectively remove the additional Wilson coefficients, leaving only those included in the worldline framework, which we affirm here.  
Field-theory approaches~\cite{Chen:2021kxt,Chen:2022clh, Aoude:2022trd, Aoude:2023vdk} based on the massive-spinor-helicity amplitudes~\cite{Arkani-Hamed:2019ymq} are a convenient means for restricting the propagation to a single irreducible quantum spin. Here we use physical state projectors~\cite{Singh:1974qz, Chowdhury:2019kaq} for the same purpose.

The results of Ref.~\cite{Bern:2022kto} raises several questions:
\begin{enumerate}

\item What is a complete description of a spinning body in general relativity? 

\item Can one construct a worldline theory that matches field-theory descriptions containing extra independent Wilson coefficients?  If so, what extra degrees of freedom are needed?

\item The field-theory construction of Ref.~\cite{Bern:2022kto} uses propagating reducible representations of the rotation group (spin representations), some with negative norm.  In the context of this construction, what happens if only a single quantum spin propagates?

\item Can one build a field theory based on positive-norm irreducible representations of the rotation group that also contain extra independent Wilson coefficients?

\item  Should a classical spin be modeled as a definite-spin field or as a superposition of fields with different spins?  A related question on the latter case is whether transitions between different spins are allowed that change the magnitude of the spin vector even in the conservative sector.\footnote{\baselineskip=14pt With dissipation and absorption included the spin magnitude is, of course, not preserved (see e.g. Refs.~\cite{Goldberger:2020fot, Saketh:2022xjb, Aoude:2023fdm} for recent discussions).} 

\item Can one build an effective two-body Hamiltonian with extra degrees of freedom whose physical observables match field-theory results containing extra Wilson coefficients? 

\item What is the physical interpretation of the operators associated with additional Wilson coefficients?

\end{enumerate}

To address these questions we turn to electrodynamics, which has been useful as a toy model for gravity~\cite{Westpfahl:1979gu, Damour:1990jh, Buonanno:2000qq, Kosower:2018adc, Saketh:2021sri,
Bern:2021xze, Bern:2023ccb}.  While electrodynamics cannot answer all questions about gravity, the overlap is more than sufficient to make this a useful test case.  In addition to the absence of photon self-interactions, electrodynamics is particularly helpful for our questions because the additional independent operators and their Wilson coefficients affect observables already at the first order spin, rather than at third order as for gravity, greatly simplifying the analysis.  

We use various field theories, worldline theories and effective two-body Hamiltonians, comparing and contrasting the results from each.  In particular, to help identify the origin of the extra Wilson coefficients we evaluate Compton amplitudes and scattering angles for three related but distinct field theories of electrodynamics coupled with higher-spin fields:

\begin{enumerate}[label=FT\arabic*:,ref=FT\arabic*]
\item \hypertarget{ft1s}{The} \hypertarget{ft1g}{setup} from Refs.~\cite{Bern:2020buy, Bern:2022kto}, except for electrodynamics instead of general relativity.
%
%
The matter states of this theory form an irreducible representation of the Lorentz group and a reducible representation of the rotation group, 
thereby as a quantum theory it carries more degrees of freedom than those of a fixed-spin particle, including negative-norm states. 
In this theory we consider \hyperlink{ft1s}{FT1s} with classical asymptotic states having spin tensors obeying the covariant spin supplementary condition (SSC), $S_{\mu \nu}p^{\nu} = 0$, and \hyperlink{ft1g}{FT1g} with classical asymptotic states having unconstrained spin tensors. This is equivalent to relaxing the covariant SSC, so that the resulting amplitudes explicitly contain factors of $S_{\mu\nu}p^{\nu}$. 
When we do not need to distinguish between \hyperlink{ft1s}{FT1s} and \hyperlink{ft1g}{FT1g}, we collectively refer to them as \ref{fieldtheory1}.
The results of \hyperlink{ft1s}{FT1s} are obtained from those of \hyperlink{ft1g}{FT1g} simply by imposing the covariant SSC on the initial and final spin tensors.
\label{fieldtheory1}
\item The higher-spin field is constrained to contain a single irreducible spin-$s$ representation of the rotation group~\cite{Singh:1974qz}. The external massive states are traceless and transverse due to the equation of motion.  In contrast to \hyperlink{ft1s}{FT1s} and \hyperlink{ft1g}{FT1g}, only positive-norm states propagate, and as we shall see, the covariant SSC is automatically imposed on the spin tensors. \label{fieldtheory2}
\item \hypertarget{ft3s}{The} \hypertarget{ft3g}{same} construction as for \ref{fieldtheory2} except that two positive-norm irreducible representations of the rotation group, one with spin-$s$ and the other with spin-$(s-1)$, are considered. While this field content allows us to reliably capture effects linear in spin,
it is sufficient to demonstrate that such field theories support more Wilson coefficients than \ref{fieldtheory2}.  
We include suitable couplings between matter fields of different spin.
Similarly to \ref{fieldtheory1}, we consider \hyperlink{ft3s}{FT3s} with asymptotic states having spin tensors obeying the covariant SSC and \hyperlink{ft3g}{FT3g}  with asymptotic states being a particular combination of the asymptotic states of the two fields.
When we do not need to distinguish between \hyperlink{ft3s}{FT3s} and \hyperlink{ft3g}{FT3g}, we collectively refer to them as \ref{fieldtheory3}. 
\label{fieldtheory3}
\end{enumerate}
The above field-theory constructions do not exhaust the ways to adjust the spectrum of propagating states. For example, one can use the chiral construction of Ref.~\cite{Ochirov:2022nqz}, based on the representation $(2s, 0)$ of the Lorentz group  leading to the same number $(2s+1)$ of propagating degrees as a quantum spin-$s$ particle.
We note that \ref{fieldtheory1} are not fully consistent as quantum theories because of the appearance of propagating negative-norm states. Because of this we use them only in the classical limit, as envisioned in Ref.~\cite{Bern:2020buy}. We moreover see that there is a close relation between them and \ref{fieldtheory3}, which is constructed using only positive-norm states.

To address the question of what kind of worldline theory has the same observables as field theories with extra Wilson coefficients we consider two worldline theories:
\begin{enumerate}[label=WL\arabic*:,ref=WL\arabic*]
\item The standard worldline construction with the covariant SSC imposed. We use the formalism of Ref.~\cite{JanSteinhoff:2015ist}. \label{worldline1}
\item A modified worldline construction with no SSC imposed and consequently with extra degrees of freedom. In the absence of an SSC we can include additional operators and Wilson coefficients equivalent to the additional ones that can be included in \hyperlink{ft1g}{FT1g} through the constructed orders.\label{worldline2}
\end{enumerate}

Finally, we construct two two-body effective field-theory Hamiltonians by matching the amplitudes of field theories with different number of internal and asymptotic degrees of freedom. This allows us to directly construct observables for these field theories and compare them with worldline theories:
\begin{enumerate}[label=EFT\arabic*:,ref=EFT\arabic*]
\item The two-body Hamiltonian of the type in Ref.~\cite{Bern:2020buy} containing only the spin vector $\clS$ for each body. The parameters of this Hamiltonian can be adjusted to match either \ref{fieldtheory2} or \ref{worldline1}. We may also match this Hamiltonian to  \ref{fieldtheory1}, \ref{fieldtheory3}, and \ref{worldline2} when the additional Wilson coefficients are set to specific values. \label{H1}
\item A two-body Hamiltonian containing both a spin vector $\clS$ and a Lorentz boost vector $\clK$, interpreted as a mass dipole and inducing an electric dipole. With suitable parameters this Hamiltonian matches \hyperlink{ft1g}{FT1g}, \hyperlink{ft3g}{FT3g} and \ref{worldline2}. \label{H2}
\end{enumerate}


\subsection{Summary of Results}

We compute and compare electrodynamics Compton amplitudes, impulses, spin kicks and scattering angles in the theories outlined above.  With $\alpha$ denoting the fine structure constant, the results of these computations through ${\cal O}(\alpha^2 S)$ for two-body observables and through  ${\cal O}(\alpha S^2)$ for Compton amplitudes yield the following findings: 
\begin{enumerate}

\item In electrodynamics with the massive propagating degrees of freedom of a single spin-$s$ particle realized as a symmetric traceless transverse $s$-index tensor, as in \ref{fieldtheory2} and following Ref.~\cite{Singh:1974qz}, the number of Wilson coefficients agrees with the standard worldline construction~\cite{Levi:2015msa}, in accord with Refs.~\cite{Kim:2023drc, Haddad:2023ylx}.

\item 
By including additional degrees of freedom either by relaxing the transversality constraint of fields or/and by replacing the $s$-index symmetric tensor by a more general $(l,r)$ representation of the Lorentz group, as in \ref{fieldtheory1}, additional Wilson coefficients can appear in the classical limit. 
Thus, the additional Wilson coefficients reflect the additional degrees of freedom present in nontransverse fields.

\item 
We demonstrate that additional propagating positive-norm degrees of freedom in the form of symmetric traceless transverse lower-rank tensor, as in \ref{fieldtheory3}, also lead to additional Wilson coefficients in the classical limit.  Thus, the additional Wilson coefficients are not tied specifically to nontransverse fields, but are a manifestation of additional propagating degrees of freedom.

\item By relaxing the SSC constraint on the worldline, the Compton amplitudes as well as two-body physical observables such as the impulse and spin kick, match the corresponding results of field theories \ref{fieldtheory1} and \ref{fieldtheory3}. 

\item  To match the worldline and field-theory amplitudes with additional asymptotic degrees of freedom  and Wilson coefficients, a two-body EFT Hamiltonian with both spin and boost degrees of freedom are required. 

\item In the systems with additional degrees of freedom and additional Wilson coefficients, the magnitudes of spin vectors are not preserved\footnote{\baselineskip=14pt
We note that the non-conservation of the magnitude of the intrinsic angular momentum of subsystems of gravitationally-interacting conservative many-body systems has been known for some time,~see~e.g.~\cite{Vines:2010ca, Naoz:2012bx, Kuntz:2022onu}. 
} 
in the scattering process while the magnitudes of spin tensors are preserved.

\item  For specific choices of Wilson coefficients, such as the root-Kerr solution~\cite{Arkani-Hamed:2019ymq}, the extra degrees of freedom decouple and the system can be described by removing the boost degrees of freedom.
\end{enumerate}

These results are rather striking.  Dropping the SSC would seem to contradict the standard interpretation of the worldline spin gauge symmetry, where local shifts in the worldline are interpreted as a symmetry~\cite{Steinhoff:2015ksa,Levi:2015msa,Vines:2016unv}.  
Here we are reinterpreting this in terms of certain degrees of freedom of extended nonrigid objects, in much the same way as the spin is interpreted as an internal degree of freedom.  
As we discuss in \sect{sec:WL}, in the electromagnetic case there is a natural explanation in terms of an induced electric dipole moment correlated to the mass dipole.

This paper is organized as follows: In \sect{sec:fieldtheory} we present the field-theory constructions \ref{fieldtheory1}. \ref{fieldtheory2} and \ref{fieldtheory3} for electrodynamics, giving a nonminimal Lagrangian that contains additional Wilson coefficients compared to the standard worldline approaches. We also describe the classical asymptotic states in terms of coherent states and discuss the effect of using different Lorentz representations. The purpose of the various field theories is to identify the source of the extra Wilson coefficients. 
\sect{sec:FourPointCompt} then gives the field-theory amplitudes associated with these theories, including the Compton tree amplitudes needed to build the one-loop two-body amplitudes, which are also presented.  
To interpret these results in the context of the more standard worldline framework, in \sect{sec:WL} we construct the two worldline theories \ref{worldline1} and \ref{worldline2} and compare their Compton amplitudes with the field-theory ones.
In \sect{sec:EFTHamiltonian} we construct two-body EFT Hamiltonians so that the scattering amplitudes of the corresponding EFTs match those of the various field theories. One Hamiltonian contains only the usual spin operator and the other also contains a boost operator. 
The impulse and spin kick derived from the latter are the same as those following from the SSC-less worldline theory. 
A remarkably compact form of physical observables is given in terms of an eikonal formula.  
\sect{sec:WilsonCoeffs} describes the link between extra Wilson coefficients and the degrees of freedom that propagate in the field theory.  In \sect{sec:Conclusion} we summarize our conclusions. 
Finally, an ancillary file is included containing mathematica expressions for Hamiltonian coefficients and observables~\cite{anc}.

\section{Field Theory}
\label{sec:fieldtheory} 


In this section we construct the field theories \ref{fieldtheory1}, \ref{fieldtheory2} and \ref{fieldtheory3} listed in \sect{sec:intro} that we use to track the source of additional degrees of freedom and Wilson coefficients.  
We begin by discussing the covariantization of the free matter Lagrangians, which we refer to as the ``minimal" Lagrangians, first in the framework of Refs.~\cite{Bern:2020buy, Bern:2022kto} where the propagating states form a reducible representation of the rotation group, and then in the framework of Ref.~\cite{Singh:1974qz}, in which the only propagating states are only the $2s+1$ physical states of a spin-$s$ field.
After summarizing the coherent-state description of the classical asymptotic states and the propagators, we then discuss nonminimal interactions which are linear in the photon field strength and the corresponding  three-point amplitudes. 
The scaling of massive momenta $p$, massless transferred momentum $q$, impact parameter $b$ and spins $S$ for obtaining the classical limit are~\cite{Bern:2020buy}
\begin{align}
p\rightarrow p, \qquad
q\rightarrow \lambda q, \qquad
b\rightarrow \lambda^{-1} b, \qquad
S\rightarrow \lambda^{-1} S \, ,
\end{align}
and the classical part of the $L$-loop two-body amplitude scales as $\lambda^{-2+L}$ while Compton amplitudes scale as $\lambda^0$.\footnote{\baselineskip=14pt
This scaling enforces the correspondence principle and the scaling parameter $\lambda$ can be related to $\hbar$, see e.g.~Ref.~\cite{Buonanno:2022pgc}.}
%
The connection of field theories \ref{fieldtheory1}, \ref{fieldtheory2} and \ref{fieldtheory3} to worldline theories will be discussed in \sect{sec:WL}.

\subsection{Minimal Lagrangian in Electrodynamics}

The extension of the construction of Refs.~\cite{Bern:2020buy} to QED and thus the definition of the covariantization of the free Lagrangians for \ref{fieldtheory1} is straightforward, with the main difference from the gravitational case being that 
the fields must be complex. The minimal coupling involves only the standard two-derivative kinetic term\footnote{\baselineskip=14pt
We are using the mostly minus signature. The $(-1)^s$ factor makes the spin-$s$ component physical.} 
\begin{equation}\label{eq:Lmin}
	\LEM = -\frac{1}{4}F_{\mu\nu}F^{\mu\nu}\,,\hskip 1.5 cm  \Lmin = - (-1)^s\phi_s (D^2 + m^2)\phib_s\,, 
\end{equation}
where $F_{\mu\nu}=\partial_{\mu}A_{\nu}-\partial_{\nu}A_{\mu}$, and the covariant derivative is defined as
\begin{equation}
	D_{\mu}\phi_s = \partial_{\mu}\phi_s - i Q A_{\mu}\phi_s\,,\hskip 1.5 cm  D_{\mu}\phib_s = \partial_{\mu}\phib_s + i Q A_{\mu}\phib_s\, .
\end{equation}
%
Without the loss of generality, we take all the massive bodies as carrying the same charge $Q$, and define the effective ``fine structure constant''\footnote{\baselineskip=14pt
Note that this differs from the standard definition of the fine structure constant in terms of the electron charge. To simplify subsequent formulae, we absorb in $\alpha$ the charge of macroscopic bodies.} as $\alpha = Q^2/(4\pi)$. 
The PL framework expands observables in powers of $\alpha$ keeping the exact velocity dependence.
In $\Lmin$, the fields $\phi_s$ and $\phib_s$ can be in generic representations of the Lorentz group as long as 
their product 
is a Lorentz-singlet. The most general choice is that both fields are in the $(l, r)$ representation, i.e. they are represented 
as
\begin{align}
	\phi_s = \phi_{\alpha_1\alpha_2\ldots\alpha_l}^{\dot\beta_1\dot\beta_2\ldots\dot\beta_r}\,,
	\hskip 1.5 cm 
	\phib_s = \phib^{\alpha_1\alpha_2\ldots\alpha_l}_{\dot\beta_1\dot\beta_2\ldots\dot\beta_r}\,,
\end{align}
where $l+r=2s$ and $\phi_s$ and $\phib_s$ are symmetric in the $\alpha_i$ and $\dot\beta_i$ indices, which transform in the two-dimensional representation of $SU(2)_L$ and $SU(2)_R$, respectively.  
The covariantized free Lagrangian $\Lmin$ in \cref{eq:Lmin} treats uniformly all the representations of the rotation group that are part of $\phi_s$. Thus, the propagator derived from $\Lmin$ is proportional to the identity operator $\identity_{(l,r)}$ in the $(l,r)$ representation. For $\phi_s$ in the $(s,s)$ representation, it is
\begin{align}\label{eq:props}
	\begin{tikzpicture}[baseline={([yshift=-0.27cm]current bounding box.center)},every node/.style={font=\footnotesize}]
		\draw [very thick] (0,0) node[above=0pt]{$\mu(s)$} -- (1.5,0) node[above=0pt]{$\nu(s)$};
	\end{tikzpicture} = \frac{(-1)^s i\,\delta_{\mu(s)}^{\nu(s)}}{p^2-m^2}\,, \qquad\qquad
        \delta_{\mu(s)}^{\nu(s)}\equiv\delta_{\mu_1}^{(\nu_1}\delta_{\mu_2}^{\nu_2}\ldots\delta_{\mu_s}^{\nu_s)} \equiv \identity_{(s,s)}  \,.
\end{align}
%
%
Consequently, there is no explicit dependence on the value of $s$ in the amplitudes that follow from  $\Lmin$, making the large-spin limit appropriate for classical physics convenient in this construction.

When evaluated on an $(s,s)$ representation, the Lagrangian \eqref{eq:Lmin} contains propagating degrees of freedom beyond the $2s+1$ associated with a 
single massive spin-$s$ particle and some of them have negative norm in Minkowskian signature. While such a theory is not consistent as a quantum theory because of difficulties with unitarity, we use this Langangian and its nonminimal extension described below to find only classical observables, so that the issue  is not directly relevant.
One may nevertheless worry that the negative norm states might cause some inconsistency even in the classical limit, and very likely they are the origin of the additional Wilson coefficients~\cite{Kim:2023drc}. 
As we will see in \cref{sec:WilsonCoeffs}, the key to the additional Wilson coefficients is the presence of propagating degrees of freedom beyond those of a single quantum spin-$s$ particle. This is independent of the sign of the norm of the extra states. Moreover, there is a direct simple map which connects amplitudes in this theory with amplitudes in a theory in which all states have positive norm. 

\ref{fieldtheory2} is designed to probe the relation between the extra Wilson coefficients and the presence of states beyond those of a spin-$s$ representation of the rotation group.
To define it and to compare straightforwardly with the Singh-Hagen Lagrangian~\cite{Singh:1974qz} for a single spin-$s$ particle it is convenient to choose $\phi_s$ in the $(s, s)$ representation, which is realized as a 
symmetric traceless rank-$s$ tensor,\footnote{\label{symmetrization_footnote}\baselineskip=14pt Throughout the paper, the symmetrization is defined as $f_{(\mu_1\mu_2\ldots\mu_s)}\equiv\frac{1}{s!}(f_{\mu_1\mu_2\ldots\mu_s}+\text{ permutations})$.}
\begin{align}
	\phi_s \equiv  \phi_{\alpha_1\alpha_2\ldots\alpha_s}^{\dot\beta_1\dot\beta_2\ldots\dot\beta_s} \propto            
	\phi^{(\mu_1\mu_2\ldots \mu_s)}(\sigma_{\mu_1}){}_{(\alpha_1}{}^{(\dot\beta_1}\dots (\sigma_{\mu_s}){}_{\alpha_s)}{}^{\dot\beta_s)} \,,
\end{align}
where as usual the parenthesis on the indices signify that they are symmetrized.
We primarily focus on this representation in subsequent sections, especially when carrying out calculations at fixed values of the spin. 

We ensure that only the $2s+1$ states of a spin-$s$ field are propagating by imposing the requisite constraints with auxiliary fields, following the strategy of Ref.~\cite{Singh:1974qz}. The net effect of imposing transversality is that the minimal Lagrangian $\Lmin$ in Eq.~\eqref{eq:Lmin} is modified to 
\begin{align}\label{eq:Lspins}
	\Lspins = - (-1)^s\Big[\phi_s (D^2+m^2)\phib_s + s (D\phi_s)(D\phib_s)+\ldots\Big] \, ,
\end{align}
where $(D\phi_s) \equiv D_{\mu}\phi^{\mu\mu_2\ldots\mu_s}$ and the ellipsis stand for terms that remove unwanted states, as explained below.

The coupling $s(D\phi_s)(D\phib_s)$ originates from integrating out an auxiliary $\phi_{s-1}$ field that impose transversality via the equation of motion. To see this, we add to free part of $\Lmin$ 
the term $a \phi_{s-1} \partial \phi_s$ as well as a standard quadratic term for $\phi_{s-1}$, where $a$ is a normalization. The equations of motion are
\begin{align}\label{eq:aux1}
	(\partial^2+m^2)\phi_s^{\mu_1\mu_2\ldots\mu_s}=a\, \partial^{(\mu_1}\phi^{\mu_2\ldots\mu_s)}_{s-1}\,,\qquad (b\,\partial^2+c\,m^2)\phi_{s-1}=(\partial\phi_s)\,,
\end{align}
where we introduced two additional normalization constants $b$ and $c$.  A solution to the equation of motion is $\phi_{s-1}=\partial\phi_s=0$. Requiring that this is the only solution gives $b=0$ and $a=scm^2$ such that
\begin{align}\label{eq:aux2}
	(\partial^2+m^2)\phi_s^{\mu_1\mu_2\ldots\mu_s} = s\,\partial^{(\mu_1}(\partial\phi_s)^{\mu_2\ldots\mu_s)}\, .
\end{align}
Covariantization with respect to the photon gauge symmetry follows as usual, by replacing the partial derivatives with the appropriate covariant derivatives, leading to the $s(D\phi_s)(D\phib_s)$ term in \cref{eq:Lspins}.

The process continues, as transversality of $\phi_s$ implies $\partial\partial\phi_s=0$, which must also be imposed 
through an equation of motion. More auxiliary fields are therefore needed, and this process can be carried out recursively~\cite{Singh:1974qz}.  The resulting couplings involving traces, multiple-divergences like $D_{\mu}D_{\nu}\phi^{\mu\nu\mu_3\ldots\mu_s}$, and auxiliary fields with lower spins are collected in the ellipsis in \cref{eq:Lspins}.
Up to $s=3$, the Lagrangians generated by this procedure are
\begingroup
\allowdisplaybreaks
\begin{subequations}\label{eq:L_low_spin}
\begin{align}
	\mathcal{L}_{s=1} &=  \phi^{\mu_1} (D^{2}+m^2) \phib_{\mu_1} + (D_{\mu}\phi^{\mu})(D^{\nu}\phib_{\nu})\,, \\
	\mathcal{L}_{s=2} &= -\phi^{\mu_1\mu_2} (D^2+m^2) \phib_{\mu_1\mu_2} - 2 (D_{\mu}\phi^{\mu\mu_2})(D^{\nu}\phib_{\nu\mu_2}) + \phi_{\mu}{}^{\mu}(D^2+m^2)\phib{}^{\nu}{}_{\nu} \nonumber\\*
	&\quad - \phi_{\mu}{}^{\mu}D^{\rho}D^{\lambda}\phib_{\rho\lambda} - \phib{}^{\mu}{}_{\mu}D_{\rho}D_{\lambda}\phi^{\rho\lambda}\,, \\
	\label{eq:Ls3}
	\mathcal{L}_{s=3} &= \phi^{\mu_1\mu_2\mu_3} (D^2+m^2)\phib_{\mu_1\mu_2\mu_3} + 3 (D_{\mu}\phi^{\mu\mu_2\mu_3})(D^{\nu}\phib_{\nu\mu_2\mu_3}) \nonumber\\* 
	&\quad - 3 \phi_{\mu}{}^{\mu\mu_3} (D^2+m^2)\phib{}^{\nu}{}_{\nu\mu_3} + 3 \phi_{\mu}{}^{\mu\mu_3}D^{\rho}D^{\lambda}\phib_{\rho\lambda\mu_3}+3\phib{}^{\mu}{}_{\mu\mu_3}D_{\rho}D_{\lambda}\phi^{\rho\lambda\mu_3}\nonumber\\*
	&\quad +\frac{3}{2}(D_{\mu}\phi^{\mu\rho}{}_{\rho})(D_{\nu}\phib{}^{\nu\lambda}{}_{\lambda}) + 2 \varphi (D^2+4m^2)\bar{\varphi} + m (\varphi D_{\mu}\phib{}^{\mu\lambda}{}_{\lambda}+\bar{\varphi}D_{\mu}\phi^{\mu\lambda}{}_{\lambda})\,,
\end{align}
\end{subequations}
\endgroup
where $\varphi$ and $\bar{\varphi}$  in $\mathcal{L}_{s=3}$ are ghost-like scalar auxiliary fields. The $\mathcal{L}_{s=1}$ and $\mathcal{L}_{s=2}$ here are the Proca~\cite{Proca:1936fbw} and Fierz-Pauli Lagrangian~\cite{Fierz:1939ix}, respectively,
and $\mathcal{L}_{s=3}$ was first obtained by Chang~\cite{Chang:1967zzc}. 
We note that the construction in Ref.~\cite{Singh:1974qz} uses only symmetric and traceless fields, and we have absorbed certain auxiliary fields into the trace of $\phi_s$. We   use \cref{eq:Lspins} --- which is the arbitrary-spin generalization of Eq.~\eqref{eq:L_low_spin} --- as the covariantization of the free Lagrangian of \ref{fieldtheory2}.

\ref{fieldtheory3} is constructed to probe whether the extra Wilson coefficients in \ref{fieldtheory1} are due to the unphysical nature of the extra states of this theory. Thus, we define the covariantization of the free part of \ref{fieldtheory3} as being given, up to nonminimal terms, by the sum of Lagrangians for physical transverse fields with spins $s$, $s-1$, $\dots, 0$.
For simplicity, here we consider a Lagrangian that involves only spin $s$ and $s-1$,
\begin{align}
\label{twospins}
	\Ltwospin = \mathcal{L}^s_{\text{min}}+\mathcal{L}^{s-1}_{\text{min}}
	&=- (-1)^s\Big[\phi_s (D^2+m^2)\phib_s + s (D\phi_s)(D\phib_s)+\ldots\Big] \\
	&\quad -(-1)^{s-1}\Big[\phi_{s-1} (D^2+m^2)\phib_{s-1} + (s-1) (D\phi_{s-1})(D\phib_{s-1})+\ldots\Big]\, .
	\nonumber
\end{align}
We show below that this Lagrangian is sufficient to describe classical physics at $\mathcal{O}(S^1)$ up to the one-loop order. 
We assume that $\phi_s$ and $\phi_{s-1}$ have the same minimal coupling to the photon. Somewhat loosely, one may interpret this Lagrangian as being obtained from Eq.~\eqref{eq:Lmin} upon separating $\phi_s$ into fields obeying transversality constraints and dropping the derivative factors that are responsible for the negative norms of the $s-(2k+1)$
components. 

The minimal Lagrangians 
of \ref{fieldtheory2} and \ref{fieldtheory3} make explicit reference to the value of $s$, as can be seen in the explicit expressions in \cref{eq:L_low_spin}, and consequently the propagators (and vertices) have the same property.
The propagators for massive $s=1$ and $s=2$ fields can be easily derived from the quadratic part of $\mathcal{L}_{s=1}$ 
and $\mathcal{L}_{s=2}$. They are
\begin{align}
	\begin{tikzpicture}[baseline={([yshift=-0.27cm]current bounding box.center)},every node/.style={font=\footnotesize}]
		\draw [very thick] (0,0) node[above=0pt]{$\mu$} -- (1.3,0) node[above=0pt]{$\nu$};
		\node at (0,0) [above=0pt]{$\phantom{\mu_1\mu_2}$};
		\node at (1.3,0) [above=0pt]{$\phantom{\nu_1\nu_2}$};
	\end{tikzpicture} &= \frac{-i\mathcal{P}_{\mu,\nu}}{p^2-m^2} = \frac{-i \Theta_{\mu\nu}}{p^2-m^2}=\frac{-i}{p^2-m^2}\left(\eta_{\mu\nu}-\frac{p_{\mu}p_{\nu}}{m^2}\right)\,,\\
	\begin{tikzpicture}[baseline={([yshift=-0.27cm]current bounding box.center)},every node/.style={font=\footnotesize}]
		\draw [very thick] (0,0) node[above=0pt]{$\mu_1\mu_2$} -- (1.3,0) node[above=0pt]{$\nu_1\nu_2$};
	\end{tikzpicture} &= \frac{i\mathcal{P}_{\mu_1\mu_2,\nu_1\nu_2}}{p^2-m^2}=\frac{i}{p^2-m^2}\,\frac{1}{2}\left[\Theta_{\mu_1\nu_1}\Theta_{\mu_2\nu_2}+\Theta_{\mu_1\nu_2}\Theta_{\mu_2\nu_1}-\frac{2}{3}\Theta_{\mu_1\mu_2}\Theta_{\nu_1\nu_2}\right],
\end{align}
where $\Theta_{\mu\nu}=\eta_{\mu\nu}-\frac{p_{\mu}p_{\nu}}{m^2}$.
The numerators are instances of the spin-$s$ state projector $\mathcal{P}$; its general closed-form expression~\cite{Singh:1974qz},
\begin{align}\label{eq:state_proj}
	\mathcal{P}_{\mu(s)}^{\nu(s)}=\sum_{j=0}^{\lfloor s/2\rfloor}\frac{(-1)^j s! (2s-2j-1)!!}{2^j j! (s-2j)! (2s-1)!!}\Theta_{(\mu_1\mu_2}\Theta^{(\nu_1\nu_2}\ldots\Theta_{\mu_{2j-1}\mu_{2j}}\Theta^{\nu_{2j-1}\nu_{2j}}\Theta_{\mu_{2j+1}}^{\nu_{2j+1}}\ldots\Theta_{\mu_s)}^{\nu_s)}\,,
\end{align}
is manifestly symmetric, transverse and traceless on-shell. 

Beyond $s=2$ the off-diagonal nature of the quadratic terms in  $\mathcal{L}^{s}_\text{min}$ makes the construction of propagators
more involved. For example, the $\mathcal{L}^{s=3}_\text{min}$ Lagrangian contains quadratic mixing between $\phi_{\mu_1\mu_2\mu_3}$ and the auxiliary scalar $\varphi$; thus to derive the propagators it is necessary to diagonalize the quadratic terms, effectively summing over all possible insertions of such two-point vertices. 
We represent the resummed propagators by a cross in the middle,
\begin{subequations}\label{eq:props3}
	\begin{align}
		\begin{tikzpicture}[baseline={([yshift=-0.25cm]current bounding box.center)},every node/.style={font=\footnotesize}]
			\draw [very thick] (0,0) node[above=0pt]{$\mu_1\mu_2\mu_3$} -- (2,0) node[above=0pt]{$\nu_1\nu_2\nu_3$};
			\node at (1,0) {$\times$};
		\end{tikzpicture}&=\frac{-i\,\mathcal{P}_{\mu_1\mu_2\mu_3}^{\nu_1\nu_2\nu_3}}{p^2-m^2}+\frac{i\,\mathcal{Q}_{\mu_1\mu_2\mu_3}^{\nu_1\nu_2\nu_3}}{40m^6} \,,
         \\[8pt]
		\begin{tikzpicture}[baseline={([yshift=-0.25cm]current bounding box.center)},every node/.style={font=\footnotesize}]
			\draw [very thick] (0,0) node[above=0pt]{$\mu_1\mu_2\mu_3$} -- (1,0);
			\draw [very thick, dashed] (1,0) -- (2,0) node[above=0pt]{$\phantom{\nu_1\nu_2\nu_3}$};
			\node at (1,0) {$\times$};
		\end{tikzpicture}&=\frac{(p^2-m^2)\eta_{(\mu_1\mu_2}p_{\mu_3)}-4p_{\mu_1}p_{\mu_2}p_{\mu_3}}{40m^5}\,, 
       \\[8pt]
		\begin{tikzpicture}[baseline={([yshift=-0.25cm]current bounding box.center)},every node/.style={font=\footnotesize}]
			\draw [very thick, dashed] (0,0) node[above=0pt]{$\phantom{\mu_1\mu_2\mu_3}$} -- (2,0) node[above=0pt]{$\phantom{\nu_1\nu_2\nu_3}$};
			\node at (1,0) {$\times$};
		\end{tikzpicture}&= \frac{i(p^2+5m^2)}{40m^4}\,.
	\end{align}
\end{subequations}
Apart from the non-local term encoding the energy-momentum relation, the propagator for the physical $s=3$ particle also 
has an additional local contribution, with tensor structure
\begin{align}
	\mathcal{Q}_{\mu_1\mu_2\mu_3}^{\nu_1\nu_2\nu_3}=\eta_{(\mu_1\mu_2}p_{\mu_3)}\eta^{(\nu_1\nu_2}p^{\nu_3)}(p^2-7m^2)-4p_{\mu_1}p_{\mu_2}p_{\mu_3}\eta^{(\nu_1\nu_2}p^{\nu_3)}-4\eta_{(\mu_1\mu_2}p_{\mu_3)}p^{\nu_1}p^{\nu_2}p^{\nu_3}\,.
\end{align}
Meanwhile, the contribution from the auxiliary field is completely local, indicating that they carry no physical (asymptotic) degrees of freedom, as expected.


\subsection{Classical Asymptotic States and Coherent States}
\label{StatesSubsection}

The standard description of the asymptotic states of (massive) spinning fields is in terms of Lorentz tensors labeled by 
the (massive) little group. Extending Ref.~\cite{Bern:2020buy}, we first consider the asymptotic state $\polM$ and its conjugate $\polMb$ for general $(l,r)$ representations of the Lorentz group with $l+r=2s$.  Subsequently, we specialize to integer $s$ and consider the $(s,s)$ representation, in which we identify general consequences of transversality. In the classical limit, these states are chosen to minimize the dispersion of the Lorentz generators, $\polM\cdot M^{\mu\nu}\cdot\polMb$, where $M^{\mu\nu}$ satisfies the Lorentz algebra
\begin{align}\label{eq:Lorentz_Alg}
	[M^{\mu\nu},M^{\rho\lambda}] = -i (\eta^{\mu\rho}M^{\lambda\nu}+\eta^{\nu\rho}M^{\mu\lambda}-\eta^{\mu\lambda}M^{\rho\nu}-\eta^{\nu\lambda}M^{\mu\rho})\,.
\end{align}
In the rest frame, the state $\polM$ generalizes the spin coherent states of $SU(2)$ implicit in the construction of Ref.~\cite{Bern:2020buy} to those of $SU(2)_L\times SU(2)_R$.
We start by representing the states in terms of spinors such that\footnote{\baselineskip=14pt We note that the $SU(2)$ indices are raised and lowered by $$\epsilon_{\alpha\beta}=\epsilon_{\dot\alpha\dot\beta}=\left(\begin{array}{cc}
	0 & -1 \\
	1 & 0
\end{array}\right)\qquad\epsilon^{\alpha\beta}=\epsilon^{\dot\alpha\dot\beta}=\left(\begin{array}{cc}
0 & 1 \\
-1 & 0
\end{array}\right). $$} 
\begin{align}\label{polM}
\polM(p)_{\alpha(l)\dot\beta(r)} & =  \xi(p)_{\alpha_1}\dots 
 \xi(p)_{\alpha_l}  \chi(p)_{\dot\beta_1}\dots  \chi(p)_{\dot\beta_r}\,,\nonumber
\\
\polMb(p)^{\alpha(l)\dot\beta(r)} & = \tilde \xi(p)^{\alpha_1}\dots 
\tilde \xi(p)^{\alpha_l} \tilde\chi(p)^{\dot\beta_1}\dots \tilde\chi(p)^{\dot\beta_r}\,.
\end{align}
Here, we choose $\polM$ to be null for convenience. We note that the final result does not rely on $\polM$ being null.
As with the spin coherent states, the spinors $\xi(p)$, $\chi(p)$, $\tilde\xi(p)$ and $\tilde\chi(p)$ are constructed by boosting their rest frame 
counterparts $\xi_{0}$, $\chi_0$, ${\tilde\xi}_{0}$ and ${\tilde \chi}_0$,
\begin{align}
\xi(p)_{\alpha} &= \exp(i\eta {\hat p}^k {\hat K}{}^k_L)_{\alpha}{}^{\beta}\xi_{0\beta}\,,
&\quad
\tilde\xi(p)_{\alpha}& = \exp(i\eta {\hat p}^k {\hat K}{}_{L}^k)_{\alpha}{}^{\beta}\tilde\xi_{0\beta} \, ,\nonumber
\\
\chi(p)^{\dot\alpha} &= \exp(i\eta {\hat p}^k {\hat K}{}^k_R)^{\dot\alpha}{}_{\dot\beta}\chi_{0}^{\dot\beta}\,,
&\quad
\tilde\chi(p)^{\dot\alpha} &= \exp(i\eta {\hat p}^k {\hat K}{}_{R}^k)^{\dot\alpha}{}_{\dot\beta}\tilde\chi_{0}^{\dot\beta} \, ,
\label{boosts}
\end{align}
where $ (\hat{K}^{k}_{L})_{\alpha}{}^{\beta}=(i/2)(\sigma^k)_{\alpha}{}^{\beta}$ and $(\hat{K}^k_{R})^{\dot\alpha}{}_{\dot\beta}=(-i/2)(\sigma^k)^{\dot\alpha}{}_{\dot\beta}$ are the left/right-handed boost operators, $\eta$ is the 
rapidity and ${\hat p}^{k}$ are the components of the unit vector along the spatial part of the momentum. 

The rest frame coherent-state spinors are~\cite{SpinCohSt} 
\begin{align}
\xi_{0\alpha} & = \exp(z_L \hat N^L_+ - z_L^* \hat N^L_-){}_\alpha{}^\beta \xi^+_{0\beta}\,,
\qquad
&\tilde\xi_{0\alpha} & = \exp(z_L \hat N^L_+ - z_L^* \hat N^L_-)_{\alpha}{}^{\beta} \xi^{-}_{0\beta}\,,
\nonumber\\
\chi_{0}^{\dot\alpha} &= \exp(z_R \hat N^R_+ - z_R^* \hat N{}^R_-)^{\dot\alpha}{}_{\dot\beta} \chi^{+,\dot\beta}_{0}\,,
\qquad
&\tilde\chi_{0}^{\dot\alpha} &= \exp(z_R \hat N^R_+ - z_R^* \hat N{}^R_-)^{\dot\alpha}{}_{\dot\beta} \chi^{-,\dot\beta}_{0}\,,
\end{align}
where $(\hat N^L_{\pm})_{\alpha}{}^{\beta}=(1/2)(\sigma^1\pm i\sigma^2)_{\alpha}{}^{\beta}$ and $(\hat N^R_{\pm})^{\dot\alpha}{}_{\dot\beta}=(1/2)(\sigma^1\pm i\sigma^2)^{\dot\alpha}{}_{\dot\beta}$ are the generators of $SU(2)_L$ and $SU(2)_R$, $\xi_{0}^{\pm}$ and 
$\chi_{0}^{\pm}$
are the eigenvectors of $\sigma^3$ with eigenvalues $\pm1$, and
\begin{equation}
z_{L,R} \equiv -(\theta_{L,R}/2)e^{-i\phi_{L,R}}\,,
\end{equation}
are coherent-state parameters. The rest frame spinors are normalized as $\xi_0^{\alpha}\tilde{\xi}_{0\alpha}=\chi_{\dot\alpha}\tilde\chi^{\dot\alpha}=-1$, such that  $\polM(p)\cdot\polMb(p) = (-1)^r$. They are related to unit vectors via
\begin{align}
\label{unitvectors}
& n_L^i = \xi_{0}^\alpha (\sigma^i){}_\alpha{}^\beta \tilde \xi_{0}{}_\beta \equiv 
\xi_{0} \sigma^i \tilde \xi_{0}\,, 
& &
\pmb{n}_L=(\sin\theta_L\cos\phi_L, \sin\theta_L\sin\phi_L, \cos\theta_L)\,,\nonumber
\\
& n_R^i =  \chi_{0}{}_{\dot\alpha} (\sigma^i){}^{\dot\alpha}{}_{\dot\beta}  \tilde \chi_{0} ^{\dot\beta}
\equiv  \chi_{0} \sigma^i  \tilde \chi_{0} \,,
& &
\pmb{n}_R=(\sin\theta_R\cos\phi_R, \sin\theta_R\sin\phi_R, \cos\theta_R)
\,.
\end{align}
The rotation and boost generators in the $(l,r)$ representation is given by
\begin{align}\label{SandK}
\hat{\pmb{S}} &= \hat{\pmb{S}}_L + \hat{\pmb{S}}_R\,,
\qquad
\hat{\pmb{K}} = \hat{\pmb{K}}_L+ \hat{\pmb{K}}_R\,,\nonumber
\\
\hat{S}^{k}_L &= \frac{1}{2}\sum_{m=1}^l \underbrace{\id\otimes \dots \id}_{m-1}\otimes\; \sigma^k \otimes \id\dots\otimes \id\,,
\qquad
\hat{S}^k_R = \frac{1}{2}\sum_{m=1}^r \underbrace{\id\otimes \dots \id}_{m-1}\otimes\; \sigma^k \otimes \id\dots\otimes \id\,,
\\
\hat{K}^k_L &= \frac{i}{2}\sum_{m=1}^l \underbrace{\id\otimes \dots \id}_{m-1}\otimes\; \sigma^k \otimes \id\dots\otimes \id\,,
\qquad
\hat{K}^k_R = -\frac{i}{2}\sum_{m=1}^r \underbrace{\id\otimes \dots \id}_{m-1}\otimes\; \sigma^k \otimes \id\dots\otimes \id\, ,\nonumber
\end{align}
where the summation is over the position of $\sigma^k$.
With these definitions, the expectation values of the rotation and boost generator under the rest frame spin coherent states are
\begin{align}
\polM_{0} \cdot \hat{\pmb{S}} \cdot \polMb_{0}  =  \frac{1}{2} (l\, \pmb{n}_L+r\,\pmb{n}_R)  \equiv \pmb{S}\,,
\qquad
\polM_{0} \cdot \hat{\pmb{K}} \cdot \polMb_{0}  =\frac{i}{2} (l\,\pmb{n}_L - r\,\pmb{n}_R)  \equiv i\pmb{K} \, .
\label{vevs}
\end{align}
We identify the former with the classical rest-frame spin vector $S^{\mu}_0=(0,\pmb{S})$ and the latter with the boost vector $K^{\mu}_0=(0,\pmb{K})$. 
If $\polM$ is not null, we view \cref{vevs} as the definition for the classical spin and boost vector.
The classical rest-frame spin tensor given by 
\begin{align}\label{nonSSCspintensor}
 \gS^{\mu\nu}_{0} = \polM_{0} \cdot M^{\mu\nu} \cdot \polMb_{0} = S^{\mu\nu}_0 + i K^{\mu\nu}_0 \,,
\end{align}
does not obey the SSC, where 
\begin{align}
\label{eq:FTSpinTDef}
S^{\mu\nu}_0=\frac{1}{m}\lc^{\mu\nu\rho\lambda}p_{0\rho}S_{\lambda}\,, 
\hskip 1.5 cm 
 K^{\mu\nu}_{0} = \frac{1}{m}(p_{0}^\mu K_0^\nu - p_0^\nu K_0^\mu)\,.
\end{align}
It contains\footnote{\baselineskip=14 pt Our convention for the Levi-Civita tensor is $\epsilon_{0123}=+1$.} an SSC-obeying component $S^{\mu\nu}_0$ and an SSC-violating one 
$K^{\mu\nu}_{0}$, where $p_0^{\mu} = (m, 0)$ is 
the rest-frame momentum.
An important feature for generic $(l,r)$ representations is that $\pmb{K}$ no longer vanishes identically, so that ${\gS}^{\mu\nu}_0$ no longer satisfies the convariant SSC condition. By suitably choosing $l$ and $r$, the norm $|\pmb{K}|$ can be subleading in the classical limit or commensurate with that of the spin vector.  In this way, the appearance of $\pmb{K}$ in the classical limit appears natural, simply by adjusting the Lorentz representation in the underlying quantum system. For generic values, the classical limit is independent of the details of the representation.  However, for the special case of the irreducible transverse $(s,s)$ representation then $\pmb{K}$ vanishes, as noted in Appendix C of Ref.~\cite{Chung:2019duq}.

The next step is to restore the momentum dependence of various quantities by boosting the particle out of the rest frame.
It is somewhat tedious but straightforward to use \cref{boosts} and the properties of the Pauli matrices to boost products of polarization tensors and Lorentz generators for any $(l,r)$ representation, as well as \cref{nonSSCspintensor} and its two components to arbitrary frames. To leading order in the classical limit, we find
\begin{align}\label{eq:generalS}
& \polM_1 \cdot \{M^{\mu_1\nu_1 } , \dots , M^{\mu_n\nu_n } \} \cdot \polMb_2
=   \gS(p_1)^{\mu_1\nu_1}\dots \gS(p_n)^{\mu_n\nu_n} 
\polM_1 \cdot \polMb_2 + {\cal O}(q^{1-n}) \, ,
\end{align} 
where $\polM_i\equiv\polM(p_i)$, $q=p_2-p_1$ is the momentum transfer 
and $\gS(p_i)^{\mu\nu}$ is the boost of \cref{nonSSCspintensor} to the frame moving with momentum $p_i$. The spin tensor scales as $\gS\sim q^{-1}$, and we neglect all the subleading $\mathcal{O}(q^{1-n})$ terms. The symmetric product of Lorentz generators $\{M^{\mu_1\nu_1 } , \dots , M^{\mu_n\nu_n } \}$ is defined as
\begin{align}\label{eq:symM}
	&\{M_{\mu_1\nu_1},M_{\mu_2\nu_2}\}=\frac{1}{2}(M_{\mu_1\nu_1}M_{\mu_2\nu_2}+M_{\mu_2\nu_2}M_{\mu_1\nu_1})\,,\nonumber\\
	&\{M_{\mu_1\nu_1},M_{\mu_2\nu_2},\ldots,M_{\mu_n\nu_n}\}=\frac{1}{n!}(M_{\mu_1\nu_1}M_{\mu_2\nu_2}\ldots M_{\mu_n\nu_n}+\text{permutations})\, .
\end{align}
They form a basis for arbitrary product of Lorentz generators under the Lorentz algebra.
The factorization \eqref{eq:generalS} of the expectation value of the product of Lorentz generators into the product of individual expectation values is a reflection of the classical nature of the asymptotic states $\polM$ and $\polMb$. In \cref{{eq:generalS}}, the product of polarization tensors is given by 
\begin{align}
	(-1)^r\polM_1 \cdot \polMb_2
	= \exp\left[-\frac{1}{m} \pmb{q}\cdot \pmb{K}   \right]
	\exp\left[ - i   \frac{\epsilon_{r
			i j k }u_1^i q^j S^k}{m(1 +\sqrt{1+\pmb{u}_1^2})}+{\cal O}(q^2) \right] +{\cal O}(q) \, ,
	\label{epdotep}
\end{align}
where $u^k_i = p^k_i/m$, generalizing the corresponding expression in Ref.~\cite{Bern:2020buy} to general $\pmb{K}$. 
Eq.~\eqref{epdotep} captures the leading terms in the classical limit and, apart from the sign on the left-hand side, it is agnostic to the $(l,r)$ representation
chosen for the fields.

\subsection{The Transverse \texorpdfstring{$(s,s)$}{(s,s)} Representation}
\label{sec:ss_rep}

We now consider the special case of the $(s,s)$ representation, which corresponds to symmetric-traceless fields. The coherent-state polarization 
tensors have an equal number of dotted and undotted indices; in the rest frame, they can be written as
\begin{align}\label{eq:spin_coherent}
	(\polM_{0}^{(s)})_{\alpha(s)\dot\beta(s)} =(\polM_{0}^{(s)})^{\mu_1\mu_2\ldots\mu_s}(\sigma_{\mu_1})_{\alpha_1\dot\beta_1}\ldots(\sigma_{\mu_s})_{\alpha_s\dot\beta_s}=\xi_{0\alpha_1}\ldots\xi_{0\alpha_s}\chi_{0\dot\beta_1}\ldots\chi_{0\dot\beta_s}\,.
\end{align}
With this definition, we can explore the additional restrictions on the coherent states required by the transversality of $\polM_{0}^{(s)}$. It suffices to analyze it in the rest frame, where it reads,\footnote{\baselineskip=14pt In this form transversality can be imposed on the polarization tensor of a general $(l\ne 0 ,r\ne 0)$ state.}
\begin{align}
	p_{0\mu} \polM_{0}^{\mu \mu_2\ldots\mu_s} = 0
	\qquad\Longleftrightarrow\qquad
	(p_{0\mu} \sigma^\mu ){}^\alpha{}_{\dot\beta} (\polM_{0})_{\alpha\alpha_2\dots\alpha_s}^{\dot \beta\dot\beta_2 \dots\dot\beta_s} = 0 \, .
\end{align}
Using the explicit form of the rest-frame momentum, $p_0=(m,0,0,0)$, and that $(\sigma^0){}^\alpha{}_{\dot\beta}$ is numerically equal to the $2\times 2$ Levi-Civita, it follows that 
\begin{align}
	0 = (p_{0\mu} \sigma^\mu ){}^\alpha{}_{\dot\beta} (\polM_{0})_{\alpha\alpha_2\dots\alpha_s}^{\dot \beta\dot\beta_2 \dots\dot\beta_s} \propto \xi_{0\alpha}\epsilon^{\alpha}{}_{\dot\alpha}\chi_{0}^{\dot\alpha}  \, .
\end{align}
The solution, accounting for normalization, is 
\begin{align}\label{eq:trans_condition}
	\xi_{0\alpha} = \chi_{0}^{\dot\alpha} \quad\text{as column vectors,}
\end{align}
which in turn implies $z_{L}=z_R$ and hence the equality of the left-handed and right-handed unit vectors $\pmb{n}_L$ and $\pmb{n}_R$ in \cref{unitvectors}.
Together with \cref{vevs} this implies that
\begin{align}
	\pmb{K} = 0
	\quad
	\Longleftrightarrow
	\quad
	{\gS}^{\mu\nu}_0 = S^{\mu\nu}_0 \, ,
	\label{transverse_SSC}
\end{align}
for the transverse $(s,s)$ representation, and therefore, cf.~\cref{nonSSCspintensor},  $\polM_{0}^{(s)}\cdot M^{\mu\nu} \cdot \polMb_{0}^{(s)}$ becomes an SSC-satisfying spin tensor. 
On the other hand, if we do not impose transversality, then the discussion for a generic $(l,r)$ representation also applies to $(s,s)$, such that $\pmb K$ does not vanish and hence that covariant SSC is not obeyed. 
We thus see that the $(s, s)$ transverse asymptotic states chosen in
Ref.~\cite{Bern:2020buy} can be replaced with more general nontransverse ones.
%
%
The polarization tensor for the transverse $(s,s)$ representation can be written as a direct product of transverse $s=1$ coherent state vectors
\begin{align}\label{eq:pol_vec}
	\polM^{(s)}(p)^{\mu_1\mu_2\ldots\mu_s} = \polm(p)^{\mu_1}\polm(p)^{\mu_2}\ldots\polm(p)^{\mu_s}\,,\qquad
	\polm(p)^{\mu}(\sigma_\mu)_{\alpha\dot\beta} = \xi_{\alpha}(p) \chi_{\dot\beta}(p)\,,
\end{align}
where the spinors are boosted from the rest frame ones that satisfy the condition~\eqref{eq:trans_condition}, and we normalize the polarization vectors as $\polm\cdot\polmb=-1$. For such external states, the expectation value of \cref{eq:generalS} becomes
\begin{align}
	K^{\mu\nu}(p) = 0 
	\quad
	\Longleftrightarrow
	\quad
	\gS^{\mu\nu}(p) = S^{\mu\nu}(p) \,,	\label{generalframeSSC}
\end{align}
which is simply the counterpart of the rest frame relation~\eqref{transverse_SSC}. 
The product~\eqref{epdotep} simplifies to
\begin{align}\label{edotep_trans}
	(-1)^s\polM^{(s)}_1\cdot\polMb^{(s)}_2=(-\polm_1\cdot\polmb_2)^{s}=\exp\left[ - i   \frac{\epsilon_{rs k }u_1^r q^s S^k}{m(1 +\sqrt{1+\pmb{u}_1^2})}+{\cal O}(q^2) \right] +{\cal O}(q)\,.
\end{align}

The transverse $(s,s)$ representation is used in \ref{fieldtheory2} and \ref{fieldtheory3}. 
Because the Lagrangian depends explicitly on $s$, we need to use the explicit form of the Lorentz generators, 
\begin{align}\label{eq:LorentzGen}
	(M^{\mu\nu})_{\alpha(s)}{}^{\beta(s)}=-2i\delta_{(\alpha_1}^{[\mu}\eta^{\nu](\beta_1}\delta_{\alpha_2}^{\beta_2}\ldots\delta_{\alpha_s)}^{\beta_s)}\,.
\end{align}
Consequently, the results are given in terms of various symmetric and antisymmetric combinations of the polarization vector $\polm$ and momenta. To convert them into spin tensors, we need to compute the left-hand side of \cref{eq:generalS} and identify the resulting structures with spin tensors. 

We first consider the transverse $\polM^{(s)}$. Starting with $\mathcal{O}(S^1)$, we have
\begin{align}\label{eq:esMes}
	\polM_1^{(s)}\cdot  M^{\mu\nu}\cdot \polMb_2^{(s)}&=-2is(\polm_1\cdot\polmb_2)^{s-1} \polm_1^{[\mu}\polmb_2^{\nu]}\,.
\end{align}
According to \cref{eq:generalS}, this combination should be identified with $S(p_1)^{\mu\nu}$, such that 
\begin{align}\label{eq:S1rep}
	(\polm_1\cdot\polmb_2)^{s-1}\polm_1^{[\mu}\polmb_2^{\nu]} = \frac{i S(p_1)^{\mu\nu}}{2s}(\polm_1\cdot\polmb_2)^{s} + \mathcal{O}(q^0)\,.
\end{align}
In amplitudes, we can use this relation to turn antisymmetric combination of polarization vectors into spin tensors. The classical amplitude is obtained by further taking the $s\rightarrow\infty$ limit. Similarly, at $\mathcal{O}(S^2)$, we can use the following identity, 
\begin{align}
	\polM_1^{(s)}\cdot\{M^{\mu\nu},M^{\rho\lambda}\}\cdot\polMb_4^{(s)} &= -4s(s-1)(\polm_1\cdot\polmb_4)^{s-2}\polm_1^{[\mu}\polmb_4^{\nu]}\polm_1^{[\rho}\polmb_4^{\lambda]}\\
	&\quad - s (\polm_1\cdot\polmb_4)^{s-1}\left(\eta^{\mu\lambda}\polm_1^{(\nu}\polmb_4^{\rho)}+\eta^{\nu\rho}\polm_1^{(\mu}\polmb_4^{\lambda)}-\eta^{\mu\rho}\polm_1^{(\nu}\polmb_4^{\lambda)}-\eta^{\nu\lambda}\polm_1^{(\mu}\polmb_4^{\rho)}\right).\nonumber
\end{align}
In the large $s$ limit, the second term is subleading, such that we have
\begin{align}\label{eq:Srep2}
	(\polm_1\cdot\polmb_2)^{s-2}\polm_1^{[\mu}\polmb_2^{\nu]}\polm_1^{[\rho}\polmb_2^{\lambda]}\xrightarrow{\;\text{large }s\;}-\frac{(\polm_1\cdot\polmb_2)^s}{4s^2}S(p_1)^{\mu\nu}S(p_1)^{\rho\lambda}+\mathcal{O}(q^{-1})\,,
\end{align}
which is equivalent to applying \cref{eq:S1rep} twice. Contracting the Lorentz indices on the spin tensors once leads to
\begin{align}\label{eq:Srep1}
	(\polm_1\cdot\polmb_2)^{s-1}\polm_1^{(\mu}\polmb_2^{\nu)}\xrightarrow{\;\text{large }s\;}-\frac{(\polm_1\cdot\polmb_2)^{s}}{2s^2}S(p_1)^{\mu\rho}S(p_1)_{\rho}{}^{\nu} + \mathcal{O}(q^{-1})\,.
\end{align}
Similar identities have been previously used in e.g. Refs.~\cite{Holstein:2008sw, Vaidya:2014kza}. They are sufficient for amplitudes up to $\mathcal{O}(S^2)$.

\subsection{The Nontransverse \texorpdfstring{$(s,s)$}{(s,s)} Representation}
\label{sec:ss_rep_non_trans}

For a nontransverse field in the $(s,s)$ representation we can use the general results obtained for an $(l,r)$ representation. In particular, it has $\pmb K \neq 0$.
It is instructive to identify the origin of $\pmb K$, and thus the structures governing its covariant version $K^{\mu\nu}$, in terms of the lower-spin (longitudinal) components of $\polM_{\mu_1\ldots\mu_s}$. This will be important when discussing \ref{fieldtheory3} which has physical lower-spin fields.

The coherent state in \ref{fieldtheory1} can be decomposed as
\begin{align}\label{eq:Edecom}
	\polM_{\mu_1\ldots\mu_s} = \polM^{(s)}_{\mu_1\ldots\mu_s} + \left(u\polM^{(s-1)}\right)_{\mu_1\ldots\mu_s} + \left(u^2\polM^{(s-2)}\right)_{\mu_1\ldots\mu_s} + \ldots\,,
\end{align}
where the spin-$(s-k)$ component is represented by
\begin{align}
\label{eq:ExplicitLowSpinState}
	\left(u^k \polM^{(s-k)}\right)_{\mu_1\ldots\mu_s} =  {s \choose k}^{1/2} u_{(\mu_1} \ldots u_{\mu_k} \polm_{\mu_{k+1}} \ldots \polm_{\mu_s)}\,.
\end{align}
The states with even $k$ have positive norm and those with odd $k$ have negative norm. We may now compute products involving these polarization tensors. We have
\begin{align}\label{eq:eSeSm1}
\polM^{(s)}_1 \cdot \left(u_2 \polMb_2^{(s-1)}\right) &= -\frac{ \sqrt{s} (\varepsilon_1 \cdot {\bar\varepsilon}_2)^{s-1} \varepsilon_1 \cdot q }{m} + \mathcal{O}(q^2)\,.
\end{align}
At $\mathcal{O}(S^1)$, we plug \cref{eq:Edecom} into \cref{eq:generalS} and find that
\begin{align}\label{eq:esMes2}
	& \polM^{(s)}_1\cdot M^{\mu\nu}\cdot \left(u_2\polMb^{(s-1)}_2\right) = i \sqrt{s} \left(\polm_1\cdot\polmb_2\right)^{s-1}\left(u_2^{\mu}\polm_1^{\nu}-u_2^{\nu}\polm_1^{\mu}\right) + \mathcal{O}(q)\,, \\
	& \left(u_1\polM^{(s-1)}_1\right)\cdot M^{\mu\nu}\cdot \polMb^{(s)}_2 = -i\sqrt{s}(\polm_1\cdot\polmb_2)^{s-1}\left(u_1^{\mu}\polmb_2^{\nu}-u_1^{\nu}\polmb_2^{\mu}\right) + \mathcal{O}(q)\,,\nonumber
\end{align}
while all the other $\polM^{(s)}_1\cdot M^{\mu\nu}\cdot \left(u^{k}_2\polMb^{(s-k)}_2\right)$ vanish at the classical order. We note that contractions like $\left(u_1^k\polM^{(s-k)}_1\right)\cdot M^{\mu\nu}\cdot \left(u_2^k\polMb^{(s-k)}_2\right)$ and $\left(u_1^k\polM^{(s-k)}_1\right)\cdot M^{\mu\nu}\cdot \left(u_2^k\polMb^{(s-k-1)}_2\right)$ give identical result as \cref{eq:esMes,eq:esMes2} in the limit $s\geqslant k$. They contribute an overall factor that can be absorbed in the normalization of the states. In this sense, we can identify the above combination with $K^{\mu\nu}$ or $K^{\mu}$. More precisely, we have 
\begin{align}\label{eq:poltoK}
	-(\polm_1\cdot\polmb_2)^{s-1} \Big[u_1^{\mu}\left(\polm_1^{\nu}+\polmb_2^{\nu}\right)-u_1^{\nu}\left(\polm_1^{\mu}+\polmb_2^{\mu}\right)\Big] &\xrightarrow{\;\text{large }s\;} \frac{\polM_1\cdot\polMb_2}{\sqrt{s}}K(p_1)^{\mu\nu} \,,\nonumber\\
	-(\polm_1\cdot\polmb_2)^{s-1}\left(\polm_1^{\mu}+\polmb_2^{\mu}\right)&\xrightarrow{\;\text{large }s\;} \frac{\polM_1\cdot\polMb_2}{\sqrt{s}}K(p_1)^{\mu} \,,
\end{align}
Using these two relations in the product $\polM_1\cdot\polMb_2$, we find that
\begin{align}
	\polM_1\cdot\polMb_2 &= (\polm_1\cdot\polmb_2)^{s} + \sqrt{s} (\polm_1\cdot u_2+\polmb_2\cdot u_1)(\polm_1\cdot\polmb_2)^{s-1} + \ldots \nonumber\\
	&= (\polm_1\cdot\polmb_2)^s + \frac{q\cdot K}{m}\polM_1\cdot\polMb_2 + \ldots 
\end{align}
where the first term comes from $\polM_1^{(s)}\cdot\left(u_2\polMb_2^{(s-1)}\right)+\left(u_1\polM_1^{(s-1)}\right)\cdot\polMb_2$ and the $\ldots$ contains the contraction between $\polM^{(s)}$ and the states with spin less than $s-1$.
Again, similar contractions between lower-spin states   contribute an overall factor that can be normalized away.
We thus get the relation between the transverse $\polM_1^{(s)}\cdot\polMb_2^{(s)}=(\polm_1\cdot\polmb_2)^s$ and the full result $\polM_1\cdot\polMb_2$ up to the first order in $q$ and $K$,
\begin{align}\label{eq:eetoEE}
	(\polm_1\cdot\polm_2)^{s} = \left(1-\frac{q\cdot K}{m}\right)(\polM_1\cdot\polMb_2) + \mathcal{O}(q^2,K^2)\,,
\end{align}
which is of course consistent with \cref{epdotep,edotep_trans}. The relations~\eqref{eq:poltoK} and \eqref{eq:eetoEE} are used to extract the $\mathcal{O}(K^1)$ terms in \ref{fieldtheory3}. 
More generally, the contractions

\begin{align}
    \polM_1^{(s)}\cdot\left(u_2^k\polMb_2^{(s-k)}\right) &= {s \choose k}^{1/2}(\polm_1\cdot\polmb_2)^{s-k}\left(-\frac{q\cdot\polm_1}{m}\right)^k \,,
\\
    \polM_1^{(s)}\cdot\left\{M^{\mu_1\nu_1},\ldots,M^{\mu_k\nu_k}\right\}\cdot\left(u_2^k\polMb_2^{(s-k)}\right) &= {s \choose k}^{1/2}(k!)(\polm_1\cdot\polmb_2)^{s-k}\prod_{j=1}^{k}\left(2i u_2^{[\mu_i}\polm_1^{\nu_i]}\right)\nonumber\\
    &\rightarrow s^{k/2}(\polm_1\cdot\polmb_2)^{s-k}\prod_{j=1}^{k}\left(2i u_2^{[\mu_i}\polm_1^{\nu_i]}\right) ,
    \label{eq:GeneralEpMtokuEpTOpolvectors}   
\end{align}
can be used to show that for the $s$ to $s-k$ amplitudes,
\begin{align}
   & (q\cdot K)^{k}  \rightarrow s^{k/2}(\polm_1\cdot\polmb_2)^{s-k}\left(\frac{q\cdot\polm_1}{m}\right)^k\,,
    \label{eq:qDotKToTheK}
\\
&	\polM_1^{(s)}\cdot\{\underbrace{M,\ldots M}_{m}\}\cdot \left(u_2^k\polMb_2^{(s-k)}\right)  \sim K^k S^{m-k} \, ,
 \label{offdiagonalKdependence}
\end{align}
which are necessary to identify the structures related to $K^k S^{m-k}$ in the amplitudes.
Recall that the notation $u_2^k\polMb_2^{(s-k)}$ includes a factor $s^{k/2}$, cf. Eq.~\eqref{eq:ExplicitLowSpinState}.
We leave for future work the detailed study of structures of higher orders in spin.

Apart from states created by the operators\footnote{\baselineskip=14pt We denote by $a^\dagger_{(s)}$ the creation operators of the field $\phi_{(s)}$ corresponding to the state labeled by the rest-frame polarization tensors in Eqs.~\eqref{polM} and \eqref{boosts}.}  
$\polM^{(s)}\cdot a^\dagger_{(s)} $ and $\polM^{(s-1)}\cdot a^\dagger_{(s-1)}$ of the fields with definite spin, in \ref{fieldtheory3} we may also choose 
asymptotic states with indefinite spin, which are a normalized linear combinations of these definite-spin states (and, in general, also of lower-spin fields). 
In a quantum theory such a choice is disfavored as it breaks the little-group symmetry. In the classical theory, effectively with a single asymptotic state, this is not an issue. We therefore also evaluate amplitudes in \ref{fieldtheory3} with asymptotic states 
\begin{align}
\label{FT3asymptoticG}
|g\rangle = \frac{1}{\sqrt{2}} \left(
\polM^{(s)}\cdot a^\dagger_{(s)}
+
\polM^{(s-1)}\cdot a^\dagger_{(s-1)} \right) |0\rangle \,.
\end{align}
Similar states have also been considered in Refs.~\cite{Aoude:2021oqj,Aoude:2023fdm}. We will refer to these amplitudes as ${\cal A}^\text{FT3g}$;
in terms of definite-spin states they are
\begin{align}
{\cal A}^\text{FT3g} = \frac{1}{2}\left(
{\cal A}^\text{FT3s}_{s\rightarrow s}
+
{\cal A}^\text{FT3s}_{s-1\rightarrow s-1}
+
{\cal A}^\text{FT3s}_{s-1\rightarrow s}
+
{\cal A}^\text{FT3s}_{s\rightarrow s-1}
\right) \, .
\end{align}
There is no simple polarization tensor that can be assigned to the state $|g\rangle$; the closest analog of the sandwich of Lorentz generators and polarization tensors is the expectation value of the (field) generator of Lorentz transformations in the state $|g\rangle$. While does not have a simple interpretation in terms of the $\bm S$ and $\bm K$ vectors, the interaction~\eqref{eq:transition_L} will supply the requisite factors of momenta for such an interpretation to be possible, cf.~\cref{eq:poltoK}. 

\subsection{Nonminimal Lagrangian}
\label{sec:nonminL}

We are primarily interested in amplitudes in the classical limit, where the spin $s$ is taken to be large. We expect that 
the relevant interaction terms do not depend on a particular representation of the spin, and thus are Lorentz singlets 
constructed from covariant derivatives, photon field strengths,  $\phi_s$ and Lorentz generators~\eqref{eq:Lorentz_Alg} in the same representation
as $\phi_s$. 
Moreover, we   consider for the time being only those interactions that survive in the classical limit. The close relation
in Eq.~\eqref{eq:generalS} between Lorentz generators and the spin tensor and the scaling of momenta in the classical limit imply that  the number of derivatives on the photons must be equal to the number of Lorentz generators. 
Under these guidelines, we can write down the following nonminimal linear-in-$F_{\mu\nu}$ interactions up to two powers of spins,
\begin{align}\label{eq:LS1}
	(-1)^s\Lnonmin 
	&= Q C_1 F_{\mu\nu} \phi_s M^{\mu\nu} \phib_s + \frac{QD_1}{m^2}F_{\mu\nu} ( D_{\rho}\phi_s M^{\rho\mu}D^{\nu}\phib_s + \cc)\\
	& \quad -\frac{iQ C_2}{2m^2}\partial_{(\mu}F_{\nu)\rho}(D^{\rho}\phi_s \S^{\mu}\S^{\nu}\phib_s-\cc)-\frac{iQ D_2}{2m^2} \partial_{\mu}F_{\nu\rho}(D_{\alpha}\phi_s M^{\alpha\mu}M^{\nu\rho}\phib_s-\cc)\,,\nonumber
\end{align}
where for later convenience we choose\footnote{\baselineskip=14pt  
Neutral particles can also have nonminimal couplings analogous to those in Eq. ~\eqref{eq:LS1}. The corresponding Lagrangian is obtained by the double-scaling limit $Q\rightarrow 0$, $C_i, D_i\rightarrow \infty$ with fixed products $QC_i$ and $QD_i$.
} to scale the Wilson coefficients by $Q$ so that at each order amplitudes display overall powers of $\alpha$, and  
the Pauli-Lubanski spin operator $\S^{\mu}$ is defined as
\begin{align}
	\S^{\mu} \equiv \frac{-i}{2m} \, \lc^{\mu\nu\rho\sigma}M_{\rho\sigma}D_{\nu}\,.
\end{align}
We note that the $C_i$ operators are the electrodynamics analogs of operators~\cite{Porto:2006bt, Porto:2007tt, Levi:2015msa} of general relativity, and the $D_i$'s are the electrodynamics analogs of the typical examples of ``extra Wilson coefficients'' of Ref.~\cite{Bern:2022kto}.
From the effective-field-theory point of view, we can write down another operator that contributes classically at the second order in spin,
\begin{align}\label{eq:LD2b}
    \mathcal{L}_{D_{2b}}=\frac{i Q D_{2b}}{2m^4}\partial_{(\mu}F_{\nu)\rho}(D_{\lambda}\phi_s M^{\lambda\mu}M^{\nu}{}_{\sigma}D^{(\sigma}D^{\rho)}\phib_s-\text{c.c})\,.
\end{align}
While $C_2$ and $D_2$ give independent contribution to three-point amplitudes at $\mathcal{O}(S^2)$ and $\mathcal{O}(S^1 K^1)$, see \cref{sec:3pamp}, the above $D_{2b}$ operator gives independent contribution at $\mathcal{O}(K^2)$. Since the purpose of our current work is to understand the existence of extra Wilson coefficients, for simplicity we will not consider this operator further.

\begin{table}
	\begin{tabular}{c|c|c|c}
		\; Field theory\; & Lagrangian & \; Amplitude \; & External state \\ \hline
		\multirow{2}{*}{FT1} & \multirow{2}{*}{$\LEM+\Lmin+\Lnonmin$} & $\mathcal{A}^{\text{FT1s}}$& spin-$s$ \\
		 & & $\mathcal{A}^{\text{FT1g}}$& generic \\ \hline
		FT2 & $\LEM+\Lspins+\Lnonmin$ & $\mathcal{A}^{\text{FT2}}$ & spin-$s$ \\ \hline
		\multirow{2}{*}{FT3} & \multirow{2}{*}{$\;\LEM+\Ltwospin+\Lnonmints\;$} & $\mathcal{A}^{\text{FT3s}}$& \; spin-$s$ \;\\
		  &   & $\mathcal{A}^{\text{FT3g}}$& \; indefinite spin\;
	\end{tabular}
	\caption{Field-theory amplitudes, corresponding Lagrangians and external states.  The Lagrangians are given in Eqs.~\eqref{eq:Lmin}, \eqref{eq:Lspins}, \eqref{twospins}, \eqref{eq:LS1} and \eqref{eq:transition_L}.}
	\label{tab:AFT}
\end{table}

\Cref{tab:AFT} collects the Lagrangians of the four effective field theories that are our focus noting also the notation we
use for their  corresponding amplitudes.
\ref{fieldtheory1} are described by the same Lagrangian,  $\LEM + \Lmin+\Lnonmin$. To compute amplitudes in these two theories, we do not need to specify a particular value for $s$. As discussed in \cref{sec:ss_rep},
the representations of the rotation group with spin $s-2k$  that are part of the field $\phi_s$ have positive norm and are therefore physical, while those with spin $s-(2k+1)$ have negative norm.
In contrast, \ref{fieldtheory2} is described by the Lagrangian $\LEM + \Lspins+\Lnonmin$, and contains only the physical spin-$s$ degrees of freedom. 
When computing the amplitude $\mathcal{A}^{\text{FT1s}}$, we restrict the external states to be the physical spin-$s$ states, which are transverse such that the resultant spin tensors satisfy the covariant SSC according to \cref{sec:ss_rep}. Meanwhile, we keep the external states generic in $\mathcal{A}^{\text{FT1g}}$. As a result, the amplitude $\mathcal{A}^{\text{FT1g}}$ contains explicit SSC-violating terms compared to $\mathcal{A}^{\text{FT1s}}$. 

We   show in \cref{sec:FourPointCompt} that despite having the same physical spin-$s$ external states, $\mathcal{A}^{\text{FT1s}}$ and $\mathcal{A}^{\text{FT2}}$ are different for four-point Compton scattering in the classical limit. In particular, the Compton amplitudes from \ref{fieldtheory2} depend only on $C_1$ and $C_2$ while $\mathcal{A}^{\text{FT1s}}$ also depend on $D_1$ and $D_2$, similar to the appearance of additional nontrivial Wilson coefficients in general relativity~\cite{Bern:2020buy}.
The differences between these amplitudes vanish for
\begin{align}
	C_1 = C_2 = 1\qquad D_1 = D_2 = 0\, ,
\end{align}
and reproduce the root-Kerr amplitudes of Ref.~\cite{Arkani-Hamed:2019ymq}, so that the additional $D_i$ operators do not contribute, in much the same way that additional operators do not contribute to the Kerr black hole.
The similarity of the root-Kerr solution in electromagnetism and the Kerr solution in general relativity follows from the double copy.

These results indicate that additional lower-spin degrees of freedom are the origin of the extra Wilson coefficients. We consider an interpolation between \ref{fieldtheory1} and \ref{fieldtheory2} in \cref{sec:WilsonCoeffs} to understand the effect of the state projector. In \ref{fieldtheory1} these degrees of freedom have negative norm; a natural question is whether lower-spin states with positive norm have similar consequences. 
\ref{fieldtheory3} explores this question. 
With some foresight which is justified in 
\cref{sec:WilsonCoeffs}, we choose the nonminimal interactions of $\phi_s$ and $\phi_{s-1}$ in \cref{twospins}, valid through the quadratic order in spin, to be
\begin{align}\label{eq:transition_L}
\Lnonmints &= Q C_1F_{\mu\nu}\phi_sM^{\mu\nu}\phib_s - \frac{2iQ \widetilde{C}_1\sqrt{s}}{m}F_{\mu\nu}\Big[(\phi_s)^{\mu}{}_{\alpha_2\ldots\alpha_s}D^{\nu}\phib_{s-1}^{\alpha_2\ldots\alpha_s}-\text{c.c}\Big] \\
& -\frac{iQ C_2}{2m^2}\partial_{(\mu}F_{\nu)\rho}(D^{\rho}\phi_s \S^{\mu}\S^{\nu}\phib_s-\cc) - \frac{2i Q \widetilde{C}_2\sqrt{s}}{m}F_{\mu\nu}\Big[(\phi_s)^{\mu}{}_{\alpha_2\ldots\alpha_s}D^{\alpha_2}\phib_{s-1}^{\nu\alpha_3\ldots\alpha_s}-\text{c.c}\Big]\,,\nonumber
\end{align}
and the Lagrangian of \ref{fieldtheory3} is given by the third line of Table~\ref{tab:AFT}.
We shall see in \cref{sec:FourPointCompt} that the Wilson coefficients $\widetilde{C}_1$ and $\widetilde{C}_2$ appear at $\mathcal{O}(K^1)$ and $\mathcal{O}(S^1 K^1)$ order of the Compton amplitudes respectively, and that there exists an effective map between the $D_i$ and $\widetilde{C}_i$ coefficients. As discussed in \cref{sec:ss_rep}, we need to include couplings between $\phi_s$ and $\phi_{s-2}$ to access the $\mathcal{O}(K^2)$ interactions, which we omit for simplicity. 

Similar to gravity, operators describing tidal deformations under the influence of external fields are necessary to describe
the electromagnetic interactions of generic spinning bodies. Simple counting of classical scaling indicates that in QED they first appear $\mathcal{O}(S^2)$. At this order in spin three independent operators are
\begin{align}
	(-1)^{s}\mathcal{L}_{F^2} &= \frac{Q^2 E_1}{m^2}F_{\mu\nu}F_{\rho\sigma}\phi_sM^{\mu\nu}M^{\rho\sigma}\phib_s +\frac{Q^2E_2}{m^2}F_{\mu\nu}F_{\rho}{}^{\mu}\phi_s M^{\nu\lambda}M_{\lambda}{}^{\rho}\phib_s \nonumber\\
	&\quad + \frac{Q^2 E_3}{m^4}F_{\mu\nu}F_{\rho\sigma}D^{\mu}\phi_sM^{\nu\lambda}M_{\lambda}{}^{\rho}D^{\sigma}\phib_s+{\cal O}(M^3)\,.
\end{align}
Including them we find that all $E_i$ Wilson coefficients vanish for the root-Kerr states in much the same way as the $D_i$ coefficients vanish for these states in \ref{fieldtheory1}.

\section{Scattering Amplitudes}
\label{sec:FourPointCompt} 


In this section we first compute the 1PL (tree) Compton amplitudes of the higher-spin effective Lagrangians introduced in the previous section and summarized in \cref{tab:AFT}. 
We then use them as the basic building blocks of the $\mathcal O(\alpha^2)$ two-body amplitudes through generalized unitarity.\footnote{\baselineskip=14pt For simplicity, we suppress a factor of $Q$ in the three-point Compton amplitudes, and a factor of $Q^2$ in the four-point Compton amplitudes, where $Q$ is the electric charge of the massive body.}
In addition, classical Compton amplitudes are also observables that can be directly compared with worldline computations along the lines of Ref.~\cite{Saketh:2022wap}. The comparison will be given in \cref{sec:WL}.

\subsection{Three-Point Amplitudes}
\label{sec:3pamp}

We start with computing and comparing the three-point Compton amplitudes from the theories in \cref{tab:AFT}. Assuming that all the momenta are outgoing, the Feynman rules for \ref{fieldtheory1} are given by
\begin{align}\label{eq:A3min}
	\mathcal{A}_{3}\!\left[\vcenter{\hbox{\scalebox{0.8}{\begin{tikzpicture}
					\draw [very thick] (0,0) node[left=0pt] {$p_1$} -- (1.5,0) node[right=0pt]{$p_2$};
					\draw [photon] (0.75,0) -- (0.75,1) node[above=0]{$\kk_3,\pol_3$};
	\end{tikzpicture}}}}\right] &= (-1)^{s}\polM_1\cdot \mathbb{M}_3(p_1,p_2,\kk_3,\pol_3)\cdot\polMb_2\,, \\
	\mathbb{M}_3(p_1,p_2,\kk_3,\pol_3) &= 2\pol_3\cdot p_1 \mathbbm{1} - 2iC_1M_{\mu\nu}\kk_3^{\mu}\pol_3^{\nu}-\frac{2iD_1}{m^2}\pol_3\cdot p_1 M_{\mu\nu}p_1^{\mu}\kk_3^{\nu} \nonumber \\
	&\quad +\frac{C_2}{m^2}\pol_3\cdot p_1\{M^{\mu\nu},M_{\mu}{}^{\rho}\}\kk_3^{\nu}\kk_3^{\rho} - \frac{C_2}{m^4}\pol_3\cdot p_1\{M_{\mu\nu},M_{\rho\lambda}\}p_1^{\mu}\kk_3^{\nu}p_1^{\rho}\kk_3^{\lambda}\nonumber\\
	&\quad + \frac{2D_2}{m^2}\{M_{\mu\nu},M_{\rho\lambda}\}p_1^{\mu}\kk_3^{\nu}\kk_3^{\rho}\pol_3^{\lambda}\,.\nonumber
\end{align}
The symmetric product between Lorentz generators is defined in \cref{eq:symM}.

In the classical limit, the massive spinning particles are described by the spin coherent states~\eqref{eq:spin_coherent}. We first consider \hyperlink{ft1g}{FT1g} with generic coherent states that do not satisfy the transversality. The expectation values of Lorentz generators are given by \cref{eq:generalS}, which lead to the classical spin tensor $\gS_{\mu\nu}$ that do not satisfy the covariant SSC.
The three-point amplitude is
\begin{align}
    \mathcal{A}_{3}^{\text{FT1g}} &= (-1)^s \polM_1\cdot\polMb_2 \left[2\pol_3\cdot p_1-2iC_1 \gS_{\mu\nu}\kk_3^{\mu}\pol_3^{\nu}+\frac{C_2}{m^2}\pol_3\cdot p_1 \gS_{\mu\nu}\kk_3^{\nu}\gS^{\mu}{}_{\lambda}\kk_3^{\lambda}\right. \nonumber\\
    &\qquad -\left.\gS_{\mu\nu}p_1^{\mu}\kk_3^{\nu}\left(\frac{2iD_1}{m^2}\pol_3\cdot p_1 + \frac{C_2}{m^4}\pol_3\cdot p_1 \gS_{\lambda\sigma}p_1^{\lambda}\kk_3^{\sigma} - \frac{2D_2}{m^2}\gS_{\lambda\sigma}\kk_3^{\lambda}\pol_3^{\sigma}\right)\right] \nonumber\\
    &=2(-1)^s \polM_1\cdot\polMb_2 \left[\pol_3\cdot p_1-iC_1 S_{\mu\nu}\kk_3^{\mu}\pol_3^{\nu}-(C_1-D_1)\pol_3\cdot p_1\frac{q\cdot K}{m}\vphantom{\left(\frac{q\cdot K}{m}\right)^2}\right.\nonumber\\
    &\qquad +\left.\frac{C_2}{2m^2}\pol_3\cdot p_1 S_{\mu\nu}\kk_3^{\nu}S^{\mu}{}_{\lambda}\kk_3^{\lambda} + iD_2 S_{\mu\nu}\kk_3^{\mu}\pol_3^{\nu}\frac{q\cdot K}{m}+D_2\pol_3\cdot p_1\left(\frac{q\cdot K}{m}\right)^2\right],
\end{align}
where in the second equal sign we have used \cref{nonSSCspintensor} to expose the SSC preserving $S$-part and the SSC violating $K$-part in $\gS^{\mu\nu}$.
As expected, the extra Wilson coefficients $D_i$ appear with the SSC-violating terms.
If we further restrict the external states to be transverse, the $K$-part becomes subleading in the classical limit and thus drops out. This leads to the three-point amplitudes $\mathcal{A}_3^\text{FT1s}$ and $\mathcal{A}_3^\text{FT2}$
\begin{align}\label{eq:3pComp}
    \mathcal{A}_3^{\text{FT1s}} = \mathcal{A}_3^{\text{FT2}} &= 2(-\polm_1\cdot\polmb_2)^s \left[\pol_3\cdot p_1-iC_1 S_{\mu\nu}\kk_3^{\mu}\pol_3^{\nu}+\frac{C_2}{2m^2}\pol_3\cdot p_1 S_{\mu\nu}\kk_3^{\nu}S^{\mu}{}_{\lambda}\kk_3^{\lambda}\right] .
\end{align}
This amplitude only depends on the $C_i$ Wilson coefficients. 
The fact that $\mathcal{A}_3^{\text{FT1s}}$  does not contain any additional Wilson coefficients is analogous to the three-point gravity amplitude of Ref.~\cite{Bern:2020buy}, which did not contain any additional Wilson coefficients either, connected to restricting the external states to traceless and transverse spin-$s$ ones.

For \ref{fieldtheory3}, we can similarly restrict the external states to be spin-$s$. The resulting amplitude $\mathcal{A}_3^{\text{FT3s}}$ is the same as \cref{eq:3pComp}, i.e.
\begin{align}
\mathcal{A}_3^{\text{FT1s}} = \mathcal{A}_3^{\text{FT2}} = \mathcal{A}_3^{\text{FT3s}} \, .
\end{align}
We may also choose the indefinite-spin states \eqref{FT3asymptoticG}; the corresponding amplitude $\mathcal{A}_3^{\text{FT3g}}$ receives contributions from both the spin-$s$ and spin-$(s-1)$ external states,
\begin{align}
    \mathcal{A}_3^{\text{FT3g}} &= \mathcal{A}_3^{\text{FT3s}} + \frac{2i\sqrt{s}}{m}(-\polm_1\cdot\polmb_2)^{s-1}(\polm_1\cdot q_3+\polmb_2\cdot \kk_3)\left[\widetilde{C}_1(\pol_3\cdot p_1) + \widetilde{C}_2(\polM_1\cdot f_3\cdot\polMb_4)\right] \nonumber\\
    &= 2(-1)^s\polM_1\cdot\polMb_2\left[\pol_3\cdot p_1 - iC_1 S_{\mu\nu}q_3^{\mu}\pol_3^{\nu} +\frac{C_2}{2m^2}\pol_3\cdot p_1 S_{\mu\nu}\kk_3^{\nu} S^{\mu}{}_{\lambda}\kk_3^{\lambda} \right.\nonumber\\
    &\qquad\qquad\qquad\qquad + \left. (i\widetilde{C}_1-1)\pol_3\cdot p_1\frac{q\cdot K}{m} +(iC_1-\widetilde{C}_2)S_{\mu\nu}\kk_3^{\mu}\pol_3^{\nu}\frac{q\cdot K}{m}\right] ,
\end{align}
where we have used \cref{eq:poltoK,eq:eetoEE} to obtain the final expression in terms of the boost vector $K$. 
The appearance of the Wilson coefficients $\widetilde{C}_1$ and $\widetilde{C}_2$ associated with SSC violation.
We find that, up to $\mathcal{O}(K^2)$ terms, the additional Wilson coefficients in \hyperlink{ft1g}{FT1g} and \hyperlink{ft3g}{FT3g} are related as
\begin{align}\label{eq:map3p}
    \mathcal{A}_3^{\text{FT3g}} = \mathcal{A}_3^{\text{FT1g}}\qquad\text{for}\quad i\widetilde{C}_1 = 1 - C_1 + D_1\;\text{ and }\;i\widetilde{C}_2 = D_2 - C_1\,.
\end{align}
The extra factor of $i$ in this map reflects the unphysical nature of the spin-$(s-1)$ states in \ref{fieldtheory1} vs. their physical nature in \ref{fieldtheory3}. While this map is not an equivalence of Lagrangians, it provides a simple relation between the amplitudes of \ref{fieldtheory1} and \ref{fieldtheory3}. A similar map also exists at $\mathcal{O}(K^2)$ if we include such interactions in both \ref{fieldtheory1} and \ref{fieldtheory3}.

\subsection{Four-Point Compton Amplitudes}
\label{Compton}

\begin{figure}
        \begin{tikzpicture}
                \draw [very thick] (0,0) node[above=0]{$p_1$} -- (4.2,0) node[above=0pt]{$p_4$};
                \draw [photon] (1.4,0) -- ++(0,1) node[above=0]{$\kk_2,\pol_2$};
                \draw [photon] (2.8,0) -- ++(0,1) node[above=0]{$\kk_3,\pol_3$};
                \begin{scope}[xshift=4.9cm]
                        \draw [very thick] (0,0) node[above=0]{$p_1$} -- (4.2,0) node[above=0pt]{$p_4$};
                        \draw [photon] (1.4,0) -- ++(0,1) node[above=0]{$\kk_3,\pol_3$};
                        \draw [photon] (2.8,0) -- ++(0,1) node[above=0]{$\kk_2,\pol_2$};
                \end{scope}
                \begin{scope}[xshift=9.8cm]
                        \draw [very thick] (0,0) node[above=0]{$p_1$} -- (4.2,0) node[above=0pt]{$p_4$};
                        \draw [photon] (2.1,0) -- (1.4,1) node[above=0]{$\kk_2,\pol_2$};
                        \draw [photon] (2.1,0) -- (2.8,1) node[above=0]{$\kk_3,\pol_3$};
                \end{scope}
        \end{tikzpicture}\,.
\caption{The three Feynman diagrams describing lowest-order Compton scattering.}
\label{fig:FeynmanCompton}
\end{figure}
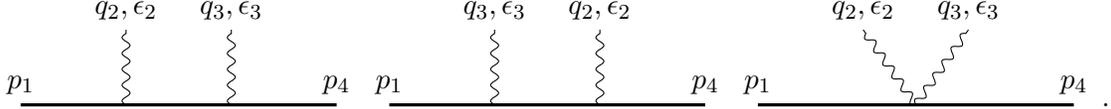

At four points, the Compton amplitudes are given by the three Feynman diagrams in \cref{fig:FeynmanCompton}, with the relevant propagators and three- and four-point vertices derived from our field theories.
While the Lagrangian of \ref{fieldtheory1} is independent of $s$ and therefore general properties of Lorentz generators and coherent states are sufficient for amplitude calculations, 
the explicit dependence on $s$ of \ref{fieldtheory2} and \ref{fieldtheory3} Lagrangians requires that for them we choose a particular representation. 
As noted in \cref{sec:ss_rep}, we choose the $(s,s)$ representation, for which the coherent states are given by \cref{eq:pol_vec} and the Lorentz generators are listed in \cref{eq:LorentzGen}.
Specifically for four-point Compton amplitudes, the Feynman rules for \ref{fieldtheory2} and  \ref{fieldtheory3}  simplify considerably because every vertex has at least one on-shell massive particle represented by the symmetric, traceless and transverse polarization tensor.
Thus, when deriving the three- and four-point vertex rules we can ignore all the interactions covered by the ellipsis in \cref{eq:Lspins} because they only include traces or/and longitudinal modes of the external on-shell particle. For the same reason, we can also ignore all the (resummed) propagators that involve lower-spin auxiliary fields.

The spin-independent part of the Compton amplitude is common to \ref{fieldtheory1} though \ref{fieldtheory3},
\begin{align}
    & \mathcal{A}_{4,\text{ cl.}}^{\text{FT1,2,3}}\Big|_{S^0}=(-1)^s \polM_1\cdot\polMb_4\frac{2(p_1\cdot f_2\cdot f_3\cdot p_1)}{(p_1\cdot\kk_2)^2}\,,
    \label{spinlessCompton}
\end{align}
where $f_i^{\mu\nu} \equiv \varepsilon_i^\mu q_i^\nu - \varepsilon_i^\nu  q_i^\mu$.  For the $(s,s)$ representation, we do not need to distinguish 
$\polM_1\cdot\polMb_4$ and $(\polm_1\cdot\polmb_4)^s$ as their difference is higher order in $S$ and $K$.
Up to the overall factor $\polM_1\cdot\polMb_4$, Eq.~\eqref{spinlessCompton} reproduces the classical limit of the scalar QED Compton amplitude given in, for example, 
Eq.~(2.8) of Ref.~\cite{Bern:2021xze}.\footnote{\baselineskip=14pt In Ref.~\cite{Bern:2021xze} higher order in $q_i$ terms are also included since they are
needed when feeding the Compton amplitudes into unitarity cuts for building higher PL two-body amplitudes.}

\subsubsection{The Linear-in-Spin Compton Amplitudes}
\label{S1_amplitudes} 

Consider now spin-dependent parts of amplitudes of the four field theories.
For \ref{fieldtheory1}, here and after we choose $\phi_s$ to be in the $(s,s)$ representation to streamline the comparison with \ref{fieldtheory2} and \ref{fieldtheory3}. 
We first consider $\mathcal{A}^{\text{FT1s}}$ in which the external states are transverse.
Evaluating the linear-in-spin part of the three Feynman diagrams in Fig.~\ref{fig:FeynmanCompton} with the propagators 
and vertices following from the Lagrangian of \ref{fieldtheory1} leads to
\begin{align}
\label{eq:s1_comp}
    \mathcal{A}_{4,\text{ cl.}}^{\text{FT1s}}\Big|_{S^1} = (-\polm_1\cdot\polmb_4)^s S(p_1)_{\mu\nu} &\left[\frac{iC_1}{(p_1\cdot \kk_2)^2}(f_2^{\mu\nu}\kk_{2\rho}f_3^{\rho\lambda}+f_3^{\mu\nu}\kk_{3\rho}f_2^{\rho\lambda})p_{1\lambda}+\frac{2iC_1^2}{p_1\cdot \kk_2}f_2^{\nu\rho}f_{3\rho}{}^{\mu}\right.\nonumber\\
	& \; +\left. \frac{2iD_1(2C_1-D_1-2)}{(p_1\cdot \kk_2)m^2}p_{1\rho}f_2^{\rho\mu}f_3^{\nu\lambda}p_{1\lambda}\right]  .
\end{align}
The amplitude $\mathcal{A}_\text{4, cl}^{\text{FT1}}\Big|_{S^1} $ depends on both the $C_1$ and $D_1$ Wilson coefficients.\footnote{\baselineskip=14pt The gravitational analog of this amplitude is of ${\cal O}(S^2)$.} We note that $D_1$ appears only together with the combination $p_{1\rho}f_2^{\rho\mu}f_3^{\nu\lambda}p_{1\lambda}$.
Repeating the calculation while relaxing the transversality on external states leads to the Compton amplitude,
\begin{align}\label{eq:s1_comp_FT2}
	& \mathcal{A}_{4,\text{ cl.}}^{\text{FT1g}}\Big|_{S^1} = (-1)^s\polM_1\cdot\polMb_4\left\{\gS(p_1)_{\mu\nu} \left[\frac{iC_1}{(p_1\cdot \kk_2)^2}(f_2^{\mu\nu}\kk_{2\rho}f_3^{\rho\lambda}+f_3^{\mu\nu}\kk_{3\rho}f_2^{\rho\lambda})p_{1\lambda}+\frac{2iC_1^2}{p_1\cdot \kk_2}f_2^{\nu\rho}f_{3\rho}{}^{\mu}\right.\right.\nonumber\\
	& \quad +\left. \frac{2iD_1(2C_1-D_1-2)}{(p_1\cdot \kk_2)m^2}p_{1\rho}f_2^{\rho\mu}f_3^{\nu\lambda}p_{1\lambda}\right] + \gS(p_1)_{\mu\nu}p_1^{\nu}\left[\frac{2iD_1(C_1+1)}{(p_1\cdot \kk_2)m^2}p_{1\rho}(f_3^{\rho\lambda}f^{}_{2\lambda}{}^{\mu}-f_2^{\rho\lambda}f^{}_{3\lambda}{}^{\mu})\right. \nonumber  \\
	& \quad +\left.\left.\frac{2iD_1}{(p_1\cdot \kk_2)^2m^2}p_{1\rho}p_{1\lambda}(f_3^{\rho\mu}\kk_{3\sigma}f_2^{\sigma\lambda}+f_2^{\rho\mu}\kk_{2\sigma}f_3^{\sigma\lambda})\right]\right\} .
\end{align}
We note that the first term is formally identical to \cref{eq:s1_comp} for \ref{fieldtheory1} except for the replacement $S\rightarrow\gS$, while the second term is proportional to the SSC condition $\gS_{\mu\nu}p_1^{\nu}$.

Proceeding to \ref{fieldtheory2} of a single transverse spin-$s$ field, we extract the classical $\mathcal{O}(S^1)$ Compton amplitude from explicit calculations for $s = 1,2,3$ using the Lagrangian given in the third row of \cref{tab:AFT}. 
Unlike \ref{fieldtheory1}, the amplitudes are now given in terms of explicit polarization vectors instead of spin tensors such that we need to convert the former into the latter. Since the classical amplitudes in terms of spin tensors scale as $\mathcal{O}(q^0)$ and the spin tensors scale as $\mathcal{O}(q^{-1})$, in a fixed-spin calculation, the classical part of the $\mathcal{O}(S^1)$ amplitude is among the $\mathcal{O}(q)$ terms of the full quantum amplitude~\cite{Bern:2020buy}.
At this order, the massive polarization vectors appear in two structures, $(\polm_1\cdot\polmb_4)^s$ and $(\polm_1\cdot\polmb_4)^{s-1}\polm_1^{[\mu}\polmb_4^{\nu]}$. The terms proportional to $(\polm_1\cdot\polmb_4)^s$ belong to the quantum spinless amplitude, which can be ignored here. We then use the relation~\eqref{eq:S1rep} to convert the second structure to spin tensors.
The final classical amplitude is obtained by extrapolating the finite-spin results to generic $s$ and taking the $s\rightarrow\infty$ of that expression. At $\mathcal{O}(S^1)$, the amplitude after the replacement~\eqref{eq:S1rep} is in fact independent of $s$, as we have explicitly checked for $s\leqslant 3$.
After identifying the classical part, the final answer for the classical Compton amplitude is 
\begin{align}\label{eq:A4FT2}
    \mathcal{A}_\text{4, cl}^{\text{FT2}}\Big|_{S^1} = (-\polm_1\cdot\polmb_4)^s S(p_1)_{\mu\nu} &\left[\frac{iC_1}{(p_1\cdot \kk_2)^2}(f_2^{\mu\nu}\kk_{2\rho}f_3^{\rho\lambda}+f_3^{\mu\nu}\kk_{3\rho}f_2^{\rho\lambda})p_{1\lambda}+\frac{2iC_1^2}{p_1\cdot \kk_2}f_2^{\nu\rho}f_{3\rho}{}^{\mu}\right.\nonumber\\
	& \; +\left. \frac{2i(C_1-1)^2}{(p_1\cdot \kk_2)m^2}p_{1\rho}f_2^{\rho\mu}f_3^{\nu\lambda}p_{1\lambda}\right] .
\end{align}
Notably, this amplitude is independent of $D_1$, and it can be obtained from \cref{eq:s1_comp} by setting $D_1$ to a special value,
\begin{equation}\label{eq:WCmap}
    \mathcal{A}_\text{4, cl}^{\text{FT1s}}\Big|_{S^1} = \mathcal{A}_{\text{4, cl}}^{\text{FT2}}\Big|_{S^1}\quad\text{for}\quad D_1=C_1-1\,.
\end{equation}
In other words, for this special value of $D_1$,  \cref{fieldtheory1} effectively propagates only the spin-$s$ states. 
Moreover, the special value $C_1=1$ and $D_1=0$ reproduces the root-Kerr 
Compton amplitudes~\cite{Arkani-Hamed:2019ymq}. 
The appearance of additional Wilson coefficients in $\mathcal{A}^{\text{FT1s}}$ compared with $\mathcal{A}^{\text{FT2}}$ can be attributed to the additional propagating degrees of freedom.\footnote{\baselineskip=14pt In Ref.~\cite{Kim:2023drc}, a similar computation was carried out for $s=1$ and observed a similar effect. Their amplitudes are equivalent to our $D_1=0$ case.}

Finally, \ref{fieldtheory3} amplitudes also receive contributions from lower-spin states. 
We first restrict the lower spins to only appear in the intermediate states. 
Repeating the same steps as for \ref{fieldtheory2} we find that spin-$(s-1)$
intermediate states contribute as 
\begin{align}
\label{ComptonFT4S1}
    \mathcal{A}_\text{4, cl}^{\text{FT3s}}\Big|_{S^1}  &=  \mathcal{A}_\text{4, cl}^{\text{FT3}} \Big|_{S^1} + (-\polm_1\cdot\polmb_4)^s S(p_1)_{\mu\nu}\left[\frac{2i\widetilde{C}_1^2}{(p_1\cdot \kk_2)m^2}p_{1\rho}f_2^{\rho\mu}f_3^{\nu\lambda}p_{1\lambda}\right] .
\end{align}
We note that the $\mathcal{O}(S^1)$ amplitude does not change if we include intermediate states with spin less than $s-1$.
Comparing with \cref{eq:s1_comp}, we find that the two amplitudes are formally related by the same map as \cref{eq:map3p},
\begin{align}
\mathcal{A}_{4,\text{ cl.}}^{\text{FT3s}}\Big|_{S^1} = \mathcal{A}_{4,\text{ cl.}}^{\text{FT1s}}\Big|_{S^1}\quad\text{for}\quad i\widetilde{C}_1 = 1 - C_1 + D_1\,.
\label{14map}
\end{align}
Furthermore, this map persists even for amplitudes with external lower-spin states. To see this, we first rewrite \cref{eq:s1_comp_FT2} using \cref{nonSSCspintensor},
\begin{align}\label{eq:A4FT1g}
    \mathcal{A}_{4,\text{ cl.}}^{\text{FT1g}}\Big|_{S^1} &= (-1)^s\polM_1\cdot\polMb_4\left\{ S(p_1)_{\mu\nu} \left[\frac{iC_1}{(p_1\cdot \kk_2)^2}(f_2^{\mu\nu}\kk_{2\rho}f_3^{\rho\lambda}+f_3^{\mu\nu}\kk_{3\rho}f_2^{\rho\lambda})p_{1\lambda}+\frac{2iC_1^2}{p_1\cdot \kk_2}f_2^{\nu\rho}f_{3\rho}{}^{\mu}\right.\right.\nonumber\\
    &  +\left. \frac{2iD_1(2C_1-D_1-2)}{(p_1\cdot \kk_2)m^2}p_{1\rho}f_2^{\rho\mu}f_3^{\nu\lambda}p_{1\lambda}\right] + 
    \frac{2K(p_1)_{\mu}p_{1\nu}}{m}\left[\frac{D_1-C_1}{(p_1\cdot q_2)^2}(q_2^{\mu}+q_3^{\mu})f_2^{\nu\rho}f_{3\rho\lambda}p_1^{\lambda}\right.\nonumber\\
    & -\left.\left.\frac{C_1(1-C_1+D_1)}{p_1\cdot q_2}(f_2^{\nu\rho}f_{3\rho}{}^{\mu}-f_3^{\nu\rho}f_{2\rho}{}^{\mu})\right]\right\} .
\end{align}

We then find $\mathcal{A}_4^{\text{FT3g}}$, with external states in Eq.~\eqref{FT3asymptoticG}, from the Lagrangian of \ref{fieldtheory3}. The momentum dependence of vertices is essential to express the contributions with spin-$(s-1)$ external states in terms of $K^{\mu}$, using \cref{eq:poltoK,eq:eetoEE}. The result is
\begin{align}\label{eq:A4FT3g}
    \mathcal{A}_4^{\text{FT3g}}\Big|_{S^1} &= (-1)^s\polM_1\cdot\polMb_4\left\{ 
    S(p_1)_{\mu\nu} \left[\vphantom{\frac{iC_1\widetilde{C}_1}{p_1\cdot q_2}} \frac{iC_1}{(p_1\cdot \kk_2)^2}(f_2^{\mu\nu}\kk_{2\rho}f_3^{\rho\lambda}+f_3^{\mu\nu}\kk_{3\rho}f_2^{\rho\lambda})p_{1\lambda}+\frac{2iC_1^2}{p_1\cdot \kk_2}f_2^{\nu\rho}f_{3\rho}{}^{\mu}\right.\right.\nonumber\\
	& +\left. \frac{2i[(C_1-1)^2+\widetilde{C}_1^2]}{(p_1\cdot \kk_2)m^2}p_{1\rho}f_2^{\rho\mu}f_3^{\nu\lambda}p_{1\lambda}\right] + \frac{2K(p_1)_{\mu}p_{1\nu}}{m}\left[\frac{i\widetilde{C}_1-1}{(p_1\cdot q_2)^2}(q_2^{\mu}+q_3^{\mu})f_2^{\nu\rho}f_{3\rho\lambda}p_1^{\lambda}\right.\nonumber\\
	& -\left.\left.\frac{iC_1\widetilde{C}_1}{p_1\cdot q_2}(f_2^{\nu\rho}f_{3\rho}{}^{\mu}-f_3^{\nu\rho}f_{2\rho}{}^{\mu})\right]\right\} .
\end{align}
Now comparing \cref{eq:A4FT1g,eq:A4FT3g}, we find that
\begin{align}
\mathcal{A}_{4,\text{ cl.}}^{\text{FT3g}}\Big|_{S^1} = \mathcal{A}_{4,\text{ cl.}}^{\text{FT1g}}\Big|_{S^1}\qquad\text{for}\quad i\widetilde{C}_1 = 1 - C_1 + D_1\,.    
\end{align}
The robustness of this map demonstrates that the terms tagged by the extra Wilson coefficients present in the amplitudes (and observables) of \ref{fieldtheory1} and \ref{fieldtheory3} carry new physical information compared to \ref{fieldtheory2}.


\subsubsection{The Quadratic-in-Spin Compton Amplitudes}

Feynman-diagram calculations using the propagators and vertices of \ref{fieldtheory1} as well as properties of transverse coherent states show that, at $\mathcal{O}(S^2)$ the Compton amplitude depends on two distinct contractions of spin tensors: 
\begin{align}\label{eq:A1S2}
	\mathcal{A}_\text{4, cl.}^{\text{FT1s}} \Big|_{S^2} &= (-\polm_1\cdot\polmb_4)^s  \big(S(p_1)_{\mu\nu}S(p_1)_{\lambda\sigma} \mathcal{X}^{\mu\nu\lambda\sigma} + S(p_1)_{\mu\lambda}S(p_1)^{\lambda}{}_{\nu} \mathcal{X}^{\mu\nu}\big)\,.
\end{align}
Their kinematic coefficients are given by
\begin{align}
	\label{eq:Xcoeff}
	\mathcal{X}^{\mu\nu\lambda\sigma} &=
        \frac{C_1^2(\kk_2\cdot\kk_3)}{2(p_1\cdot\kk_2)^2}f_2^{\mu\nu}f_3^{\lambda\sigma}+\frac{C_1D_1+D_2(C_1-D_1-1)}{(p_1\cdot\kk_2)m^2}p_{1\rho}(f_3^{\rho\mu}\kk_2^{\nu}f_2^{\lambda\sigma}-f_2^{\rho\mu}\kk_3^{\nu}f_3^{\lambda\sigma})
        \,,\\[8pt]
	\label{eq:Ycoeff}
	\mathcal{X}^{\mu\nu} &=
        \frac{C_2}{m^2}\left[\frac{p_{1\rho}p_{1\alpha}(f_2^{\rho\mu}\kk_2^{\nu}f_3^{\alpha\beta}\kk_{2\beta}+f_3^{\rho\mu}\kk_3^{\nu}f_2^{\alpha\beta}\kk_{3\beta})}{(p_1\cdot\kk_2)^2} + \frac{p_{1\rho}(f_2^{\rho\sigma}f_{3\sigma}{}^{\mu}_{\vphantom{2}}\kk_3^{\nu}-f_3^{\rho\sigma}f_{2\sigma}{}^{\mu}_{\vphantom{2}}\kk_2^{\nu})}{(p_1\cdot\kk_2)}\right. 
        \\
	  &\qquad\quad + \left. \frac{2C_1\, p_{1\rho}(f_3^{\rho\mu}f_2^{\nu\sigma}\kk_{3\sigma}-f_2^{\rho\mu}f_3^{\nu\sigma}\kk_{2\sigma})}{(p_1\cdot\kk_2)} + \frac{2(C_1-1)p_{1\rho}f_2^{\rho\mu}f_3^{\nu\sigma}p_{1\sigma}}{m^2} + 2C_1 f_{2\rho}{}^{\mu}_{\vphantom{2}}f_3^{\nu\rho}\right]
        .\nonumber
\end{align}
The dependence on Wilson coefficients indicates that both terms originate from both $\Lmin$ and $\Lnonmin$, and the Lorentz algebra was used to reduce a product of three Lorentz generators to a sum of irreducible (symmetric) products. We also note that the $D_1$ and $D_2$ dependence only appear in $\mathcal{X}^{\mu\nu\lambda\sigma}$.

Choosing general asymptotic states instead of transverse ones leads to the amplitude $\mathcal{A}^{\text{FT1g}}$. Apart from the replacement $S\rightarrow\gS$ in $\mathcal{A}_\text{4, cl}^{\text{FT1s}} \Big|_{S^2}$,
the amplitude contains terms proportional to the covariant SSC conditions: 
\begin{align}
    \label{eq:s2_comp_FT2}
    \mathcal{A}_{4,\text{ cl.}}^{\text{FT1g}} \Big|_{S^2} &= (-1)^s \polM_1\cdot\polMb_4\Big[\gS(p_1)_{\mu\nu}\gS(p_1)_{\lambda\sigma}\mathcal{X}^{\mu\nu\lambda\sigma}+\gS(p_1)_{\mu\lambda}\gS(p_1)^{\lambda}{}_{\nu}\mathcal{X}^{\mu\nu}\\
    & \quad + \gS(p_1)_{\mu\nu}p_{1}^{\nu}\gS(p_1)_{\lambda\sigma}\mathcal{Y}^{\mu\lambda\sigma}+\gS(p_1)_{\mu\nu}p_{1}^{\nu}\gS(p_1)_{\lambda\sigma}p_{1}^{\sigma}\mathcal{Y}^{\mu\lambda} + \gS(p_1)_{\mu\lambda}\gS(p_1)^{\lambda}{}_{\nu}p_1^{\nu}\mathcal{Y}^{\mu}\Big],\nonumber
\end{align}
where the additional kinematic coefficients are given by
\begingroup
\allowdisplaybreaks
\begin{align}
        \mathcal{Y}^{\mu\lambda\sigma}&=
        \frac{C_1(D_1-D_2)}{(p_1\cdot \kk_2)m^2}(\kk_{2\rho}f_3^{\rho\mu}f_2^{\lambda\sigma}-\kk_{3\rho}f_2^{\rho\mu}f_3^{\lambda\sigma}) + \frac{C_1D_1(\kk_2\cdot \kk_3)}{(p_1\cdot \kk_2)^2m^2}p_{1\rho}(f_3^{\rho\mu}f_2^{\lambda\sigma}-f_2^{\rho\mu}f_3^{\lambda\sigma})\nonumber\\*
	&\quad + \frac{2C_1D_2}{(p_1\cdot \kk_2)m^2}(\kk_2+\kk_3)^{\mu}f_2^{\sigma\alpha}f^{}_{3\alpha}{}^{\lambda} + \frac{2D_1^2}{(p_1\cdot \kk_2)m^4}p_{1\rho}p_{1\alpha}(f_2^{\rho\mu}f_3^{\alpha\lambda}\kk_2^{\sigma}-f_3^{\rho\mu}f_2^{\alpha\lambda}\kk_3^{\sigma}) \nonumber\\*
	&\quad  + \frac{2D_2}{(p_1\cdot \kk_2)^2m^2}p_{1\alpha}(\kk_{2}^{\lambda}f_2^{\sigma\mu}f_3^{\alpha\beta}\kk_{2\beta}+\kk_{3}^{\lambda}f_3^{\sigma\mu}f_2^{\alpha\beta}\kk_{3\beta}) \nonumber\\
	&\quad  + \frac{2(C_2(C_1-D_1-1)-D_1D_2)}{(p_1\cdot \kk_2)m^4}(\kk_2+\kk_3)^{\mu}p_{1\rho}f_3^{\rho\lambda}f_2^{\sigma\alpha}p_{1\alpha} \nonumber\\*
	&\quad + \frac{C_2(C_1-D_1-1)}{m^4}(f_2^{\mu\lambda}f_3^{\sigma\rho} + f_3^{\mu\lambda}f_2^{\sigma\rho})p_{1\rho} + \frac{D_1D_2}{m^4}p_{1\rho}(f_3^{\rho\mu}f_2^{\lambda\sigma}+f_2^{\rho\mu}f_3^{\lambda\sigma})
        \,,
\\[10pt]
	\mathcal{Y}^{\mu\lambda}&=\frac{2D_1^2}{(p_1\cdot \kk_2)m^4}p_{1\rho}(f_2^{\rho\mu}\kk_{2\alpha}f_3^{\alpha\lambda}-f_3^{\rho\mu}\kk_{3\alpha}f_2^{\alpha\lambda})+\frac{2D_1^2(\kk_2\cdot \kk_3)}{(p_1\cdot \kk_2)^2m^4}p_{1\rho}f_2^{\rho\mu}p_{1\alpha}f_3^{\alpha\lambda} \nonumber\\*
	&\quad + \frac{C_2}{(p_1\cdot \kk_2)^2m^4}p_{1\rho}(\kk_2^{\mu}f_2^{\rho\lambda}p_{1\alpha}f_3^{\alpha\beta}\kk_{2\beta}+\kk_{3}^{\mu}f_3^{\rho\lambda}p_{1\alpha}f_2^{\alpha\beta}\kk_{3\beta}) \nonumber\\*
	&\quad + \frac{C_2-2D_1D_2}{(p_1\cdot \kk_2)m^4}p_{1\alpha}(\kk_3^{\mu}f_2^{\alpha\beta}f^{}_{3\beta}{}^{\lambda}-\kk_2^{\mu}f_3^{\alpha\beta}f^{}_{2\beta}{}^{\lambda})  + \frac{2C_1C_2}{(p_1\cdot \kk_2)m^4}p_{1\alpha}(\kk_2^{\mu}f_2^{\alpha\beta}f^{}_{3\beta}{}^{\lambda}-\kk_3^{\mu}f_3^{\alpha\beta}f^{}_{2\beta}{}^{\lambda})\nonumber\\*
	&\quad - \frac{2C_1C_2}{m^4}f_2^{\mu\alpha}f^{}_{3\alpha}{}^{\lambda} - \frac{2C_2D_1}{m^6}p_{1\rho}p_{1\alpha}(f_2^{\rho\mu}f_3^{\alpha\lambda}+\eta^{\mu\lambda}f_2^{\rho\beta}f_{3\beta}{}^{\alpha})\,,\\[10pt]
	\mathcal{Y}^{\mu}&=\frac{2C_2D_1}{(p_1\cdot \kk_2)m^4}p_{1\rho}p_{1\alpha}(f_2^{\rho\mu}f_3^{\alpha\beta}\kk_{2\beta}-f_3^{\rho\mu}f_2^{\alpha\beta}\kk_{3\beta})\nonumber\\*
	&\quad - \frac{C_2(C_1+D_1-1)}{m^4}p_{1\rho}(f_3^{\rho\alpha}f^{}_{2\alpha}{}^{\mu}+f_2^{\rho\alpha}f^{}_{3\alpha}{}^{\mu}) \,.
\end{align}
\endgroup

We proceed next to the $\mathcal{O}(S^2)$ tree-level Compton amplitude of \ref{fieldtheory2}. Repeating at this order the classical scaling argument we described at ${\cal O}(S^1)$ shows 
that, in a fixed-spin calculation, the classical tree-level Compton amplitude is contained in the $\mathcal{O}(q^2)$ terms 
of the quantum tree-level Compton amplitude. Thus, we extract these terms from explicit $s=1,2,3$ calculations, extrapolate 
them to large spin and keep only the leading term.
The massive polarization vectors now appear in four structures,
\begin{align}
  \hskip -.2 cm 
  (\polm_1\cdot\polmb_4)^s\,, \hskip .9 cm
  (\polm_1\cdot\polmb_4)^{s-1}\polm_1^{[\mu}\polmb_4^{\nu]}\,, \hskip .9 cm 
  (\polm_1\cdot\polmb_4)^{s-1}\polm_1^{(\mu}\polmb_4^{\nu)}\,, \hskip .9 cm 
  (\polm_1\cdot\polmb_4)^{s-2}\polm_1^{[\mu}\polmb_4^{\nu]}\polm_1^{[\rho}\polmb_4^{\lambda]}\,, \hskip .2 cm 
\end{align}
where the first two correspond to the quantum spinless and $\mathcal{O}(S^1)$ amplitudes that can be ignored here.
We use the replacement~\eqref{eq:Srep2} and~\eqref{eq:Srep1} for the last two structures. It turns that the dependence on $s$ is simple so that we can extrapolate it to obtain the general $s$ dependence and take $s\rightarrow\infty$ limit. The kinematic coefficient of $S^{\mu\rho}S_{\rho}{}^{\nu}$ can be accessed by any $s\geqslant 1$; in the large $s$ limit, it exactly reproduces the $\mathcal{X}^{\mu\nu}$ shown in \cref{eq:Xcoeff}.
The structure $S^{\mu\nu}S^{\rho\lambda}$ appears for $s\geqslant 2$. A careful analysis with $s=2$ and $3$ gives identical results, so that we postulate that the coefficient of $S^{\mu\nu}S^{\rho\lambda}$ is independent of $s$.
Thus we find that the tree-level Compton amplitude of \ref{fieldtheory2} is
\begin{align}
    \mathcal{A}_\text{4, cl.}^{\text{FT2}}\Big|_{S^2}  &= (-\polm_1\cdot\polmb_4)^s  \left(S_{\mu\nu}S_{\lambda\sigma} \widetilde{\mathcal{X}}^{\mu\nu\lambda\sigma} + S_{\mu\lambda}S^{\lambda}{}_{\nu} \mathcal{X}^{\mu\nu}\right)\,,\\
    \widetilde{\mathcal{X}}^{\mu\nu\lambda\sigma} &= \frac{C_1^2(\kk_2\cdot\kk_3)}{2(p_1\cdot\kk_2)^2}f_2^{\mu\nu}f_3^{\lambda\sigma}+\frac{C_1(C_1-1)}{(p_1\cdot\kk_2)m^2}p_{1\rho}(f_3^{\rho\mu}\kk_2^{\nu}f_2^{\lambda\sigma}-f_2^{\rho\mu}\kk_3^{\nu}f_3^{\lambda\sigma})\,,\nonumber
\end{align}
where $\mathcal{X}^{\mu\nu}$ is defined in \eqn{eq:Ycoeff}.
These coefficients depend only on $C_1$ and $C_2$ and are independent of $D_1$ and $D_2$. We again observe the same pattern as in the linear-in-spin case,
\begin{align}
    \mathcal{A}_{4,\text{ cl.}}^{\text{FT1s}}\Big|_{S^2}  = \mathcal{A}_\text{4, cl.}^{\text{FT2}}\Big|_{S^2} \qquad\text{for}\quad D_1=C_1-1\,.
\end{align}
We note that the special value of $D_1$ also removes the dependence on $D_2$.

Similar to $\mathcal{O}(S^1)$, the $\mathcal{O}(S^2)$ Compton amplitudes of \ref{fieldtheory3} receive contributions 
from lower-spin intermediate states. 
Keeping the external states transverse, we get
\begin{align}\label{eq:A4S2}
    \mathcal{A}_\text{4, cl}^{\text{FT3s}}\Big|_{S^2}  &=  \mathcal{A}_\text{4, cl}^{\text{FT2}}\Big|_{S^2}  + (-\polm_1\cdot\polmb_4)^s S(p_1)_{\mu\nu}S(p_1)_{\lambda\sigma}\left[\frac{\widetilde{C}_1\widetilde{C}_2}{(p_1\cdot q_2)m^2} p_{1\rho}(f_3^{\rho\mu}\kk_2^{\nu}f_2^{\lambda\sigma}-f_2^{\rho\mu}\kk_3^{\nu}f_3^{\lambda\sigma})\right].
\end{align}
Just like the previous cases, the same formal relations hold between the additional Wilson coefficients in \ref{fieldtheory1} and \ref{fieldtheory3}. Indeed, comparing \cref{eq:A4S2} and \cref{eq:A1S2}, it is easy to see that
\begin{align}
    \mathcal{A}_\text{4, cl}^{\text{FT1s}}\Big|_{S^2} = \mathcal{A}_\text{4, cl}^{\text{FT3s}}\Big|_{S^2}\qquad\text{for}\quad i\widetilde{C}_1 = 1 - C_1 + D_1\;\text{ and }\;i\widetilde{C}_2 = D_2 - C_1 \,,
    \label{14mapS2}
\end{align}
which is identical to \cref{eq:map3p}. This demonstrates that the relation between extra Wilson coefficients and extra propagating degrees of freedom holds also at $\mathcal{O}(S^2)$.
A comparison between $\mathcal{A}_{4,\text{ cl.}}^{\text{FT1g}}$ and $\mathcal{A}_{4,\text{ cl.}}^{\text{FT3g}}$ at $\mathcal{O}(S^1 K^1)$ and $\mathcal{O}(K^2)$  requires that we include in the Lagrangian of \ref{fieldtheory3} a spin-$(s-2)$ field $\phi_{s-2}$, and additional operators that contribute independently at $\mathcal{O}(K^2)$, for example \cref{eq:LD2b} for \ref{fieldtheory1}. This is because the effect of $\mathcal{O}(K^2)$ operators show up at $\mathcal{O}(S^1 K^1)$ in the four-point Compton amplitudes due to the commutator $[K^2,K]\sim SK$.
Finally, we note that the spin-transition Compton amplitude $\mathcal{A}_4^{s\rightarrow s-1}$ under a fixed-spin calculation may superficially contain a super-classical contribution that does not cancel between the two matter channels. Consistency of the theory requires however that in the large-spin limit this term is subleading. We will assume that this cancellation holds as $s \rightarrow \infty$. It is nontrivial to carry out explicit calculations to demonstrate this, but would be worth investigating.


\subsection{Two-Body Amplitudes}
\label{sec:TwoBodyAmpl}

In previous subsections, we have explored and understood the effect of various types of interactions between
higher-spin fields and photons on Compton amplitudes, and the number of Wilson coefficients necessary to describe such interactions.
We found that, under suitable conditions like allowing spin magnitude change, this number is indeed larger than that required to describe the 
interactions of SSC preserving spins. 
The rationale of this exercise is to eventually understand their effects on two-body observables, such as the momentum impulse and
the spin kick.
It was originally suggested in the context of gravity that a larger number of Wilson coefficients may be required to describle more general interactions~\cite{Bern:2022kto}.
We therefore proceed to expose the photon-mediated two-body amplitudes and, in later sections, the observables
that follow from them as well as their comparison with a wordline perspective. 
We use the generalized unitarity  method \cite{Bern:1994zx, Bern:1994cg, Bern:1997sc} to construct the relevant integrands, while taking advantage of the simplifications introduced in Ref.~\cite{Kosmopoulos:2020pcd}.
To reduce the encountered loop integrals to known ones we make use of integration by parts~\cite{Chetyrkin:1981qh, Laporta:2000dsw} as implemented in FIRE~\cite{Smirnov:2008iw,Smirnov:2019qkx}.

We use the momentum and mass variables
\begin{align}
& \mb_1 = m_1^2 - q^2/4 \,, \hskip 1.6 cm 
\mb_2^2 = m_2^2 -q^2/4 \,, \hskip 1.6 cm 
y=\frac{\pb_1\cdot\pb_2}{\mb_1\mb_2}\,,
\nonumber \\
&
\pb_1 = p_1+q/2 = -p_4 - q/2\,, \hskip 1.6 cm
\pb_2 = p_2-q/2 = -p_3 +q/2\,, 
\label{eq:barVariables}
\end{align}
which are originally used for the expansion in the soft region of gravitational amplitudes in~\cite{Parra-Martinez:2020dzs}. We primarily focus on \ref{fieldtheory1} because this is what we compare with a worldline theory. 
Unitarity guarantees that the two-body amplitudes of \ref{fieldtheory2} can be obtained from those of \ref{fieldtheory1} by setting $D_1=C_1-1$ and imposing the covariant SSC, 
while the two-body amplitudes of \ref{fieldtheory3} can be obtained using the map between the extra Wilson coefficients proposed in the previous subsection.

\subsubsection{Tree Level}

The structure of the two-body amplitude at tree-level and in the classical limit is
\begin{align}
	i\Mtree = (4\pi\alpha)(\polM_1\cdot\polMb_4)(\polM_2\cdot\polMb_3)\left(\frac{d_{\TreeSymbol}}{q^2}\right) ,
	\label{tree_general}
\end{align}
where $q$ is the photon momentum, and the numerator $d_{\TreeSymbol}$ is a function of momenta and spin tensors which scales as $d_{\TreeSymbol}\sim q^0$ in the classical limit. 
Our results for \hyperlink{ft1g}{FT1g} through the quadratic order in spin are as follows:
\begin{align}\label{eq:dtree}
    d_{\TreeSymbol}\Big|_{\text{spinless}} &= 4iy\mb_1\mb_2\,,\nonumber \\[8pt]
    d_{\TreeSymbol}\Big|_{S_1^1S_2^0} ^{\text{FT1g}} &= -4\mb_2\gS_1{}_{\mu\nu}\left(C_{1(1)}\ub_2^{\mu}q^{\nu}-D_{1(1)}y \ub_1^{\mu}q^{\nu}\right) ,\nonumber\\[8pt]
    d_{\TreeSymbol}\Big|_{S_1^1S_2^1}^{\text{FT1g}} &= 4iC_{1(1)}C_{1(2)}\gS_1^{\mu\nu}q_{\nu}\gS_{2\mu\rho}q^{\rho}\\
    &\quad -4i\gS_1^{\mu\nu}q_{\nu}\gS_2^{\rho\lambda}q_{\lambda}\left(C_{1(1)}D_{1(2)}\ub_{2\mu}\ub_{2\rho}+C_{1(2)}D_{1(1)}\ub_{1\mu}\ub_{1\rho}-D_{1(1)}D_{1(2)}y\ub_{1\mu}\ub_{2\rho}\right), \nonumber\\[8pt]
    d_{\TreeSymbol}\Big|_{S_1^2S_2^0}^{\text{FT1g}} &= \frac{2i\mb_2}{\mb_1}\left[yC_{2(1)}\gS_1^{\mu\nu}q_{\nu}\gS_{1\mu\rho}q^{\rho}-yC_{2(1)}(\gS_1^{\mu\nu}\ub_{1\mu}q_{\nu})^2 - 2D_{2(1)}\gS_1^{\mu\nu}\ub_{1\mu}q_{\nu}\gS_1^{\rho\lambda}\ub_{2\rho}q_{\lambda}\right],\nonumber
\end{align}
where the spinless case agrees with Ref.~\cite{Holstein:2008sw, Kosower:2018adc,Bern:2021xze}.  The notation $C_{i (j)}$ and $D_{i (j)}$ refers to the $C_i$ and $D_i$ coefficients associated with body $j$.
If the external states are transverse, $(\polM_1\cdot\polMb_4)(\polM_2\cdot\polMb_3)=(\polm_1\cdot\polmb_4)^s(\polm_2\cdot\polmb_3)^s$, then the spin tensor obeys the covariant SSC, such that,
\begin{align}
    & d_{\TreeSymbol}\Big|_{S_1^1S_2^0}^{\text{FT1s}} = d_{\TreeSymbol}\Big|_{S_1^1S_2^0}^{\text{FT2}} = d_{\TreeSymbol}\Big|_{S_1^1S_2^0}^{\text{FT3s}} = -4C_{1(1)}\mb_2 S_1^{\mu\nu}\ub_{2\mu}q^{\nu}\,,\nonumber\\[8pt]
    & d_{\TreeSymbol}\Big|_{S_1^1S_2^1}^{\text{FT1s}} = d_{\TreeSymbol}\Big|_{S_1^1S_2^1}^{\text{FT2}} = d_{\TreeSymbol}\Big|_{S_1^1S_2^1}^{\text{FT3s}} = 4iC_{1(1)}C_{1(2)}S_1^{\mu\nu}q_{\nu}S_{2\mu\rho}q^{\rho}\,,\\[8pt]
    & d_{\TreeSymbol}\Big|_{S_1^2S_2^0}^{\text{FT1s}} = d_{\TreeSymbol}\Big|_{S_1^2S_2^0}^{\text{FT2}} = d_{\TreeSymbol}\Big|_{S_1^2S_2^0}^{\text{FT3s}} = \frac{2iC_{2(1)}y\mb_2}{\mb_1}S_1^{\mu\nu}q_{\nu}S_{1\mu\rho}q^{\rho}\,.\nonumber
\end{align}
The small velocity expansion of the first two expressions agrees with the results of Ref.~\cite{Holstein:2008sw}. 
They are related to \cref{eq:dtree} through the replacement $\gS_i\rightarrow S_i$ and $S_{i\mu\nu}\ub_i^{\nu}=0$, which holds to all orders in spin at tree level. We note that, to first order on $K^i$, the amplitudes of \hyperlink{ft3g}{FT3g} can also be obtained from \cref{eq:dtree} through the Wilson coefficient map \cref{eq:map3p}.

In \cref{sec:EFTHamiltonian} we compare observables from the amplitudes ${\cal A}^\text{FT1g}$ of \ref{fieldtheory1} and those from worldline calculations in the absence of an SSC. We find a perfect match both at ${\cal O}(\alpha)$, which follow from the amplitudes above, and at ${\cal O}(\alpha^2)$ which follow from the one-loop amplitudes we now summarize.

\subsubsection{One Loop}

While four-point Compton amplitudes are not relevant for the tree-level two-body scattering, they are an integral part of two-body scattering at one loop. The generalized unitarity  method \cite{Bern:1994zx, Bern:1994cg, Bern:1997sc} provides a means 
to construct the classically-relevant parts of the latter in terms of the former. 
We should therefore expect that the precise intermediate states contributing to Compton amplitudes have observable consequences for the scattering of two matter particles. In particular, we note that intermediate states of spin different 
from the external spin can be projected out either by using only transverse spin-$s$ fields \emph{or} by choosing particular values for the extra Wilson coefficients, see \cref{eq:WCmap}.
Since before loop integration, the part of the one-loop two-body amplitude that is relevant in the classical limit is literally the product of two Compton amplitudes summed over states, the latter observation must have hold at one loop as well. Thus, we may follow this strategy to compute the  one-loop two-body amplitude of \hyperlink{ft1g}{FT1g}.

The complete one-loop amplitude exhibits classically-singular, classical and quantum terms. The former two are
\begin{align}
	i\Moneloop = (4\pi\alpha)^2\Big[C_{\text{box}}(I_{\BoxSymbol}+I_{\CboxSymbol})+i\Mtri\Big]\, ,
\end{align}
where the first one, given by the box and crossed-boxed integrals 
\begin{align}
	I_{\BoxSymbol} &=\int\frac{\dd^{d}\lm}{(2\pi)^d}\frac{1}{\lm^2(\lm-q)^2(2\pb_1\cdot\lm+\ie)(-2\pb_2\cdot\lm+\ie)} \nonumber\\
	I_{\CboxSymbol} &=\int\frac{\dd^d\lm}{(2\pi)^d}\frac{1}{\lm^2(\lm-q)^2(2\pb_1\cdot\lm+\ie)(2\pb_2\cdot\lm+\ie)} \, ,
\end{align}
is the classically-singular part, while the second term, containing the triangle integral
\begin{align}
I_{\bigtriangleup} &=\int\frac{\dd^{d}\lm}{(2\pi)^d}\frac{1}{\lm^2(\lm-q)^2(2\pb_1\cdot\lm+\ie)} \, ,
\end{align}
is the classical part~\cite{Bjerrum-Bohr:2018xdl, Cheung:2018wkq}.

The spin-independent part of the amplitude is
\begin{align}
\label{1loopspinindependent}
    C_{\text{box}}\Big|_{\text{spinless}}&=-(\polM_1\cdot\polMb_4)(\polM_2\cdot\polMb_3)\left(d_{\TreeSymbol}\Big|_{\text{spinless}}\right)^2 \, ,\nonumber\\
    i\Mtri\Big|_{\text{spinless}}&=(\polM_1\cdot\polMb_4)(\polM_2\cdot\polMb_3)\frac{i(\mb_1+\mb_2)}{4\sqrt{-q^2}}\,.
\end{align}
As its tree-level counterpart, it agrees with Ref.~\cite{Bern:2021xze, Holstein:2023yxw} and it is the same in all three field theories.
%

The linear in spin part of the classically singular  term is
\begin{align}
	C_{\text{box}}\Big|_{S_1^1S_2^0}^{\text{FT1g}} = -(\polM_1\cdot\polMb_4)(\polM_2\cdot\polMb_3)\left(d_{\TreeSymbol}\Big|_{\text{spinless}}\times d_{\TreeSymbol}\Big|_{S_1^1S_2^0}^{\text{FT1g}}\right) \,,
\end{align}
As the spin-independent part (\ref{1loopspinindependent}), it is given by the product of tree-level amplitudes, in agreement with the expected exponential structure of the amplitude in the classical limit~\cite{Cheung:2018wkq, Bern:2019nnu, Bern:2021dqo}. The corresponding expression at higher powers of the spins should be given by the IBP reduction of such products of trees summed over all the possible ways of distributing the spins in the two factors.

The classical part of the one-loop two-body amplitude can be organized in terms of the various possible contractions of
spin tensors. As at tree level, we write explicitly the amplitude ${\cal A}^{\text{FT1g}}$ for \ref{fieldtheory1} and obtain 
the amplitudes in other theories via $\gS\rightarrow S$ and other limits on Wilson coefficients.
The structure of $i\Mtri\Big|_{S_1^{n_1}S_2^{n_2}}$ is
\begin{align}
	i\Mtri\Big|_{S_1^{n_1}S_2^{n_2}} = \frac{(\polm_1\cdot\polmb_4)^s(\polm_2\cdot\polmb_3)^s}{4\sqrt{-q^2}}\sum_i \alpha^{(n_1,n_2,i)}\mathcal{O}^{(n_1,n_2,i)} \,;
	\label{1loopclassical_general_form}
\end{align}
through second order in spin, the spin-tensor contractions are $\mathcal{O}^{(n_1,n_2,i)}$ are:

\begin{itemize}

\item Linear in spin:
\begin{align}
	\mathcal{O}^{(1,0,1)} = \gS_1^{\mu\nu}\ub_{2\mu}q_{\nu}\,,\qquad \mathcal{O}^{(1,0,2)} = \gS_1^{\mu\nu}\ub_{1\mu}q_{\nu}
\end{align}

\item Bilinear in spin:
\begin{align}
	\begin{array}{lll}
	\mathcal{O}^{(1,1,1)} = \gS_1^{\mu\nu}q_{\nu}\gS_{2\mu\rho}q^{\rho}\,, \hskip .5 cm \null &\mathcal{O}^{(1,1,2)} = \gS_1^{\mu\nu}\ub_{2\nu}\gS_{2\mu\rho}\ub_1^{\rho}\,,\quad &  \mathcal{O}^{(1,1,3)} = \gS_1^{\mu\nu}\ub_{2\mu}q_{\nu}\gS_2^{\lambda\sigma}\ub_{1\lambda}q_{\sigma}\,, \\
	\mathcal{O}^{(1,1,4)} = \gS_1^{\mu\nu}\gS_{2\mu\nu}\,, \quad & \mathcal{O}^{(1,1,5)} = \gS_1^{\mu\nu}\ub_{1\nu}\gS_{2\mu\rho}\ub_1^{\rho} \,, \quad & \mathcal{O}^{(1,1,6)} = \gS_1^{\mu\nu}\ub_{1\nu}\gS_{2\mu\rho}\ub_2^{\rho}\,, \\
	\mathcal{O}^{(1,1,7)} = \gS_1^{\mu\nu}\ub_{2\nu}\gS_{2\mu\rho}\ub_2^{\rho}\,, \quad & \mathcal{O}^{(1,1,8)} = \gS_1^{\mu\nu}\ub_{1\mu}q_{\nu}\gS_2^{\lambda\sigma}\ub_{1\lambda}q_{\sigma} \,, \hskip .3 cm \null  & \mathcal{O}^{(1,1,9)} = \gS_1^{\mu\nu}\ub_{1\mu}q_{\nu}\gS_2^{\lambda\sigma}\ub_{2\lambda}q_{\sigma} \,, \\
	\multicolumn{3}{c}{\mathcal{O}^{(1,1,10)} = \gS_1^{\mu\nu}\ub_{2\mu}q_{\nu}\gS_2^{\lambda\sigma}\ub_{2\lambda}q_{\sigma}\,, \hskip 1cm \mathcal{O}^{(1,1,11)} = \gS_1^{\mu\nu}\ub_{1\mu}\ub_{2\nu}\gS_2^{\lambda\sigma}\ub_{1\lambda}\ub_{2\sigma} \,,} 
\end{array}
\end{align}

\item Quadratic in spin:
\begin{align}
	\begin{array}{lll}
		\mathcal{O}^{(2,0,1)} = \gS_1^{\mu\nu}q_{\nu}\gS_{1\mu\rho}q^{\rho} \,, \hskip .5 cm \null &\mathcal{O}^{(2,0,2)} = (\gS_1^{\mu\nu}\ub_{2\mu}q_{\nu})^2\,, \quad &  \mathcal{O}^{(2,0,3)} = \gS_1^{\mu\nu}\gS_{1\mu\nu}\,,\\
		\mathcal{O}^{(2,0,4)} = \gS_1^{\mu\nu}\ub_{1\nu}\gS_{1\mu\rho}\ub_1^{\rho}\,,\quad & \mathcal{O}^{(2,0,5)} = \gS_1^{\mu\nu}\ub_{1\nu}\gS_{1\mu\rho}\ub_2^{\rho}\,,\quad & \mathcal{O}^{(2,0,6)} = \gS_1^{\mu\nu}\ub_{2\nu}\gS_{1\mu\rho}\ub_2^{\rho}\,, \\
		\mathcal{O}^{(2,0,7)} = (\gS_1^{\mu\nu}\ub_{1\mu}q_{\nu})^2\,, \quad & \mathcal{O}^{(2,0,8)} = \gS_1^{\mu\nu}\ub_{1\mu}q_{\nu}\gS_1^{\lambda\sigma}\ub_{2\lambda}q_{\sigma}\,, \hskip .4 cm \null & \mathcal{O}^{(2,0,9)} = (\gS_1^{\mu\nu}\ub_{1\mu}\ub_{2\nu})^2\,,
	\end{array}
\end{align}
\end{itemize}
All contractions that contain the covariant SSC constraints, $\gS_i^{\mu\nu}\ub_{i\mu}$, vanish for ${\cal A}^{\text{FT1s}}$, ${\cal A}^{\text{FT2}}$ and ${\cal A}^{\text{FT3s}}$.
The coefficients of $\mathcal{O}^{(n_1,n_2,i)}$ at linear order in spin are:
\begin{align}
	\alpha^{(1,0,1)} &= -\frac{y}{(y^2-1)\mb_1}\Big[2C_{1(1)}\mb_1 + (C_{1(1)}^2 - 2C_{1(1)}D_{1(1)} + D_{1(1)}^2 + 2D_{1(1)})\mb_2\Big]\,,\nonumber\\[8pt]
	\alpha^{(1,0,2)} &= \frac{1}{(y^2-1)\mb_1}\Big\{\Big[(y^2+1)C_{1(1)}+(y^2-1)D_{1(1)}\Big]\mb_1\nonumber\\
	&\qquad\qquad + \Big[C_{1(1)}^2 - (y^2+1)C_{1(1)}D_{1(1)}+y^2D_{1(1)}^2+(3y^2-1)D_{1(1)}\Big]\mb_2\Big\}\,,
\end{align}
and we collect the coefficients up to quadratic order in spin in an ancillary file.
As we reduce $\mathcal{A}^{\text{FT1g}}$ to $\mathcal{A}^{\text{FT1s}}$, the coefficients of the surviving spin structures under the covariant SSC are unchanged.
To obtain the amplitude for \ref{fieldtheory2} we further impose $D_1=C_1-1$, which also makes the $D_2$ dependence vanish up to the quadratic order in spin as in the Compton amplitudes.
Similarly, to obtain the amplitudes ${\cal A}^\text{FT3s}$ we use the relations~\eqref{eq:map3p} to replace the coefficients $D_1$ and $D_2$ by $\widetilde{C}_1$ and $\widetilde{C}_2$ after imposing the covariant SSC.
Last but not least, we can also obtain $\mathcal{A}^{\text{FT3g}}$ up to linear order in $K$ from $\mathcal{A}^{\text{FT1g}}$ by simply using the relations~\eqref{eq:map3p}.


\section{Worldline Theories}
\label{sec:WL}


What worldline theory can reproduce the field-theory results of the previous sections?  In the field theories where multiple spin states propagate, the spin vectors magnitude is no longer conserved so to match this one needs to introduce additional degrees of freedom on the worldline.  Because these additional degrees of freedom are constrained by the Lorentz generator algebra, the natural choice is to find these degrees of freedom in the spin tensor itself.  In \sect{sec:EFTHamiltonian}, we construct a two-body Hamiltonian that explicitly exhibits these additional dynamical variables.  In this section our task is to find a modified worldline that produces the same results as the field theory. 

We start from a standard worldline construction~\cite{JanSteinhoff:2015ist} with the SSC corresponding to \ref{worldline1}, listed in \sect{sec:intro}.  We see that the results we obtain for this theory then match the field theory containing only a single massive quantum spin state~\cite{Kim:2023drc, Haddad:2023ylx}, which is related to the fact that both necessarily preserve the spin-vector magnitude.  To match field-theory results  when multiple quantum spin states are present,  we introduce additional degrees of freedom on the worldline by releasing the SSC, corresponding to \ref{worldline2}.  As in general relativity this has {\em no physical effect} at first order in the coupling~\cite{Bern:2022kto}, but starting at second order in the coupling, physical differences can appear; in general relativity physical effects start at cubic order in the spin tensor, but in electrodynamics this occurs at linear order.


Specifically, we compute the tree-level Compton amplitude to quadratic order in spin and probe-limit ${\cal O}(\alpha^2)$ two-body impulse and spin kicks to linear order in spin with a scalar source. We do so initially using \ref{worldline1} with the covariant SSC imposed via a Lagrange multiplier. Then, we switch to \ref{worldline2} by removing the Lagrange multiplier terms enforcing the SSC constraint. This Compton amplitude of the modified worldline formalism has the same spin tensor dependence as found in the classical limit of the amplitude ${\cal A}^{\text{FT1g}}$ of field theory \ref{fieldtheory1}
without a physical state projector limiting it to the states of a single quantum spin. 
We find that not only do the equations of motion consistently evolve all the degrees of freedom,  but that it is possible to match the observables of the modified worldline with the field theory, with a direct correspondence between the Wilson coefficients of the two formalisms. The key consequence is that both have a larger number of independent Wilson coefficients than the conventional worldline approach in which the SSC is imposed. We emphasize that the match is rather nontrivial.  

\subsection{Worldline Action with Dynamical Mass Function}

We begin with a brief review of the worldline formalism, following Ref.~\cite{JanSteinhoff:2015ist}. The worldline formalism seeks to describe the evolution of a body of matter in terms of its spacetime location and internal degrees of freedom. We refer to the spacetime location of the body ``center'' in coordinates as $z^\mu(\lambda)$ where $\lambda$ is a real parameter which parameterizes the worldline, called the worldline time. For now we denote the internal degrees of freedom of the body as $\phi^a(\lambda)$ where $a$ is an index running over all of those internal degrees of freedom. Below we take these degrees of freedom to track the orientation of the body but for now the particular structure of these degrees of freedom is not important. The body's evolution is described by an action which is reparameterization invariant under monotonic redefinitions of the worldline time $\lambda' = \lambda'(\lambda)$. The reparametrization invariance can be imposed directly through the introduction of an einbein field $\mathtt{e}(\lambda)$. The einbein is defined to transform under reparameterizations as:
\begin{equation}
	\mathtt{e}'(\lambda') = \frac{d\lambda}{d\lambda'}\, \mathtt{e}(\lambda) \,,
\end{equation}
A generic reparameterization invariant action is then of the form:
\begin{equation}
	S[\mathtt{e}, z, \phi] = \int_{-\infty}^\infty \mathcal{L}\left(z, \frac{\dot z}{\mathtt{e}}, \phi, \frac{\dot\phi}{\mathtt{e}}\right) \mathtt{e} \, d\lambda \,,
\end{equation}
where dots indicate differentiation with respect to $\lambda$. Defining the conjugate momenta as usual:
\begin{equation}
	p_\mu = -\frac{\partial(\mathcal{L} \mathtt{e})}{\partial\dot z^\mu}\,, \hskip 1.5 cm 
        \pi_a = -\frac{\partial(\mathcal{L}\mathtt{e})}{\partial \dot \phi^a} \,,
\end{equation}
the Hamiltonian form of the action can be written as:
\begin{equation}
	S[\mathtt{e}, z, p, \phi, \pi] = \int_{-\infty}^\infty\left(-\pi_a \dot \phi^a - p_\mu \dot z^\mu - \mathtt{e} H(z, p, \phi, \pi)\right)d\lambda\,,
\end{equation}
and $p$, $\pi$, and $H$ are reparameterization invariant.
It is useful to introduce the notation:
\begin{equation}
 |p| = \sqrt{p^\mu p_\mu}, \qquad \hat p^\mu = \frac{p^\mu}{|p|} \,.
\end{equation}
For a free particle, the Hamiltonian $H = -|p| + m$ produces the geodesic equation of motion. In general, $H = -|p| + m + \delta H(z, p, \phi, \pi)$ for some function $\delta H$ containing all additional couplings. The on-shell constraint imposed by the einbein's equation of motion is always $H = 0$, which then determines $|p| = m + \delta H$. So, it is useful to introduce the dynamical mass function $\mathcal{M}(z, \hat p, \phi, \pi)$ as the solution for $|p|$ imposed by the einbein equation of motion: $|p| = \mathcal{M}(z, \hat p, \phi, \pi)$. Then, we can take the Hamiltonian:
\begin{equation}
    H(z, p, \phi, \pi) = -|p| + \mathcal{M}(z, \hat p, \phi, \pi).
\end{equation}
Note that this is equivalent to taking $H=p^2-\mathcal M^2$ as in \cite{JanSteinhoff:2015ist}, up to a redefinition of the Lagrange multiplier $\mathtt e$.

In the context of electrodynamics it is possible to add the minimal coupling through the dynamical mass function but then the conjugate momentum of the body is not gauge invariant. Instead, by taking $p_\mu$ to be the kinetic momentum (the conjugate momentum plus $Q A_\mu$), we can have $p_\mu$ and consequently $\mathcal{M}$ be gauge invariant at the cost of shifting $p_\mu\dot z^\mu$ to $(p_\mu-QA_\mu) \dot z^\mu$. Thus, to couple the worldline particle to electromagnetism it is simplest to use the action:
\begin{equation}
	S[\mathtt{e}, z, p, \phi, \pi] = \int_{-\infty}^\infty\left(-p_\mu\dot z^\mu + QA_\mu \dot z^\mu - \pi_a\dot \phi^a + \mathtt{e}\left(\sqrt{p^\mu p_\mu} -\mathcal{M}(z,\hat p, \phi, \pi)\right)\right) d\lambda \,,
\label{WLAction}
\end{equation}
with a gauge and reparameterization invariant Lorentz-scalar dynamical mass $\mathcal{M}$. 

\subsection{Worldline Theory with SSC}

\subsubsection{Worldline Spin Degrees of Freedom}
\label{sec:WLspin}

The standard worldline formulation incorporates spin in a way reminiscent of rigid bodies in classical mechanics.  For a moving body, there is some point defined as the ``center'' of that body, tracked by the worldline, which moves in spacetime and we assume that other points of the body move along with that center in ``quasirigid'' motion, as defined in Ref.~\cite{Ehlers:1977rud}, requiring that the internal structure is essentially unchanged.\footnote{\baselineskip=14pt More precisely, quasirigidity is the requirement that the multipole moments of the body's current density and stress tensor evolve only by translating along the worldline and Lorentz transforming according to the orientation tracking tetrad.}  The orientation of the body is tracked by a tetrad ${e^\mu}_A(\lambda)$ that represents the change of internal body displacements undergone during the motion with respect to some arbitrary default frame. Capital Latin indices are used for the body's internal Lorentz indices while lowercase Greek indices are used as spacetime indices. As usual, the tetrad satisfies: 
\begin{equation}
	{e^\mu}_A{e^\nu}_B \eta^{AB} = g^{\mu\nu} , \hskip 1.5 cm
        g_{\mu\nu}{e^\mu}_A{e^\nu}_B = \eta_{AB} \,.
	\label{eq:WLtetradconditions}
\end{equation}
Internal body displacements are defined in the body's center of momentum frame, so that $\hat p^\mu$ is instantaneously taken as the time direction. Thus by definition we take:
\begin{equation}
{e^\mu}_0 = \hat p^\mu \, .
\end{equation}

Beyond this condition ${e^\mu}_A$ may be any tetrad satisfying Eq.~\eqref{eq:WLtetradconditions}. Any such tetrad can be decomposed into (1) a tetrad which is parallel transported along the worldine, then boosted by a standard boost so that its timelike element is boosted to $\hat p^\mu$, and (2) an arbitrary little-group element of $\hat p^\mu$. The three little-group parameters of $\hat p^\mu$ can then be taken to be the $\phi^a$ coordinates. The spin angular momentum of the body is the generator of Lorentz transformations of the body orientation about the body center, and so is given by:
\begin{equation}
\gS_{\mu\nu} = -\pi_a \left.\frac{d\phi^a}{d\theta^{\mu\nu}}\right|_{\theta = 0}\,,
\end{equation}
with Lorentz transformation parameters $\theta^{\mu\nu}$. A short computation with the above definitions reveals that they enforce the covariant SSC:
\begin{equation}
\gS_{\mu\nu}p^\nu = 0 \,.
\end{equation}
In addition:
\begin{equation}
	-\frac{1}{2}\gS_{\mu\nu}\Omega^{\mu\nu} = \pi_a\dot\phi^a\,,
\end{equation}
where the angular velocity tensor ${\Omega^{\mu\nu}}$ is defined by:
\begin{equation}
	{\Omega^{\mu\nu}} = \eta^{AB}{e^\mu}_A\frac{D{e^\nu}_B}{D\lambda} \,.
\end{equation}

Using the spin tensor and the arbitrariness of the default frame of the body, the action for a spinning body takes the form:
\begin{align}
        S[\mathtt{e}, \xi, \chi, z, p, e, \gS] = \int_{-\infty}^\infty&\left(-(p_\mu-QA_\mu)\dot z^\mu + \frac{1}{2}\gS_{\mu\nu}\Omega^{\mu\nu} + \mathtt{e}(|p| - \mathcal{M}(z, \hat p, \gS))\right.\nonumber \\
	&\phantom{\qquad}\left.\vphantom{\frac{1}{2}}+ \xi_\mu \gS^{\mu\nu} p_\nu + \chi_\mu({e^\mu}_0-\hat p^\mu)\right)d\lambda \,.
        \label{eq:WLactbig}
\end{align}
Lagrange multipliers $\xi_\mu$ and $\chi_\mu$ enforce $\gS^{\mu\nu}\hat p_\nu = 0$ and ${e^\mu}_0 = \hat p^\mu$. This formulation of the action imposes the covariant SSC $\gS_{\mu\nu}p^\nu = 0$, corresponding to the \ref{worldline1} theory.

One can shift the definition of the worldline $z^\mu$ and in doing so one finds that the definition of the spin changes as does the constraint satisfied by the spin. Thus, one can change to a new SSC through a shift of the worldline. In this formalism, the ability to locally shift the definition of the worldline in this way may be thought of as a gauge transformation \cite{Steinhoff:2015ksa, Levi:2015msa, Vines:2016unv} and the Lagrange multipliers supplied to enforce the covariant SSC and ${e^\mu}_0 = \hat p^\mu$ correspond to a gauge fixing.  Because the SSC removes the $\gS_{0a}$ components of the spin tensor in the body's center of momentum frame, these timelike components are not physical degrees of freedom.  (Even when an SSC other than the covariant SSC is considered, these timelike components are determined by the other degrees of freedom using the appropriate SSC.)

\subsubsection{Equations of Motion with SSC}

The variation of Eq.~\eqref{eq:WLactbig} in Minkowski space gives an electromagnetic version of the Mathisson–Papapetrou–Dixon (MPD)~\cite{Dixon:1970mpd,Mathisson:1937zz,Papapetrou:1951pa} equations:
\def\F{F}
\begin{align}
	&\dot z^\mu = \hat p^\mu - \frac{1}{\mathcal{M}}{\frac{\partial\mathcal{M}}{\partial\hat p_\mu}} + \frac{\hat p_\sigma}{\mathcal{M}}\frac{\partial\mathcal{M}}{\partial\hat p_\sigma}\hat p^\mu + \gS^{\mu\nu} \frac{\frac{\partial\mathcal{M}}{\partial z^\nu} - 2\frac{\partial\mathcal{M}}{\partial \gS^{\nu\rho}} p^\rho - Q\F_{\nu\rho}\left(\hat p^\rho - \frac{1}{\mathcal{M}}\frac{\partial\mathcal{M}}{\partial\hat p_\rho} + \frac{\hat p_\sigma}{\mathcal{M}}\frac{\partial\mathcal{M}}{\partial\hat p_\sigma}\hat p^\rho\right)}{\mathcal{M}^2 - \frac{Q}{2}\gS^{\alpha\beta}\F_{\alpha\beta}} \,, \nonumber \\
	&\dot p_\mu = -Q\F_{\mu\nu} \dot z^\nu + \frac{\partial\mathcal{M}}{\partial z^\mu}\,, \nonumber \\
	&\dot \gS_{\mu\nu} = p_\mu \dot z_\nu - p_\nu\dot z_\mu - 2 {\gS_\mu}^\rho \frac{\partial \mathcal{M}}{\partial \gS^{\rho\nu}} + 2 {\gS_{\nu}}^\rho\frac{\partial\mathcal{M}}{\partial \gS^{\rho\mu}} - \frac{\partial \mathcal{M}}{\partial\hat p^\mu} \hat p_\nu + \frac{\partial\mathcal{M}}{\partial\hat p^\nu}\hat p_\mu \,.
\label{EqOfMotionSSC}
\end{align}
In varying to find these equations of motion we find that the equations of motion are consistent with simply taking $\chi_\mu = 0$ and so if $\hat p^\mu = {e^\mu}_0$ is imposed as an initial condition then never adding the $\chi_\mu$ term to the action still preserves this condition for later times. 

At linear order in spin the generic symmetry consistent dynamical mass function is:
\begin{equation}
	\mathcal{M} = m - \frac{Q C_1}{2m} \gS^{\mu\nu} \F_{\mu\nu} \,,
\label{DynamicalMassSSC}
\end{equation}
for constant free mass $m$ and Wilson coefficient $C_1$. 
With this form of the dynamical mass function, the equations of motion to linear in spin order are:	
\begin{align}
	&\dot z^\mu = \left(1 + \frac{Q C_1}{2m^2} \gS^{\alpha\beta}\F_{\alpha\beta}\right)\frac{p^\mu}{m} + \frac{Q(C_1-1)}{m^3} \gS^{\mu\nu}\F_{\nu\rho}p^\rho + \mathcal{O}(\gS^2)\,, \nonumber \\
	&\dot p_\mu = -Q\F_{\mu\nu}\dot z^\nu - \frac{QC_1}{2m} \gS^{\rho\sigma}\partial_\mu \F_{\rho\sigma} + \mathcal{O}(\gS^2)\,, \nonumber \\
	&\dot \gS_{\mu\nu} = p_\mu\dot z_\nu - p_\nu \dot z_\mu + \frac{QC_1}{m}\left(\gS_{\mu\rho}{\F^\rho}_\nu - \gS_{\nu\rho}{\F^\rho}_\mu\right) + \mathcal{O}(\gS^2) \,. \label{eq:WLSSCEOM}
\end{align}
These linear in spin equations of motion depend only on a single Wilson coefficient following from the fact that with the SSC imposed, the only independent linear in spin operator is the one in \eqn{DynamicalMassSSC}.  This is similar to the situation in general relativity where the SSC allows only a single independent Wilson coefficient at the linear in spin level~\cite{Levi:2015msa}.   The appearance of two Wilson coefficients in the field theory (cf.~Eqs.~(\ref{eq:LS1}), (\ref{eq:s1_comp}) and (\ref{eq:s1_comp_FT2})) and one coefficient in the worldline with the SSC imposed is the analog of the similar appearance of a different number of Wilson coefficients in general relativity between the field-theory and worldline descriptions starting at the spin-squared level in the action~\cite{Bern:2022kto}. 
\subsection{Worldline Theory with no SSC}
\label{sec:WLnoSSC}

In Sect.~\ref{sec:WLspin} we reviewed that an SSC (and particularly the covariant SSC) is natural for the worldline formalism for quasirigid bodies. Here we consider a modified version of the worldline formalism in which we ``remove'' the SSC. This corresponds to our worldline theory \ref{worldline2}. It explicitly introduces additional physical degrees of freedom into the theory. Remarkably we find that this modified worldline theory cleanly matches the field-theory results of \ref{fieldtheory1}g at 2PL $\mathcal O (\gS^1)$, including its extra independent Wilson coefficients.  This then allows us to interpret the appearance of extra Wilson coefficients purely on the worldline, tying them to additional dynamical degrees of freedom. A similar construction was described in Ref.~\cite{Kim:2023drc}. We find that these extra degrees of freedom allow for the magnitude of the spin vector to change.

\subsubsection{Removing the SSC}

Consider the worldline action,
\begin{align}
	S[\mathtt{e}, \xi, \chi, z, p, e, \gS] = \int_{-\infty}^\infty&\left(-(p_\mu-QA_\mu)\dot z^\mu + \frac{1}{2}\gS_{\mu\nu}\Omega^{\mu\nu} + \mathtt{e}(|p| - \mathcal{M}(z, \hat p, \gS))\right) d\lambda \,,
        \label{eq:WLactbigNoSSC}
\end{align}
which is identical to \eqn{eq:WLactbig}, except that the Lagrange multiplier terms that enforce the SSC are dropped. By not including these, the interdependence between the definition of the body center degrees of freedom $(z, p)$ and the body orientation degrees of freedom $(e, \gS)$ is removed.
As already noted, in \eqn{eq:WLactbig} the SSC implies that the $\gS_{0a}$ components of the spin tensor are not independent physical degrees of freedom. 
In contrast, in \eqn{eq:WLactbigNoSSC} with no SSC imposed we are explicitly promoting these timelike components to be treated as physical. As we shall see, this does not lead to inconsistencies in the equations of motion, but instead adds dynamical degrees of freedom.

The variation of Eq.~\eqref{eq:WLactbigNoSSC} with no SSC imposed results in equations of motion,
\begin{align}
	&\dot z^\mu = \hat p^\mu - \frac{1}{\mathcal{M}}{\frac{\partial\mathcal{M}}{\partial\hat p_\mu}} + \frac{\hat p_\sigma}{\mathcal{M}}\frac{\partial\mathcal{M}}{\partial\hat p_\sigma}\hat p^\mu\,, \nonumber \\
	&\dot p_\mu = -q\F_{\mu\nu} \dot z^\nu + \frac{\partial\mathcal{M}}{\partial z^\mu}\,, \nonumber \\
	&\dot \gS_{\mu\nu} = p_\mu \dot z_\nu - p_\nu\dot z_\mu - 2 {\gS_\mu}^\rho \frac{\partial \mathcal{M}}{\partial \gS^{\rho\nu}} + 2 {\gS_{\nu}}^\rho\frac{\partial\mathcal{M}}{\partial \gS^{\rho\mu}} - \frac{\partial \mathcal{M}}{\partial\hat p^\mu} \hat p_\nu + \frac{\partial\mathcal{M}}{\partial\hat p^\nu}\hat p_\mu \,.
\label{EqOfMotionNoSSC}
\end{align}
Comparing to Eq.~\eqref{EqOfMotionSSC} we see that only the equation of motion for the worldline trajectory $z^\mu$ differs from the case with the SSC imposed. Moreover, the absence of the SSC Lagrange multiplier term results in this equation being simpler.

At linear order in spin the generic symmetry consistent dynamical mass function is:
\begin{equation}
	\mathcal{M} = m - \frac{Q C_1}{2m}\gS^{\mu\nu}\F_{\mu\nu} - \frac{Q D_1}{m} \hat p_\mu \gS^{\mu\nu}\F_{\nu\rho} \hat p^\rho \,,
        \label{eq:WLDMF}
\end{equation}
for constants $m, C_1, D_1$. In this case, instead of the single Wilson coefficient $C_1$ we have the additional coefficient $D_1$ analogous to the appearance of a second coefficient in the field theory \ref{fieldtheory2}.  To give a physical meaning to $D_1$ it is useful to define the spin vector $S^\mu$ and mass moment vector $K^\mu$:
\begin{equation}
	S^\mu = \frac{1}{2}\epsilon^{\mu\nu\rho\sigma} \hat p_\nu \gS_{\rho\sigma}\,, \hskip 1.5 cm
        K^\mu = -\gS^{\mu\nu}\hat p_\nu \,,
\label{eq:K_WL_def}
\end{equation}
where $\epsilon_{0123} = +1$. The boost vector $K^\mu$ is precisely what is eliminated when the covariant SSC is imposed, or equivalently what is algebraically constrained when a different SSC is used. The complete information in the spin tensor is recovered from these two vectors by:
\begin{equation}
\label{eq:WLSpinTDef}
	\gS^{\mu\nu} = \hat p^\mu K^\nu - K^\mu \hat p^\nu + \epsilon^{\mu\nu\rho\sigma} \hat p_\rho S_\sigma \,.
\end{equation}
Directly, $K^\mu$ is the generator of ``intrinsic'' Lorentz boosts (where by ``intrinsic'' we mean acting only on the internal degrees of freedom). As well, $-\frac{K^\mu}{|p|}$ can be interpreted as the displacement between the actual worldline $z^\mu(\lambda)$ being used and the worldline $z^\mu_{\text{COM}}(\lambda)$ that would trace out the center of mass of the body. To see this, look at the total angular momentum $J^{\mu\nu}$:
\begin{equation}
	J^{\mu\nu} = z^\mu p^\nu - p^\mu z^\nu + \gS^{\mu\nu} = \left(z^\mu - \frac{K^\mu}{|p|}\right) p^\nu - p^\mu \left(z^\nu - \frac{K^\nu}{|p|}\right) + \epsilon^{\mu\nu\rho\sigma}\hat p_\rho S_\sigma \,.
\label{WLDefK}
\end{equation}
We can see that if the definition of the worldline is shifted by $-\frac{K^\mu}{|p|}$ to a new worldline $z'^\mu = z^\mu - \frac{K^\mu}{|p|}$ then the resulting new spin tensor $\gS'^{\mu\nu}$ would satisfy the covariant SSC. In the conventional worldline formalism this is considered as an allowed redefinition which should lead to a physically equivalent theory (in that whether the spin and coupling expansions are performed about one or the other should not affect observables). Here we do not require it to be so.

In the previous discussion, the bodies were treated as point-like.  It is useful to remind ourselves of the meaning $ \clK$
in the context of classical extended bodies.
A familiar analysis of the spin vector can allow further insight into the meaning of $K^\mu$. Let $\textbf{J}$ be the generator of rotations about the origin (not body centered) acting on a matter distribution with energy density $\mathcal{E}(\textbf{x})$ and linear momentum density $\boldsymbol{\wp}(\textbf{x})$ in a region $V$ of space,
\begin{equation}
    \textbf{J} = \int_V \textbf{x}\times\boldsymbol{\wp}(\textbf{x})d^3\textbf{x}.
\end{equation}
When the center of the body is identified with the point $\textbf{z}$, the orbital generator of Lorentz boosts is of course,
\begin{equation}
    \textbf{L} = \textbf{z}\times\textbf{p}.
\end{equation}
Thus, the ``intrinsic" generator of rotations (the spin) of the body is given by a familiar formula,
\begin{equation}
    \clS = \textbf{J} - \textbf{L} = \int_V(\textbf{x} - \textbf{z})\times\boldsymbol{\wp}(\textbf{x})d^3\textbf{x}.
\end{equation}
Now performing the same analysis for the generator of Lorentz boosts, let $\clK_\text{total}$ be the generator of Lorentz boosts about the origin acting on the matter distribution,
\begin{equation}
    \clK_\text{total} = \int_V(t \boldsymbol{\wp}(\textbf{x})-\textbf{x} \mathcal{E}(\textbf{x})) d^3\textbf{x}.
\end{equation}
The ``orbital" generator of Lorentz boosts is then,
\begin{equation}
    \clK_\text{orbital} = \textbf{p} t -\text{E}\textbf{z}
\end{equation}
where E and $\textbf{p}$ are the total energy and momentum of the body. Thus, the ``intrinsic" generator of Lorentz boosts of the body is:
\begin{equation}
    \clK = \clK_\text{total} - \clK_\text{orbital} = \text{E} \textbf{z} -\int_V\textbf{x} \mathcal{E}(\textbf{x})d^3\textbf{x}
\end{equation}
Let $\textbf{z}_\text{COM}$ be the center of momentum position of the body in the center of momentum frame ($\text{E} = |p|$). Then, automatically:
\begin{equation}\label{zCOM}
    \textbf{z}_\text{COM} = \frac{1}{\text{E}}\int_V\textbf{x} \mathcal{E}(\textbf{x}) d^3\textbf{x} \implies \textbf{z}_\text{COM} = \textbf{z} - \frac{\clK}{|p|}.
\end{equation}
This precisely establishes the interpretation of $-\frac{K^\mu}{|p|}$ as a displacement between the worldline around which the spin and coupling expansions are performed and the worldline which tracks the center of mass of the body. 

Note that we use a different convention in this section compared to \sect{sec:fieldtheory}. In particular, the worldline $\clK$ and the field-theory $\pmb{K}$ are related by an analytic continuation, $i \pmb{K} \mapsto \clK$ with both $\pmb{K}$ and $\clK$ being real, while the rest-frame spin vectors are simply equal, $\pmb{S}\leftrightarrow \clS$. We comment further in \sect{sec:KReality} on the rationale behind this analytic continuation.

Writing $K^\mu$ as a spatial integral moment of the energy-momentum tensor as above identifies it as a mass dipole moment of the body about the worldine position $z^\mu$. This identification can be made directly from Dixon's formalism~\cite{Dixon:1970mpd}. For a body with a charge density proportional to its mass density then $-\frac{Q}{|p|}K^\mu$ would be the electric dipole moment of the body. However, for a generic object it is not necessarily the case that these densities are proportional and so we need not assume that the electric dipole moment is $-\frac{Q}{|p|}K^\mu$. In particular, in the body's center of momentum frame its energy is simply its dynamical mass function minus $Q A_0$ and in that frame \eqref{eq:WLDMF} becomes:
\begin{equation}
	\mathcal{M} = m + \frac{Q (C_1 - D_1)}{m}\textbf{E}\cdot \clK - \frac{Q C_1}{m} \textbf{B}\cdot \clS + \mathcal{O}(\F^2) \,.
\end{equation}

Thus the induced electric dipole moment $\textbf{d}$ and magnetic dipole moment $\boldsymbol{\mu}$ relative to the worldline center $\textbf{z}$:
\begin{equation}
	\textbf{d} = -\frac{Q(C_1-D_1)}{m}\clK, \quad\quad \boldsymbol{\mu} = \frac{Q C_1}{m}\clS \,.
\label{eq:InducedDipoles}
\end{equation}

Immediately, $2C_1$ is the gyromagnetic ratio  of the body (which should take the value 1 for a classical distribution of mass and charge which are proportional). For a distribution in which mass and charge are proportional, $C_1 - D_1 = 1$. Here we consider the possibility that it takes a generic value different from 1. The value of $C_1 - D_1 = 1$ is explicitly required by a worldline formalism which is assumed to have worldline shift symmetry \cite{Steinhoff:2015ksa,Levi:2015msa,Vines:2016unv} because the definition of the electric dipole moment immediately implies a shift of the dipole moment by $-\frac{Q}{|p|} K^\mu$ whenever the worldline is shifted by $-\frac{K^\mu}{|p|}$. Thus, $C_1 - D_1 \neq 1$ breaks the worldline shift symmetry.

Of course, to have a proper description of extended bodies that fits into the \ref{worldline2} framework one should understand the constraints on the energy and momentum distributions arising from the Lorentz algebra.  It would also be very interesting to directly connect extended objects with appropriate distributions of energy and momentum to the extra Wilson coefficient of  \ref{worldline2}.

\subsubsection{Equations of Motion with no SSC}

With the dynamical mass function \eqref{eq:WLDMF} we find equations of motion in \ref{worldline2} to linear order in spin:
\begin{align}
	\dot z^\mu =& \frac{p^\mu}{m}\left(1 + \frac{QC_1}{2m^2} \gS^{\rho\sigma}\F_{\rho\sigma} - \frac{QD_1}{m^4} p_\nu \gS^{\nu\rho}\F_{\rho\sigma}p^\sigma\right)
        + \frac{Q D_1}{m^3} p^\rho \gS_{\rho\sigma}\F^{\sigma\mu}+ \frac{QD_1}{m^3} \gS^{\mu\nu}\F_{\nu\rho}p^\rho + \mathcal{O}(\gS^2) \,, \hskip .3 cm 
        \nonumber \\
	\dot p_\mu =& -Q \F_{\mu\nu}\dot z^\nu - \frac{QC_1}{2m} \gS^{\rho\sigma}\partial_\mu \F_{\rho\sigma} - \frac{Q D_1}{m^3} p_\rho \gS^{\rho\sigma}\partial_\mu\F_{\sigma\alpha} p^\alpha + \mathcal{O}(\gS^2)
        \nonumber \\
	\dot \gS^{\mu\nu} =& -\frac{Q C_1}{m}\left({\F^\mu}_\rho \gS^{\rho\nu} - {\F^\nu}_\rho \gS^{\rho\mu}\right) - \frac{Q D_1}{m^3}\left(\F^{\mu\rho} p_\rho \gS^{\nu\sigma}p_\sigma - \F^{\nu\rho}p_\rho \gS^{\mu\sigma}p_\sigma\right)
        \nonumber \\
	&+ \frac{Q D_1}{m^3}\left(p^\mu \gS^{\nu\rho}\F_{\rho\sigma}p^\sigma - p^\nu \gS^{\mu\rho}\F_{\rho\sigma}p^\sigma\right) + \mathcal{O}(\gS^2). 
	\label{eq:WLnoSSCEOM}
\end{align}

If one begins the time evolution with initial conditions satisfying the covariant SSC and $C_1 - D_1 = 1$, the covariant SSC is preserved dynamically. $C_1 - D_1\neq1$ produces violations of the covariant SSC. In light of this, notice that if $C_1 - D_1$ is set to 1 in \eqref{eq:WLnoSSCEOM} and covariant SSC satisfying initial conditions are chosen, then the equations of motion \eqref{eq:WLnoSSCEOM} reduce to the equations of motion \eqref{eq:WLSSCEOM}. Consequently, this modified worldline formalism is strictly more general than the conventional \ref{worldline1} as it contains the \ref{worldline1} as a special case when appropriate initial conditions and Wilson coefficient values are selected. In order to ``turn on'' the SSC and reduce to the conventional worldline formalism we can set $D_1$ to the special value $D_1 = C_1 - 1$ at any stage of calculation and use initial conditions satisfying the covariant SSC. 

\subsubsection{Worldline Compton Amplitude}
\label{sec:WLCompton}

Using the \ref{worldline2} equations of motion we compute the classical Compton amplitude to order ${\cal O}(\alpha \gS^2)$ for general values of $C_1, D_1$  The classical Compton is computed by computing the coefficient of the outgoing spherical electromagnetic wave produced by the response of the spinning body to in an incoming electromagnetic plane wave as in Appendix D of~\cite{Kosower:2018adc} or as is done for gravity in~\cite{Saketh:2022wap}. In particular, we consider an incoming plane wave vector potential in Lorenz gauge,
\begin{equation}
    A^{\text{in}}_\mu(X) = e^{ik\cdot x}\xi_\mu
\end{equation}
and the response of a spinning particle to this potential using the equations of motion of \ref{worldline2}. The ${\cal O}(\alpha)$ perturbative solutions can be returned to the current,
\begin{align}
    J^\mu(X) &= \frac{\delta S}{\delta A_\mu} \\
    &= \int_{-\infty}^\infty\left(Q \dot z^\mu \delta(X-Z) + \frac{\mathtt{e}Q}{m} \left(C_1 \gS^{\mu\nu} + D_1\left(\hat p^\mu \gS^{\mu\rho}\hat p_\rho - \gS^{\mu\rho}\hat p_\rho \hat p^\nu\right)\right)\partial_\nu\delta(X-Z)\right)d\lambda.\nonumber
\end{align}
Then, treating that current as a source we compute the perturbation of the vector potential. The large distance behavior of the perturbed vector potential allows one to read off the Compton amplitude $\mathcal{A}^{\mu\nu}$ by:
\begin{equation}
    A^\mu(X) = e^{ik\cdot x}\xi^\mu + \frac{e^{ikr - i\omega t}}{4\pi r} \mathcal{A}^{\nu\mu}\xi_\nu + O(\frac{1}{r^2}).
\end{equation}
The Compton amplitude can then be extracted directly from the current by using the Lorenz gauge solution to the wave equation at large distances. Doing so one finds:
\begin{equation}
    \widetilde{J}^\mu = 2\pi \mathcal{A}^{\nu\mu}\xi_\nu
\end{equation}
where $\widetilde{J}^\mu$ is the Fourier transform of the current evaluated at the outgoing photon momentum. 

Using the current computed from the worldline equations of motion, the resulting classical Compton amplitude is found to fully agree with the ${\cal A}^\text{FT1g}$ Compton amplitude in \eqref{spinlessCompton}, \eqref{eq:s1_comp_FT2}, \eqref{eq:s2_comp_FT2} (with the $\gS^2$ terms matching up to contact terms, which we did not explicitly include either on the field theory or in the worldline theory).

\subsubsection{Worldline Impulses}
\label{sec:WLImpulses}

For computing observables with these equations of motion we consider the probe limit of a spinning particle of mass $m$ scattering off of a stationary scalar source. 
%
For simplicity, we consider only the probe limit; even so, the result 
is sufficiently complex
to demonstrate a rather nontrivial comparison with the field-theory calculations.
 The source -- a point charge moving with four-velocity $u_2$ -- has vector potential,
\begin{equation}
    A_\mu(x) = \frac{Q u_{2\mu}}{4\pi\sqrt{(x\cdot u_2)^2 - x\cdot x}} \, .
\end{equation}
The solutions to the equations of motion of the probe in powers of $\alpha = {Q^2}/{(4\pi)}$ are of the form:
\begin{align}
    & z^\mu(\lambda) = b^\mu + u_1^\mu \lambda + \alpha \delta z_{(1)}^\mu(\lambda) + \alpha^2 \delta z_{(2)}^\mu(\lambda) + \mathcal{O}(\alpha^3) \\
    & p^\mu(\lambda) = m u_1^\mu + \alpha \delta p_{(1)}^\mu(\lambda) + \alpha^2 \delta p_{(2)}^\mu(\lambda) + \mathcal{O}(\alpha^3) \\
    & \gS^{\mu\nu}(\lambda) = \gS_1^{\mu\nu} + \alpha \delta \gS^{\mu\nu}_{(1)}(\lambda) + \alpha^2 \delta \gS^{\mu\nu}_{(2)} + \mathcal{O}(\alpha^3).
\end{align}
The impact parameter $b^\mu$ is defined to be transverse on the initial momentum, $b\cdot p_1 = 0$. The initial momentum $m u_1^\mu$ defines the initial four-velocity $u_1^\mu$.  All perturbations of $p^\mu$ and $\gS^{\mu\nu}$ asymptotically vanish for $\lambda \to \pm \infty$ while the trajectory perturbations are logarithmically divergent with the worldline time due to the long range nature of the Coulomb potential. Due to this logarithmic divergence, in order to treat the $\mathcal{O}(\alpha^2)$ and higher solutions correctly, all the perturbations may be set to 0 at an initial cutoff time $\lambda = -T$. Impulse observables are then computed by taking the difference in observables at time $T$ and $-T$ and at the end taking the limit $T\to\infty$. Equivalently, the perturbations may be given representations in terms of standard Feynman integrals and computed using dimensional regularization, such as in Ref.~\cite{Kalin:2020mvi}. 

Computing the momentum impulse and spin kick to $\mathcal{O}(\alpha^2)$ and $\mathcal{O}(\gS^1)$ in this way gives a perfect match to the corresponding observables obtained from ${\cal A}^\text{FT1g}$ when the worldline Wilson coefficients $C_1$ and $D_1$ are identified with their field-theory counterparts, as detailed in Sec.~\ref{subsec:WLComparison} below. The results of \ref{worldline1} can be recovered from the more general results of \ref{worldline2} by setting the special value $D_1 = C_1 -1$. To express the impulses, it is useful to define:
\begin{gather}
\label{eq:GammaDef}
    \gamma = u_1\cdot u_2, \quad\quad\quad v = \frac{\sqrt{\gamma^2-1}}{\gamma}, \\
    \check u_1^\mu=u_1^\mu-\gamma u_2^\mu, \qquad \check u_2^\mu=u_2^\mu-\gamma u_1^\mu,
\end{gather}
and to decompose the impulses according to:
\begin{alignat}{3}\label{eq:finalkicksWL}
\Delta p_1^\mu&=\alpha\Delta p_{1(\alpha^1)}^\mu+\alpha^2\Delta p_{1(\alpha^2)}^\mu+\mathcal O(\gS_1^2)+\mathcal O(\alpha^3),
\\\nn
\Delta \gS_1^{\mu\nu}&=\alpha\Delta \gS_{1(\alpha^1)}^{\mu\nu}+\alpha^2\Delta \gS_{1(\alpha^2)}^{\mu\nu}+\mathcal O(\gS_1^2)+\mathcal O(\alpha^3).
\end{alignat}
Then at order ${\cal O}(\alpha)$ and with the notation $|b|=\sqrt{-b^\mu b_\mu}$, we find the impulse 
\begin{alignat}{3}
\Delta p_{1(\alpha^1)}^\mu=&\frac{2 b^\mu}{v|b|^2}+\frac{2}{m_1\gamma v|b|^2}\bigg(2\frac{b^\mu b^\nu}{|b|^2}\gS_{1\nu\rho}+\gS_1{}^\mu{}_\rho\bigg)\Big(D_1\gamma u_1^\rho-C_1 u_2^\rho\Big)
\\\nn
&-\frac{2\gS_{1\nu\rho}u_1^\nu u_2^\rho}{m_1\gamma^3v^3|b|^2}\Big[(C_1-D_1\gamma^2)u_1^\mu+(D_1-C_1)\gamma u_2^\mu\Big]
\end{alignat}
and the spin kick
\begin{alignat}{3}
\Delta \gS_{1(\alpha^1)}^{\mu\nu}&=\frac{4}{m_1\gamma v|b|^2}\Big(\gS_1{}^{[\mu}{}_\sigma\delta^{\nu]}{}_\rho b^\sigma-\gS_1{}^{[\mu}{}_\rho b^{\nu]}\Big)\Big(D_1\gamma u_1^\rho-C_1 u_2^\rho\Big).
\end{alignat}
At order ${\cal O}(\alpha^2)$, we find the impulse,
\begin{alignat}{3}
\Delta p_{1(\alpha^2)}^\mu=&-\frac{\pi b^\mu}{2 m_1 \gamma v|b|^3}-\frac{2\check u_1^\mu}{m_1\gamma^2v^4|b|^2}
\\\nn
&+\pi\frac{3b^\mu b^\nu \gS_{1\nu\rho}+|b|^2\gS_1{}^\mu{}_\rho}{2m_1^2\gamma^3v^3|b|^5}\Big[(C_1^2-C_1D_1-D_1)\check u_1^\rho +D_1(C_1-D_1-3)\gamma\check u_2^\rho\Big]
\\\nn
&+2\frac{C_1^2-D_1(2C_1-D_1-2)\gamma^2}{m_1^2\gamma^2v^2|b|^4}b^\nu \gS_{1\nu\rho}\bigg[\eta^{\rho\mu} +
\frac{ \check u_1^\rho u_1^\mu+\check u_2^\rho u_2^\mu}{\gamma^2v^2}\bigg]
\\\nn
&-4\frac{ b^\nu \gS_{1\nu\rho} }{m_1^2\gamma^6v^6|b|^4}\Big[(D_1-C_1)\gamma^2 \check u_2^\rho \check u_1^\mu+(D_1\gamma^2-C_1)\check u_1^\rho\check u_2^\mu\Big] \\\nn
&+\pi\frac{\gS_{1\nu\rho}\check u_1^\nu\check u_2^\rho}{2 m_1^2\gamma^7v^7|b|^3}\Big\{\big[(C_1-D_1)^2+2 D_1\big]\gamma\check u_1^\mu\\\nn
&\phantom{+\pi\frac{\gS_{1\nu\rho}\check u_1^\nu\check u_2^\rho}{2 m_1^2\gamma^7v^7|b|^3}\Big\{}+\big[C_1^2-C_1D_1-D_1-D_1(C_1-D_1-3)\gamma\big]\check u_2^\mu\Big\}
\end{alignat}
and the spin kick,
\begin{alignat}{3}
\Delta \gS^{\mu\nu}_{1(\alpha^2)}&=\frac{\pi}{m_1^2\gamma^3 v^3 |b|^3}\bigg[(C_1^2-C_1D_1-D_1)\Big(b^{[\mu}\gS_{1\rho}{}^{\nu]}\check u_1^\rho-\check u_1^{[\mu} \gS_{1\rho}{}^{\nu]}b^\rho\Big)
\\\nn
&\qquad\qquad\qquad\;\;+D_1(C_1-D_1-3)\gamma\Big(b^{[\mu}\gS_{1\rho}{}^{\nu]}\check u_2^\rho-\check u_2^{[\mu} \gS_{1\rho}{}^{\nu]}b^\rho\Big)\bigg]
\\\nn
&\quad+\frac{4}{m_1^2\gamma^2 v^2|b|^4}\bigg\{2\Big(C_1 b^{[\mu} u_2^{\nu]}-D_1\gamma b^{[\mu} u_1^{\nu]}\Big)b^\rho \gS_{1\rho\sigma}\big(C_1 u_2^\sigma-D_1\gamma u_1^\sigma\big)
\\\nn
&\qquad\qquad\qquad\quad\;\;-\big[C_1^2-D_1(2C_1-D_1)\gamma^2\big]b^{[\mu}\gS_{1\rho}{}^{\nu]}b^\rho\bigg\}
\\\nn
&\quad+\frac{4}{m_1^2\gamma^6 v^6|b|^2}\bigg\{(C_1-D_1)^2\gamma^2 \check u_1^{[\mu}\gS_{1\rho}{}^{\nu]}\check u_1^\rho+(C_1-D_1\gamma^2)^2\check u_2^{[\mu}\gS_{1\rho}{}^{\nu]}\check u_2^\rho
\\\nn
&\qquad\qquad\quad\;\;+(C_1-D_1\gamma^2)\gamma\Big[(C_1-D_1-1)\check u_2^{[\mu}\gS_{1\rho}{}^{\nu]}\check u_1^\rho+(C_1-D_1+1)\check u_1^{[\mu}\gS_{1\rho}{}^{\nu]}\check u_2^\rho\Big]\bigg\}.
\end{alignat}

The nontrivial nature of the above results give us confidence that we have indeed identified a worldline model whose results match those of the field theory. It would of course be useful to carry out further comparisons to field theory, not only beyond the probe limit but also more importantly to higher orders in the spin, especially for the case of general relativity.   Given the rather different setups, a direct proof that the field-theory and worldline descriptions will always yield equivalent results appears nontrivial.

\section{Effective Hamiltonian Including Lower-Spin States}
\label{sec:EFTHamiltonian}

Refs.~\cite{Bern:2020buy, Aoude:2020ygw, Kosmopoulos:2021zoq, Bern:2022kto} extend the spinless Hamiltonian of Ref.~\cite{Cheung:2018wkq} to the case of spinning bodies.  
This corresponds to the two-body effective description \ref{H1}, which is composed of the collection of all independent operators containing up to a given power of spin, each with arbitrary coefficients determined by matching to either field-theory or worldline results.  Here we  explicitly consider operators up to linear in spin.  We also construct a second EFT Hamiltonian, referred to as \ref{H2}, extending the degrees of freedom of $S^{\mu \nu}$ to include the intrinsic boost, $K^\mu$.  Interpreting the Hamiltonians as quantum operators allows us to obtain scattering amplitudes, which we then match to the quantum-field-theory amplitudes found in \sect{sec:TwoBodyAmpl}.  This determines the coefficients in the Hamiltonians.   
A suitable expectation value of the Hamiltonian operators are then reinterpreted as classical Hamiltonians. The corresponding equations of motion can be solved to give the impulse, spin and boost kick along a scattering trajectory, which we then compare to the corresponding observables obtained from the worldlines \ref{worldline1} and \ref{worldline2}, described in \sect{sec:WL}. 
We find that the extra Wilson coefficients that appear in \ref{worldline2}, \ref{fieldtheory1}g and \ref{fieldtheory3}g are naturally accounted for in \ref{H2}. Finally, we find a compact eikonal formula~\cite{Amati:1990xe, Amati:2007ak,  Laenen:2008gt, Akhoury:2013yua}, extending the spin results of Ref.~\cite{Bern:2020buy} to account for the appearance of the intrinsic boost operator, that matches the results obtained from the equations of motion and worldline. Eikonal representations are automatically compact because they encode the physical information in a single scalar function.

\subsection{Hamiltonian 1: Solely Spinning Degrees of Freedom}

We consider an effective description of the binary containing only
spin degrees of freedom, which leads to equations of motion that preserve
the magnitude of the spin vector. We refer to this effective description as \ref{H1}. This is the same treatment as the one of
Refs.~\cite{Bern:2020buy,Kosmopoulos:2021zoq,Bern:2022kto} except that
here we consider electrodynamics instead of general relativity.
We briefly describe this Hamiltonian and then proceed with a 
more extensive description of a modified Hamiltonian which contains a 
boost operator and allows for spin-magnitude~change.

\ref{H1} contains the usual spin-vector degrees of
freedom, along with the usual commutation or Poisson-bracket relations
for spin. 
In terms of the quantum-mechanical states that describe the bodies,
this construction implies that we may take them to
belong to a single irreducible representation of the rotation
group. In particular, we choose the asymptotic scattering states to be spin coherent states~\cite{SpinCohSt}, which are labeled by an integer $s$ and a direction given by a unit vector $\hat{\bm n}$~\cite{Bern:2020buy}, as in the field-theory discussion in Sec.~\ref{sec:fieldtheory}. 
To build the most generic Hamiltonian that accommodates these spin degrees of freedom, we need only consider the spin operator $\opS$. This is in accordance with the classical description of these particles, where one describes such a spinning object in terms of the spin three-vector~$\clS$.\footnote{\baselineskip=14pt For
compactness, for $\clr$ and $\clp$ we do not distinguish a quantum operator from the
corresponding classical value by using a different symbol, as in Refs.~\cite{Bern:2020buy,Kosmopoulos:2021zoq}.}

For simplicity, here we limit the discussion to a Hamiltonian for one scalar and one spinning particle valid to linear order in spin. This center-of-mass (CoM) Hamiltonian is given by (see Ref.~\cite{Bern:2020buy} for the corresponding one in general relativity),
\begin{align}
\spinH = \null &  \sqrt{\opp ^2 + m_1^2}+\sqrt{\opp^2 + m_2^2}+
V^{(0)}(\opr^2, \opp^2) \hamOperatorId +
V^{(1)}(\opr^2, \opp^2) \hamOperatorS
\,,
\label{Eq:Hamiltonian_Spin_Only_Position_Space}
\end{align}
where the potentials are
\begin{align}
V^{(a)}(\opr^2, \opp^2)&=\frac{\coupling}{|\opr|} c_1^{(a)}(\opp^2)+ \left(\frac{\coupling}{|\opr|} \right)^2
 c_2^{(a)}(\opp^2)+ {\cal O}(\coupling^3) \,,
\label{Eq:Potentials_position_space}
\end{align}
and we have taken the particle 1 to carry spin $\clS_1$, with the binary system carrying angular momentum $\clL = \clr \times \clp$. For these operators we have the commutation relations,
\begin{equation}
[\opcS_{1,i},\opcS_{1,j}] = \imUnit \epsilon_{ijk} \opcS_{1,k}\,, \hskip 1.5 cm 
[r_{i},\opcS_{1,j}] = [p_{i},\opcS_{1,j}] = 0\,, \hskip 1.5 cm  [r_i,p_j] = i\delta_{ij}\,.
\label{Eq:completing_algebra}
\end{equation}
For any operator that is a function solely of $\opr,\,\opp,\,\opS_1$ the
spin magnitude is preserved, since all such operators commute with the spin Casimir, i.e.,
\begin{equation}
[\opS_1^2, \mathcal{O}]=0\,, \hskip 1.5 cm \mathcal{O} = \{ \opr,\,\opp,\,\opS_1\}\,.
\end{equation}
Similarly, at the level of the classical equations of motion the above implies that the spin magnitude is a conserved quantity.  Further details may be found in Ref.~\cite{Bern:2020buy}.

\subsection{Hamiltonian 2: Inclusion of Boost Operator}
\label{Sect:Boost_Hamiltonian}

In this subsection we expand the degrees of freedom so that we are able to properly describe the field theories and worldline theories that also contain additional degrees of freedom, allowing
the spin magnitudes to vary by their interaction with the electromagnetic field.
To this end, inspired by the worldline construction in \sect{sec:WLnoSSC}, we extend the above Hamiltonian
to include  the generator of intrinsic boosts $\opK$.
We start by motivating this choice. We proceed to describe how does one build the most general two-body Hamiltonian out of the available operators for the problem at hand.  Finally, as an explicity illustration, we build the Hamiltonian linear in the spin and boost of one of the particles.

In order to have a Hamiltonian whose amplitudes match those of \hyperlink{ft1g}{FT1g} and \ref{fieldtheory3} we are prompted to consider additional operators. The natural choice is operators built out of the vector $\clK_1$, already encountered in Eq.~\eqref{WLDefK}. The operator $\opK_1$ should act on the intrinsic degrees of freedom of the body, hence it 
commutes with both $\opr$ and $\opp$. Accordingly, the commutation relations are, 
\begin{equation}
[\opcS_{1,i},\opcK_{1,j}] = \imUnit \epsilon_{ijk} \opcK_{1,k}\,, \hskip 1.5 cm
[r_{i},\opcK_{1,j}] = [p_{i},\opcK_{1,j}] = 0\,.
\label{Eq:SKcommutator}
\end{equation}
where the first relation simply implies that $\opK_1$ is a vector operator.
To fully characterize the operator $\opK_1$ we need to specify the commutation relations with itself.  Motivated by the connection to the worldline we take:
\begin{equation}
[\opcK_{1,i},\opcK_{1,j}] = -\imUnit \epsilon_{ijk} \opcS_{1,k} \,,
\label{Eq:KKcommutator}
\end{equation}
which identifies $\opK_1$ with the generator of intrinsic boosts.
The operator algebra is completed by the commutators familiar from the case without the $\opK_1$ operator given in Eq.~\eqref{Eq:completing_algebra}.

Alternatively, the introduction of $\opK_1$ may be motivated by the requirement that the spin magnitude should change under time evolution via the constructed Hamiltonian. In the quantum-mechanical language, this requires an operator that does not commute with $\opS_1^2$. It follows that it must also not commute with $\opS_1$, i.e. it must have tensor structure under intrinsic rotations. The simplest object that satisfies this criterion is a vector under intrinsic rotations that commutes with both $\opr$ and $\opp$. This reasoning leads to the introduction of an operator obeying the commutation relations \eqref{Eq:SKcommutator}, while \eqref{Eq:KKcommutator} still needs to be motivated by the interpretation of $\opK_1$ as the boost generator. We indeed find that inclusion of $\opK_1$ leads to scattering amplitudes between states of different spin magnitude, similar to our field-theory constructions above. Furthermore, while in these scattering amplitudes the change in spin is minute, $s\rightarrow s-1$, the effect is resummed to a finite change via Hamilton's equations, as we see in \sect{sec:EoMinH}.

We proceed to construct the effective Hamiltonian. The first question is to find the complete set of terms that can appear.  We constrain these based on symmetry considerations: the Hamiltonian is invariant under parity and time reversal (see, however, the discussion in \sect{sec:KReality}). To take advantage of these constraints we list how our operators transform under the action of these symmetries:
\begin{alignat}{13}
\label{eq:ParityTime}
&\text{Parity:} \quad
&& P^\dagger \opr && P &&= - \opr\,, \quad
&& P^\dagger \opp && P &&= - \opp \,, \quad
&& P^\dagger \opS_1 && P && = \opS_1 \,, \quad
&& P^\dagger \opK_1 && P &&= - \opK_1 \,, \quad
\nn \\
&\text{Time Reversal:} \quad
&& T^\dagger \opr && T &&= \opr\,, \quad
&& T^\dagger \opp && T &&= - \opp \,, \quad
&& T^\dagger \opS_1 && T &&= - \opS_1 \,, \quad
&& T^\dagger \opK_1 && T &&= \opK_1 
\,, \quad
\end{alignat}
see e.g. Sect.~2.6 of Ref.~\cite{Weinberg:1995mt}. Furthermore, we construct terms that have classical scaling.  The scaling of our operators in the classical limit is 
\begin{align}
\opr \sim \frac{1}{\lambda} \opr \,, \quad 
\opp \sim \lambda^0 \, \opp \,, \quad
\opS_1 \sim \frac{1}{\lambda} \opS_1 \,, \quad
\opK_1 \sim \frac{1}{\lambda} \opK_1 \,,
\end{align}
where $\lambda$ is a small parameter that characterizes the classical limit (usually associated with $\hbar$).  

Two additional properties that reduce the number of operators are on-shell conditions and Schouten identities.
The former capture the freedom of field redefinitions in the quantum-mechanical context or the freedom of canonical transformations in the classical context. The latter stem from the fact that we work with more than three three-dimensional vectors, hence there must be linear relations among them.  While these considerations are not important for the purposes of this paper, they can significantly reduce the number of terms one needs to consider when looking at higher orders in spin and boost (see e.g. Ref.~\cite{Bern:2022kto}).

Using the above one may systematically construct independent terms in the Hamiltonian. At linear order in spin and boost we have:
\begin{equation}
\hamO_1 = \hamOperatorS
\,, \hskip 1.5 cm
\hamO_2 = \hamOperatorK \,.
\label{SandKOperators}
\end{equation}
The Hamiltonian valid to linear order in spin and boost is then,
\begin{equation}
\boostH =  \sqrt{\opp ^2 + m_1^2}+\sqrt{\opp^2 + m_2^2}+
V^{(0)}(\opr^2, \opp^2) \hamOperatorId +
V^{(1)}(\opr^2, \opp^2) \hamOperatorS + 
V^{(2)}(\opr^2, \opp^2) \hamOperatorK
\,,
\label{Eq:Hamiltonian_Boost_Position_Space}
\end{equation}
where we used the operators in Eq.~\eqref{SandKOperators} and  
 the potential coefficients given in Eq.~\eqref{Eq:Potentials_position_space}.
The Hamiltonian has an additional operator containing $\opK_1$  compared to the one in Eq.~\eqref{Eq:Hamiltonian_Spin_Only_Position_Space}.

\subsection{Amplitudes from the Effective Hamiltonian}

Having identified the general form of the Hamiltonian that can capture the classical physics of our field theories with additional degrees of freedom, Eq.~\eqref{Eq:Hamiltonian_Boost_Position_Space},
we proceed to determine its coefficient functions $V^{(i)}$. We follow closely Refs.~\cite{Bern:2020buy, Kosmopoulos:2021zoq, Bern:2022kto} where analogous calculations were carried out for Hamiltonians 
depending only on the spin operator.
As in that case, we consider scattering of spin-coherent states~\cite{SpinCohSt}. These states may be superpositions of fixed-spin-magnitude states or more general superpositions that involve states of different spin magnitude, similar to the field-theory construction in Eq.~(\ref{eq:Edecom}).  For our purposes it is sufficient to consider incoming and outgoing states whose spin parts are identical. However, we note that it is possible to also consider different incoming and outgoing states. Since the incoming and outgoing states are taken to be the same, the amplitudes are expressed in terms of diagonal matrix elements of $\opS$ and $\opK$. A coherent state $| \bm s \rangle \equiv |s, \hat{\bm n}\rangle$ with fixed spin magnitude $s$ and direction $\hat{\bm n}$ is the state of highest weight along the direction $\hat{\bm n}$. Similarly with the field-theory discussion in Sec.~\ref{StatesSubsection}, for such a state we have,
\begin{align}
\langle \bm s | \opS | \bm s \rangle = \clS = s\, \hat{\bm n}\,, 
\quad
\text{and}
\quad
\langle \bm s | \opK | \bm s \rangle = 0\,.
\label{eq:sExpectationValue}
\end{align}
We build a generalized coherent state $| \Psi \rangle$ by superimposing states $| \bm s \rangle$ with different values of $s$,
such that,
\begin{align}
\langle \Psi | \opS | \Psi \rangle = \clS \,, 
\quad
\text{and}
\quad
\langle \Psi | \opK | \Psi \rangle = \clK \,,
\label{eq:psiExpectationValue}
\end{align}
where on the right-hand side of the above equation we have the classical values of $\opS$ and $\opK$.
These classical values depend on the details of the construction of $| \Psi \rangle$, but the exact dependence is not important for our purposes.
Finally, these states are built such that they obey the property
\begin{align}
\langle \Psi | \{ \opcS_{i_1} \ldots \opcS_{i_n} \} | \Psi \rangle = S_{i_1} \ldots S_{i_n} \,, 
\quad
\text{and}
\quad
\langle \Psi | \{ \opcK_{i_1} \ldots \opcK_{i_n} \} | \Psi \rangle = K_{i_1} \ldots K_{i_n} \,, 
\end{align}
up to terms that do not contribute in the classical limit, where the $\{\}$ brackets signify symmetrization and division by the number of terms (see also the discussion in \sect{StatesSubsection}).

We may now proceed to compute the EFT amplitudes. For the details of such a computation we refer the reader to Refs.~\cite{Bern:2020buy, Kosmopoulos:2021zoq}, where corresponding computations are carried out for the purely-spin case.
We give here the result for the amplitude obtained from $\boostH$.  The corresponding amplitude from $\spinH$ follows by setting the coefficients of any operators containing $\opK$ to zero.
The EFT amplitude may be organized as
\begin{equation}
\ampEFT = \ampEFT^\text{\PL{1}} + \ampEFT^\text{\PL{2}} + \ldots \,,
\end{equation}
where we have explicitly written the first and second \PL{} contributions and the ellipsis denote higher \PL{} orders. 
We have
\begin{align}
\ampEFT^{\PL{1}} =
\frac{4\pi \coupling }{\clq^2}
\Big[
a^{(0)}_1 + a^{(1)}_1 \ampOperatorS + a^{(2)}_1 \ampOperatorK 
\Big]\,,
\label{eq:MEFT_1PM}
\end{align}
and 
\begin{align}
\ampEFT^{\PL{2}} &= \ampEFT^{\PL{2}}_\triangle
+(4\pi \coupling)^2\, a_{\rm iter}\int \frac{d^{D-1}\bm \ell}{(2\pi)^{D-1}}
\frac{2\xi E}{\bm \ell^2 (\bm \ell+\clq)^2 (\bm \ell^2+2\clp\cdot \bm \ell)} \,,
\nn \\
\ampEFT^{\PL{2}}_\triangle &=
\frac{2\pi^2 \coupling^2}{|\clq|}
\Big[
a^{(0)}_2 + a^{(1)}_2 \ampOperatorS + a^{(2)}_2 \ampOperatorK 
\Big] \,,
\label{eq:MEFT_2PM}
\end{align}
where the triangle subscript in $\ampEFT^{\PL{2}}_\triangle$ indicates that the origin of the contribution is an one-loop triangle integral. Here $\clp$ and $\clp-\clq$ are the incoming and outgoing spatial momenta of particle 1 in the \com frame respectively and $\momL = i \clp \times \clq$. We also use 
\begin{align}
E = E_1 + E_2
\,, \quad \text{and} \quad
\xi = \frac{E_1 E_2}{E^2} \,,
\end{align}
where $E_{1,2}$ are the energies of particles 1 and 2, which are conserved in the \com frame (see also Eq.~\eqref{eq:COMdef}).
The vectors $\clS_1$ and $\clK_1$ that appear in the above two equations are the classical values of the corresponding quantum operators. They depend on whether one chooses to scatter the $|\bm s\rangle$ or $|\Psi\rangle$ state as shown in Eqs.~\eqref{eq:sExpectationValue} and \eqref{eq:psiExpectationValue}. The $\PL{1}$ amplitude coefficients take the form
\begin{align}
a^{(0)}_1 = -c^{(0)}_1 \,, \quad a^{(1)}_1 = c^{(1)}_1 \,, \quad a^{(2)}_1 = -c^{(2)}_1 \,,
\label{cs_to_as_1PM}
\end{align}
while for the $\PL{2}$ amplitude we have
\begin{align}
a^{(0)}_2 &= -c^{(0)}_2 + 2 E \xi  c^{(0)}_1 \pder c^{(0)}_1+\frac{(1-3 \xi ) \left(c^{(0)}_1\right)^2}{2 E \xi }\,, \nn \\
a^{(1)}_2 &= \frac{c^{(1)}_2}{2} - E \xi  c^{(1)}_1 \pder c^{(0)}_1 - E \xi  c^{(0)}_1 \pder c^{(1)}_1 + \frac{(3 \xi -1) c^{(0)}_1 c^{(1)}_1}{2 E \xi }+ \frac{E \xi \left(\left(c^{(2)}_1\right)^2-2 c^{(0)}_1 c^{(1)}_1\right)}{2 \clp^2} \,, \nn \\
a^{(2)}_2 &= -\frac{c^{(2)}_2}{2} -\frac{1}{2} E \xi  c^{(2)}_1 \left(c^{(1)}_1-2 \pder c^{(0)}_1 \right) + E \xi c^{(0)}_1 \pder c^{(2)}_1 + \frac{(1-3 \xi ) c^{(0)}_1 c^{(2)}_1}{2 E \xi } \,,
\label{cs_to_as_2PM}
\end{align}
and
\begin{align}
a_{\rm iter} = \left(c^{(0)}_1\right)^2 - c^{(0)}_1 c^{(1)}_1 \ampOperatorS + c^{(0)}_1 c^{(2)}_1 \ampOperatorK \,.
\end{align}
In the above we have used the shorthands $c^{(a)}_n \equiv c^{(a)}_n\left( \clp^2 \right)$ and $\pder \equiv \frac{d}{d \clp^2}$.

\subsection{Hamiltonian Coefficients from Matching to Field Theory}
\label{Sec:EFT_matching}

We are now in position to determine the Hamiltonian coefficients that capture the same classical physics as the field theories discussed in \sect{sec:fieldtheory}; we do so by matching the corresponding scattering amplitudes including their full mass dependence. We start by specializing the field-theory amplitudes to the \com frame. We then match each field-theory construction to an appropriate Hamiltonian and we discuss our findings. 

The \com frame is defined by the kinematics 
\begin{align}
p_1=-(E_1, \,\clp) \,,
\qquad
p_2=-(E_2, \,-\clp) \, ,
\qquad
q=(0, \,\clq)\,,
\qquad
\clp\cdot \clq = \clq^2/2\,,
\label{eq:COMdef}
\end{align}
together with $q=p_2+p_3$ and $p_1+p_2+p_3+p_4=0$. To align with the field-theory construction, we express the barred variables defined in Eq.~(\ref{eq:barVariables}) in this frame,
\begin{align}
\bar{p}_1=-(E_1, \,\clpb)\,,
\qquad
\bar{p}_2=-(E_2, \,-\clpb)\,,
\qquad
\clpb = \clp-\clq/2 \,,
\qquad
\clpb\cdot \clq = 0 \,.
\end{align}
For the asymptotic spin variables we have
\begin{align}
\gS_1^{\mu\nu} = \frac{1}{m_1} \Big( \epsilon^{\mu\nu\rho\lambda} \bar{p}_{1\rho} S_{1\lambda} + \bar{p}_1^\mu K_1^\nu - \bar{p}_1^\nu K_1^\mu \Big) \,,
\label{Eq:S_tensor_SK_vectors}
\end{align}
with 
\begin{align}
S_1^\mu = \bigg(\, \frac{{\clpb}  \cdot {\clS_1}}{m_1}, {\clS_1} + \frac{{\clpb} \cdot {\clS_1}}{m_1(E_1+m_1)} {\clpb} \, \bigg)
\, , \quad
K_1^\mu = \bigg(\, \frac{{\clpb}  \cdot {\clK_1}}{m_1}, {\clK_1} + \frac{{\clpb} \cdot {\clK_1}}{m_1(E_1+m_1)} {\clpb} \, \bigg) \,,
\label{Eq:standard_boosts}
\end{align}
where $\clS_1$ and $\clK_1$ correspond to the values in the rest frame of particle 1.
Finally, we may use Eq.~\eqref{epdotep} to express the wave-function products $\polM_1 \cdot \polMb_4$ and $\polM_2 \cdot \polMb_3$ as
\begin{align}
\polM_1 \cdot \polMb_4 = \exp \left[ -\frac{\ampOperatorS}{m_1 (E_1+m_1)} \right] \exp \left[ \frac{\ampOperatorK}{m_1} \right]\,,
\quad \text{and} \quad
\polM_2 \cdot \polMb_3 = 1\,,
\label{Eq:epsilon_dot_epsilon_withK}
\end{align}
up to terms that do not contribute to the classical limit. The second product in the above equation follows from the fact that we take the corresponding particle to be a scalar. Note that the $\clK_1$ used here agrees with that from the worldline (\ref{eq:K_WL_def}).

Using the above relations we express the field-theory amplitudes in terms of the same variables as the EFT ones. Then, we may match them and extract the Hamiltonian coefficients. In particular, we have
\begin{align}
\ampEFT^{\PL{1}} = \frac{\Mtree}{4 E_1 E_2} 
\,, \qquad \text{and} \qquad
\ampEFT^{\PL{2}} = \frac{\Moneloop}{4 E_1 E_2} \,.
\end{align}

We use the above equations to match to the field theories as follows:
\begin{align}
\text{ \ref{H1} } \leftrightarrow \text{ \ref{fieldtheory2} } \,,
\quad \text{and} \quad
\text{ \ref{H2} } \leftrightarrow \text{ \hyperlink{ft1g}{FT1g} } \,.
\label{eq:mathcingDescription}
\end{align}
Our first EFT Hamiltonian \ref{H1} contains only operators that preserve the spin magnitude. Hence, it can describe the field theory that contains a single particle of spin $s$ (\ref{fieldtheory2}).  Our second Hamiltonian allows for transitions between particles of different spin magnitude, and hence can describe a field theory that contains particles of different spin magnitude (\hyperlink{ft1g}{FT1g}).  
Regarding \ref{fieldtheory3}, the amplitudes we have computed may be mapped to those of \hyperlink{ft1g}{FT1g} via appropriate relabeling.  We expect this to be true for all amplitudes that may be computed in the two theories, in which case the same should be true for the Hamiltonian coefficients. Finally, \hyperlink{ft1s}{FT1s} may be thought of as a restriction of \hyperlink{ft1g}{FT1g} where we only allow for spin-$s$ external states. We discuss the possible matching of \hyperlink{ft1s}{FT1s} to our two Hamiltonians separately.

For the \PL{1} matching of \ref{H2} to \hyperlink{ft1g}{FT1g} we find
\begin{align}
c^{(0)}_1 = \frac{m_1 m_2 \gamma }{4 E_1 E_2} 
\,, \quad
c^{(1)}_1 = \frac{m_1 m_2 \gamma -E C_1 \left(m_1+E_1\right)}{4 E_1 E_2 m_1 \left(m_1+E_1\right)}
\,, \quad
c^{(2)}_1 = \frac{m_2 \gamma  \left(-C_1+D_1+1\right)}{4 E_1 E_2} \,,
\label{eq:1PLMatchingCoefficients}
\end{align}
where $\gamma$ is defined in Eq.~(\ref{eq:GammaDef}).
We give the \PL{2} coefficients in an accompanying ancillary file~\cite{anc}. Importantly, we find $c^{(2)}_{1}=c^{(2)}_{2}=0$ if $D_1 = C_1-1$, such that all $\clK_1$ dependence in the Hamiltonian vanishes for this choice. For this reason, both the \PL{1} and \PL{2} coefficients
related to the matching of \ref{H1} and \ref{fieldtheory2} follows from the above by setting $D_1 = C_1-1$,  hence we do not report them separately.

We conclude this subsection by commenting on \ref{fieldtheory1}s.  Given that \ref{fieldtheory1}s is defined as a collection of amplitudes that are a subset of the ones of \ref{fieldtheory1}g,  the most appropriate matching procedure is to extend to \ref{fieldtheory1}g and follow the analysis given above to match to \ref{H2}. Alternatively, one can also match \ref{H1} to \ref{fieldtheory1}s as was carried out in Ref.~\cite{Bern:2022kto} following similar steps. In this case, the effects of the lower-spin states propagating in the field-theory amplitude are captured by the vertices of the Hamiltonian. By examining the resulting Hamiltonian, we find that some of the coefficients (in particular $c^{(1)}_2 (\clp^2)$) 
admit only a Laurent series 
around $\clp^2=0$. This is a familiar phenomenon in QFT where one integrates out a state that may go on-shell in the processes of interest, and, borrowing the terminology of that context, we refer to it as a 
non locality.\footnote{\baselineskip=14pt
We stress that not every Hamiltonian which contains some coefficient that does not admit a Taylor expansion around $\clp^2=0$ is non local in the sense described here. Indeed, it is certainly possible to alter the Hamiltonian coefficients by performing a field redefinition in the quantum-mechanical context or a canonical transformation in the classical context, which may potentially remove such a behavior. In addition,  when dealing with more than three three-dimensional vectors there exist Schouten identities that might cause the coefficients of the Hamiltonian to have apparent singularities in the $\clp^2 \rightarrow 0$ limit.}
A non-local quantum description may be consistent as long as one always considers amplitudes with appropriate external states. However, we find that the observables computed from this Hamiltonian match the corresponding ones from \ref{worldline1} or those from \ref{worldline2}  only for the choice $D_1 = C_1-1$, for which the non-locality vanishes.


\subsection{Observables from the Equations of Motion}
\label{sec:EoMinH}

\def\doe{\partial}
\def\bs{\boldsymbol}
\def\mc{\mathcal}

Having analyzed the implications of interpreting our Hamiltonians as quantum operators, we proceed to consider them as generating functions of the classical evolution of the system.  In particular, given a classical Hamiltonian $\boostH(\clr(t),\clp(t),\clS_1(t),\clK_1(t))$ of the form (\ref{Eq:Hamiltonian_Boost_Position_Space}), the classical time evolution of any quantity $f(\clr(t),\clp(t),\clS_1(t),\clK_1(t))$ is determined by $\dot f=\mathrm df/\mathrm dt=\{f,\boostH\}$, where the classical Poisson brackets $\{f,g\}$ are given directly by the quantum-operator algebra of Eqs.~(\ref{Eq:completing_algebra}), (\ref{Eq:SKcommutator}) and (\ref{Eq:KKcommutator}) with $\hat{f} \rightarrow f$ and $[\hat{f},\hat{g}] \rightarrow \imUnit \{f,g\}$.  This leads to the explicit equations of motion
\begin{alignat}{3}
\dot{\clr}&=\frac{\doe \boostH}{\doe\clp},
&
\dot{\clS}_1&=\frac{\doe \boostH}{\doe\clS_1}\times\clS_1+\frac{\doe \boostH}{\doe\clK_1}\times\clK_1 \,,
\nn \\
\dot{\clp}&=-\frac{\doe \boostH}{\doe\clr},
\qquad\qquad&
\dot{\clK}_1&=\frac{\doe \boostH}{\doe\clS_1}\times\clK_1-\frac{\doe \boostH}{\doe\clK_1}\times\clS_1 \,.
\end{alignat}
The addition of $\clK_1$ as a dynamical quantity changes basic properties of the equations.  Specifically, the magnitude of the spin $\clS_1$  is no longer conserved.

We solve the equations of motion order by order in $\coupling$. Given that $\boostH=E_1+E_2+\mc O(\alpha)$ at zeroth order in the coupling, with $E_{1,2}=\sqrt{m_{1,2}^2+\clp^2}$, we see that perturbative solutions to these equations take the form
\begin{alignat}{3}
\clr(t)&=\bs b^{(0)}+\frac{E_1+E_2}{E_1E_2}\clp^{(0)} t+\alpha\bs r^{(1)}(t)+\alpha^2 \bs r^{(2)}(t)+\ldots\,,
\nn\\
\clp(t)&=\clp^{(0)}+\alpha\clp^{(1)}(t)+\alpha^2 \clp^{(2)}(t)+\ldots\,,
\nn\\
\clS_1(t)&=\clS_1^{(0)}+\alpha\clS_1^{(1)}(t)+\alpha^2 \clS_1^{(2)}(t)+\ldots\,,
\nn\\
\clK_1(t)&=\clK_1^{(0)}+\alpha \clK_1^{(1)}(t)+\alpha^2 \clK_1^{(2)}(t)+\ldots\,,
\end{alignat}
where $\bs b^{(0)}$, $\clp^{(0)}$, $\clS_1^{(0)}$, and $\clK_1^{(0)}$ are constants determined from the initial conditions.  The constant $\bs b^{(0)}$ is the usual impact parameter for the scattering process.
Substituting these expansions into the equations of motion, using the explicit Hamiltonian \eqref{Eq:Hamiltonian_Boost_Position_Space} and separating orders in $\alpha$, we obtain integral expressions for
\begin{equation}
 \mathcal{O}^{(n)}(t) = \left \{ \bs r^{(n)}(t),\clp^{(n)}(t),\clS_1^{(n)}(t),\clK_1^{(n)}(t)\right \} \,.
\end{equation}
These depend on lower-order solutions $\mathcal{O}^{(\tilde{n})}(t)$, with $0\le \tilde{n} < n$, as well as  the Hamiltonian coefficients $c_n^{(a)}(\clp^2)$ and their derivatives evaluated at $\clp^2=\left( \clp^{(0)} \right)^2$.  Working iteratively, we obtain explicit expressions for $\mathcal{O}^{(n)}(t)$ by performing simple one-dimensional integrals with respect to $t$.  We choose the integration constants by enforcing $\mathcal{O}^{(n)}(t) \to \{0,0,0,0\}$ as $t\to-\infty$ for all $n\ge 1$, ensuring that $\bs b^{(0)}$, $\clp^{(0)}$, $\clS_1^{(0)}$, and $\clK_1^{(0)}$ characterize the initial conditions.  Without loss of generality, we can choose $\bs b^{(0)}\cdot \clp^{(0)}=0$ and identify $\bs b^{(0)}$ as the incoming impact parameter vector.  
In particular, we choose
\begin{align}
\bm b^{(0)} = (-b, 0,0)\,,
\quad
\clp^{(0)} = (0, 0,p_\infty)\,,
\quad
\clS_1^{(0)} = (S_{1x}^{(0)},S_{1y}^{(0)},S_{1z}^{(0)})\,,
\quad
\clK_1^{(0)} = (K_{1x}^{(0)},K_{1y}^{(0)},K_{1z}^{(0)})\,.
\end{align}
Following the above procedure, we finally obtain $\left(\clp,\clS_1,\clK_1\right)$ in the outgoing state from the limit $t\to+\infty$, given as functions of the incoming $\left\{\bs b^{(0)},\clp^{(0)},\clS_1^{(0)},\clK_1^{(0)}\right\}$.

As we emphasized, a key consequence of including $\clK_1$ in the Hamiltonian is that the magnitude of $\clS_1$ is not conserved under time evolution. Indeed, it is a straightforward consequence of the equations of motion that 
\begin{align}
\frac{d}{d t} \left( \clS_1^2 - \clK_1^2 \right) = 0 \,,
\label{eq:SKConservation}
\end{align}
which reduces to the equation for spin-magnitude conservation only if $\clK_1$ is constant throughout the trajectory, as would hold for a rigid object with no internal degrees of freedom other than the spin. Explicitly,  solving the equations of motion we find that the spin magnitude does indeed change. 
We define the change of the spin and boost magnitude as
\begin{equation}
\Delta \clS_1^2 \equiv \clS_1^2(t=\infty) - \clS_1^2(t=-\infty) \,,
\quad 
\Delta \clK_1^2 \equiv \clK_1^2(t=\infty) - \clK_1^2(t=-\infty) \,.
\end{equation} 
We have that through \PL{1} they are given by 
\begin{align}
\Delta \clS_1^2 = \Delta \clK_1^2 = \frac{4 \alpha  E_1 E_2 \left(K_{1z}^{(0)} S_{1y}^{(0)}-K_{1y}^{(0)} S_{1z}^{(0)}\right) 
   c_1^{(2)}(p_\infty^2) }{b\, p_\infty (E_1+E_2)} + \mathcal{O}(\coupling^2) \,,
\label{eq:spinMagnitude1PL}
\end{align}
in accordance with Eq.~(\ref{eq:SKConservation}).
Thus the spin magnitude is conserved to \PL{1} order if we choose the initial condition $\clK_1^{(0)} = 0$.  Similarly, the boost magnitude is also conserved if $\clS_1^{(0)} = 0$. 
However, starting at \PL{2} order, this is no longer true. In particular,
\begin{align}
\Delta \clS_1^2 \Big\rvert_{\clK_1^{(0)} \rightarrow 0} = \Delta \clK_1^2 \Big\rvert_{\clK_1^{(0)} \rightarrow 0} = \frac{4 \alpha ^2 E_1^2 E_2^2 \left( \left( S_{1y}^{(0)}\right)^2 + \left( S_{1z}^{(0)} \right)^2 \right) \left(c_1^{(2)}(p_\infty^2)\right)^2}{b^2 \, p_\infty^2
   (E_1+E_2)^2} + \mathcal{O}(\coupling^3) \,.
\label{eq:spinMagnitude2PL}
\end{align}
As expected, the spin magnitude is conserved if we choose $D_1 = C_1 - 1$, as can be seen by combining the above equations with Eq.~(\ref{eq:1PLMatchingCoefficients}).

The above equations further imply that for an object with $D_1 \neq C_1 - 1$ the intrinsic boost, and hence the induced electric dipole moment (see Eq.~(\ref{eq:InducedDipoles})), is not a constant of motion.  In particular, even if a body has $\clK_1=0$ at some moment in time, time evolution induces non-zero values for $\clK_1$. In other words, a body which satisfies the covariant SSC at the initial time violates it at later times. 

It is interesting to ask whether we could instead remove $\clS_1$ and have a system that is described only by $\clK_1$.  Up to \PL{1} order it is consistent to have $\clS_1=0$ with $\clK_1\neq 0$, as can be seen in Eq.~(\ref{eq:spinMagnitude1PL}). However, at \PL{2} order we find
\begin{align}
\Delta \clS_1^2 \Big\rvert_{\clS_1^{(0)} \rightarrow 0} = \Delta \clK_1^2 \Big\rvert_{\clS_1^{(0)} \rightarrow 0} = \frac{4 \alpha ^2 E_1^2 E_2^2 \left( \left( K_{1y}^{(0)}\right)^2 + \left( K_{1z}^{(0)} \right)^2 \right) \left(c_1^{(2)}(p_\infty^2)\right)^2}{b^2 \, p_\infty^2
   (E_1+E_2)^2} + \mathcal{O}(\coupling^3) \,.
\end{align}
As for Eq.~(\ref{eq:spinMagnitude2PL}), this only vanishes for the special value  $D_1 = C_1 - 1$.
Hence, without the special choice, a non-rotating body starts spinning via the electromagnetic interaction if it starts with non-zero intrinsic boost $\clK_1$. 

The dynamics that we consider here are an extension of those that satisfy an SSC along their evolution. Indeed, at any step of the calculation one is free to set $D_1 = C_1 - 1$ and retrieve the evolution of an SSC-satisfying body. Such a restriction would remove all $\clK_1$ dependence from the Hamiltonian as we mentioned below Eq.~(\ref{eq:1PLMatchingCoefficients}) and render $\clK_1$ to be a constant of motion that does not affect the dynamics.

The complete results of solving the equations of motion through $\mathcal O(\alpha^2)$ as outlined above are quite lengthy, hence we give the explicit solutions in the ancillery file~\cite{anc}.  A much more compact way to represent the amplitude is through an eikonal formula, which we give below.

\subsection{Observables from an Eikonal Formula}

Analyzing the results of the perturbative integration of Hamilton's equations as described in the previous section, we find that the outgoing-state observables can be simply expressed in terms of derivatives of an eikonal phase, which is a scalar function of the incoming-state variables.  This is motivated by the analagous eikonal formula found in Ref.~\cite{Bern:2020buy} for the pure spin case, except now there are additional degrees of freedom from the intrinsic boost.  At the order to which we are working here, the eikonal phase coincidences with a two-dimensional Fourier transform of the EFT amplitude.

For convenience we rename the incoming-state quantities, called $\left\{ \bs b^{(0)},\clp^{(0)},\clS_1^{(0)},\clK_1^{(0)}\right\}$ above, now simply as $\left\{ \bs b,\clp,\clS_1,\clK_1\right\}$.  Then we denote the outgoing-state observables by $\left\{ \clp+\Delta\clp, \clS_1+\Delta\clS_1, \clK_1+\Delta\clK_1\right\}$. 

We find empirically that the changes in the observables $\clp$, $\clS_1$, and $\clK_1$ are given in terms of an eikonal phase $\chi(\bs b,\clp,\clS_1,\clK_1)$ as follows: The impulse is given by
\begin{equation}
\Delta\clp =\frac{\doe\chi}{\doe\bs b}
+\frac{1}{2}\{\chi,\frac{\doe\chi}{\doe\bs b}\}+
\mathcal D_{L}(\chi,\frac{\doe\chi}{\doe\bs b})-\frac{1}{2}\frac{\doe}{\doe\bs b}\mathcal D_{L}(\chi,\chi)-\frac{\clp}{2\clp^2}\bigg(\frac{\doe\chi}{\doe\bs b}\bigg)^2+\mathcal O\left(\chi^3\right) \,,
\label{eq:impulsefromchi}
\end{equation}
which simultaneously gives contributions orthogonal and along $\clp$.
In this formula  $\clp\cdot\bs b=0$ so all the $\bs b$-derivatives are projected orthogonal to the incoming momentum $\clp$.
The spin and boost kicks are given by 
\begin{align}
\Delta\clS_1 & =\{\chi,\clS_1\}
+\frac{1}{2}\{\chi,\{\chi,\clS_1\}\}+
\mathcal D_{L}(\chi,\{\chi,\clS_1\})-\frac{1}{2}\{\mathcal D_{L}(\chi,\chi),\clS_1\}+ \mathcal O\left(\chi^3\right) \,,
\label{eq:kicksfromchi} \\ \nn
\Delta\clK_1 &=\{\chi,\clK_1\}
+\frac{1}{2}\{\chi,\{\chi,\clK_1\}\}+
\mathcal D_{L}(\chi,\{\chi,\clK_1\})-\frac{1}{2}\{\mathcal D_{L}(\chi,\chi),\clK_1\}+ \mathcal O\left(\chi^3\right) \,.
\end{align}
The brackets here are given by the Lorentz algebra,
\begin{equation}
\{S_{1i},S_{1j}\}=\epsilon_{ijk} S_{1k}\,,
\quad 
\{S_{1i},K_{1j}\}=\epsilon_{ijk} K_{1k}\,,
\quad
\{K_{1i},K_{1j}\}=-\epsilon_{ijk} S_{1k}\,,
\end{equation}
with all others vanishing. We also define
\begin{equation}\label{defineDL}
\mathcal D_L(f,g)\equiv -\epsilon_{ijk}\bigg(S_{1i}\frac{\doe f}{\doe S_{1j}}
  + K_{1i}\frac{\doe f}{\doe K_{1j}}\bigg)\frac{\doe g}{\doe L_k}\, ,
\end{equation}
which is a $K$-dependent extension of the operator ${\cal D}_{SL}$ of Ref.~\cite{Bern:2020buy}.
The angular momentum $\bs L$ and the incoming impact parameter $\bs b$ are related by $\bs L=\bs b\times\clp$ and $\bs b=\clp\times \bs L/\clp^2$, implying $(\doe/\doe L_i)=\epsilon_{ijk}(p_k/\clp^2)(\doe/\doe b_j)$ in Eq.~(\ref{defineDL}).    An all-orders generalization may follow along the lines of Eq.~(7.21) of Ref.~\cite{Bern:2020buy} by including $\clK$, although at higher orders in $\alpha$ the radial action may be more natural than the eikonal phase~\cite{Bern:2021dqo}.


The appropriate eikonal function is proportional to the two-dimensional Fourier transform (from $\bs q$ space to $\bs b$ space) of the EFT amplitude as given in Eqs.~\eqref{eq:MEFT_1PM} and \eqref{eq:MEFT_2PM}, while keeping only the triangle contribution in Eq.~\eqref{eq:MEFT_2PM}~\cite{Bern:2020buy},
\begin{equation}
\chi=\frac{1}{4E|\clp|}\int\frac{d^2\bs q}{(2\pi)^2}e^{-i\bs q\cdot\bs b}(\mathbb M^{\PL{1}}+\mathbb M^{\PL{2}}_\triangle)+\mathcal O(\alpha^3)\, ;
\label{Eq:eikonal_Fourier}
\end{equation}
the box contribution to the amplitude is effectively included in the exponentiation of the tree-level amplitude $\mathbb M^{\PL{1}}$. Explicitly, we have
\begin{alignat}{3}
\chi&=\alpha\frac{\xi E}{|\clp|}\bigg[{-}a_{1}^{(0)}\log |\bs b|^2-\frac{2a_{1}^{(1)} }{|\bs b|^2}\bs b\times\clp\cdot\clS_1+ \frac{2a_{1}^{(2)}}{|\bs b|^2}\bs b\cdot\clK_1\bigg]
\label{eq:eikonalexplicit}
\\\nn
&+\pi\alpha^2 \frac{\xi E}{|\clp|}\bigg[ \frac{ a_{2}^{(0)}}{|\bs b|}-\frac{a_{2}^{(1)}}{|\bs b|^3}\bs b\times\clp\cdot\clS_1+ \frac{a_{2}^{(2)}}{|\bs b|^3}\bs b\cdot\clK_1\bigg]+\mathcal O(\alpha^3)\,,
\end{alignat}
where the amplitude coefficients $a_n^{(m)}(\clp^2)$ are given in terms of the Hamiltonian coefficients $c_n^{(m)}(\clp^2)$ via the same relations (\ref{cs_to_as_1PM}) and (\ref{cs_to_as_2PM}) found from the EFT matching, here all evaluated at the incoming momentum $\clp$.  The above relations hold for general values of the Hamiltonian coefficients $c_n^{(m)}(\clp^2)$.

\subsection{Comparison to Observables from the Worldline Theory}
\label{subsec:WLComparison}

Having in hand the observables $\Delta\clp$, $\Delta\clS_1$, and $\Delta\clK_1$ obtained from Hamilton's equations resulting from an EFT matching to a QFT amplitude, we are in a position to ask how these compare to equivalent observables obtained from a worldline theory as in Sec.~\ref{sec:WL}.  We find that the observables of the spinning-probe worldline theory without an SSC match precisely onto those from the probe limit of \hyperlink{ft1g}{FT1g} via the transformations of variables detailed bellow --- these are in one-to-one correspondence with the transformations used to relate the EFT amplitudes to the covariant forms of the field-theory amplitudes in \sect{Sec:EFT_matching}. As discussed in \sect{sec:WLImpulses}, the probe limit provides a nontrivial check.

In the worldline theory, we considered a probe/test particle with mass $m_1$, initial momentum $p_1^\mu=m_1 u_1^\mu$, and initial spin tensor $\gS_1^{\mu\nu}$, scattering off the field of a background Coulomb source with velocity $u_2^\mu$.  The changes $\Delta p_1^\mu$ and $\Delta \gS_1^{\mu\nu}$ from the initial to the final state were expressed in terms of these quantities and the initial impact parameter $b^\mu$.  

Using three-dimensional vectors in the rest frame of the background source, we identify
\begin{equation}
u_2^\mu=(1,0,0,0)\,,\qquad\qquad p_1^\mu=m_1u_1^\mu=(m_1\gamma,\clp)\,,
\label{Eq:big_U_little_u}
\end{equation}
so $\clp$ here is the spatial momentum of the probe in the background frame, with $\clp^2=m_1^2(\gamma^2-1)$, and $m_1\gamma$ is its energy, where $\gamma=u_1\cdot u_2$ is the relative Lorentz factor.  For the spin tensor in the probe limit, just as in (\ref{Eq:S_tensor_SK_vectors}) and (\ref{Eq:standard_boosts}), we decompose it into components $S_1^\mu$ and $K_1^\mu$ in the probe's rest frame,
\begin{align}\label{Eq:decompS1WL}
\gS_1^{\mu\nu} =  \epsilon^{\mu\nu\rho\lambda} u_{1\rho} S_{1\lambda} + u_1^\mu K_1^\nu - u_1^\nu K_1^\mu \,,
\end{align}
and we then relate these, respectively, to three-dimensional vectors $\clS$ and $\clK$ in the background frame by the standard boost taking $u_2^\mu$ into $u_1^\mu$,
\begin{align}
S_1^\mu = \bigg(\, \frac{\clp  \cdot \clS_1}{m_1}, \clS_1 + \frac{\clp \cdot \clS_1}{m_1^2(\gamma +1)} \clp \, \bigg)
\, , \quad
K_1^\mu = \bigg(\, \frac{\clp  \cdot \clK_1}{m_1}, \clK_1 + \frac{\clp \cdot \clK_1}{m_1^2(\gamma +1)} \clp \, \bigg) \,.
\label{Eq:boost_again}
\end{align}
Note that for the complete translation of the observables, we must consider all of (\ref{Eq:big_U_little_u})--(\ref{Eq:boost_again}) applied to both the initial state quantities and to the final state quantities.  Finally, for the impact parameter, we have $b^\mu=(0,\bs b_\mathrm{cov})$, where this should be related to the vector $\bs b$ appearing in the solution of Hamilton's equations by
\begin{equation}
\bs b=\bs b_\mathrm{cov}+\frac{\clp\times\clS_1}{m_1^2(\gamma+1)}+\frac{1}{m_1}\bigg(\clK_1-\frac{\clp\cdot\clK_1}{\clp^2}\clp\bigg)\,,
\end{equation}
which is the Fourier conjugate, under (\ref{Eq:eikonal_Fourier}), of multiplication by the factor $\polM_1 \cdot \polMb_4$ in Eq.~(\ref{Eq:epsilon_dot_epsilon_withK}), in the probe limit.

Taking the solutions for $\Delta p_1^\mu$ and $\Delta \gS_1^{\mu\nu}$ from solving the worldline equations of motion, given in (\ref{eq:finalkicksWL}), and converting them into 3-vector forms using the translations given in the previous paragraph (again, being careful to apply (\ref{Eq:decompS1WL}) and (\ref{Eq:boost_again}) separately to both the initial and final states, using the initial and final momenta), we find expressions for $\Delta \clp$, $\Delta \clS_1$, and $\Delta\clK_1$ which precisely match those coming from solving the equations of motion coming from the Hamiltonian matched to \ref{fieldtheory1}g, given by (\ref{eq:impulsefromchi}) and (\ref{eq:kicksfromchi}) with (\ref{eq:eikonalexplicit}), (\ref{eq:1PLMatchingCoefficients}), and~\cite{anc}.

\subsection{On the Reality of \clK}
\label{sec:KReality}

We conclude this section by commenting on the reality properties of $\clK_1$. In  the quantum theory $\opK_1$ is an antihermitian operator for any finite-dimensional representation\footnote{\baselineskip=14pt Here we refer to the size of the spin space available to the particle (e.g. the states $|1/2, \pm1/2\rangle$ for a spin-1/2 particle), in other words the size of the little-group representation. In contrast, the complete Hilbert space of a particle is always infinite due to the momentum assuming continuous values.}
of the Lorentz group, which implies that its expectation value $\clK_1$ in any such state is imaginary. On the other hand, if we allow for an infinite-dimensional representation, $\opK_1$ may be taken to be hermitian, which would result in $\clK_1$ being real (see e.g. Sect.~10.3 of Ref.~\cite{Tung:1985na}).

We first consider the implications of choosing a finite-dimensional representation, given that these are the representations employed by our field-theory constructions. 
In this case, for the Hamiltonian to be a hermitian operator, we need the coefficients of all Hamiltonian terms that contain an odd number of factors of the boost operator to be imaginary. 
This is indeed so for \ref{fieldtheory3}g, while for \ref{fieldtheory1}g the coefficients are real. For the \PL{1} coefficients, this can be seen by combining Eqs.~(\ref{14map}) and~(\ref{eq:1PLMatchingCoefficients}). In this way, the unphysical nature of the lower-spin states in \ref{fieldtheory1}g results into a non-hermitian Hamiltonian. Interestingly, the hermitian Hamiltonian corresponding to \ref{fieldtheory3}g breaks time-reversal symmetry, which can be seen by combining Eq.~\eqref{eq:ParityTime} with the fact that time-reversal is an antiunitary operator (see e.g. Sect.~2.6 of Ref.~\cite{Weinberg:1995mt}).

Secondly, we examine the case of infinite-dimensional representations. For these, all Hamiltonian coefficients may be taken to be real. This implies that time-reversal symmetry is satisfied. Furthermore, this case meshes well with the classical interpretation of $\clK_1$ as a mass moment, which implies that $\clK_1$ is real. 

While the above seem to suggest the use of a field theory for an infinite-dimensional representation, we do not attempt such a construction in the present paper. Instead, we find that the analytical continuation below Eq.~\eqref{zCOM} is sufficient for our purposes. In particular, such an analytical continuation allows for the matching between our field-theory and worldline constructions, and also results in a hermitian and time-reversal-symmetric Hamiltonian. We defer further analysis of this issue to the future.

\section{Wilson coefficients and propagating degrees of freedom}
\label{sec:WilsonCoeffs}

We have seen in the previous section that the Compton amplitudes computed in \hyperlink{ft1s}{FT1s} depend on additional Wilson coefficients compared to those of \ref{fieldtheory2} (see \cref{tab:AFT} for the Lagrangians for these field theories). 
In \cref{sec:WL} we showed that the number of Wilson coefficients of \ref{fieldtheory2} matches the usual worldline formulation \ref{worldline1} with an SSC imposed.  We also found a modified worldline theory, \ref{worldline2}, containing the same number of additional Wilson coefficients as found in \hyperlink{ft1s}{FT1s}, \hyperlink{ft1g}{FT1g} and \ref{fieldtheory3}. 
Thus, additional Wilson coefficients (relative to e.g. \ref{fieldtheory2} or \ref{worldline1}) are a reflection of additional degrees of freedom in the short-distance theory. In \ref{fieldtheory1} some of these extra states are unphysical, having negative norm, see Sec.~\ref{sec:ss_rep_non_trans}.
In this section we elaborate on the rationale behind \ref{fieldtheory3}, which 
may be thought of as a rewriting of  \ref{fieldtheory1} such that all states have positive norm, and demonstrate that the same outcome---physically-relevant extra Wilson coefficients---can also result when all states have positive norm.  

As in previous discussions of \ref{fieldtheory3}, we focus on fields in the $(s,s)$ representation. We  begin by separating such a field into components with definite spin. While the external states of the amplitudes ${\cal A}^\text{FT1s}$ are transverse and thus spin $s$, the intermediate states may contain lower-spin components, some of which are unphysical. 
We use factorization and gauge invariance to study the exchanges of lower-spin particles in amplitudes with spin-$s$ external states in \ref{fieldtheory1}. We find that 
the map given in Eq.~(\ref{14map}) yields the results of \hyperlink{ft3s}{FT3s} from those of \hyperlink{ft1s}{FT1s}; the imaginary unit in Eq.~(\ref{14map}) is indicative of the negative-norm nature of the exchanged states of \ref{fieldtheory1}.
We also discuss from a general perspective the intermediate-state spins that can contribute in the classical limit and construct their contribution to the Compton amplitude. 
This analysis sets on firm footing the field content we chose for the Lagrangian of \ref{fieldtheory3}. 
Because of the structure of the Lorentz generators in the $(s,s)$ representation~\eqref{eq:LorentzGen}, the trace part of intermediate states can be projected out by simply choosing traceless external states, such as 
the coherent states in Eq.~\eqref{eq:spin_coherent}. We  therefore focus on the consequences of transversality or lack thereof.
%

\subsection{Resolution of the Identity and Amplitudes with Lower-Spin States}
\label{Sect:AmpsAndProjectors}

As reviewed earlier, a field in the representation $(s,s)$ of the Lorentz group contains states of all spins 
between 0 and $s$. To develop a general picture of the interplay and couplings of these states it is useful to formally 
expose them in the Lagrangian of \ref{fieldtheory1}. 
We use the resolution of the identity operator in this representation,
\begin{align}
\label{identityresolution}
	\delta_{\mu(s)}^{\nu(s)}=\sum_{n=0}^{s}{s\choose n}u_{(\mu_1}\ldots u_{\mu_n}u^{(\nu_1}\ldots u^{\nu_n}\mathscr{P}_{\mu_{n+1}\ldots\mu_s)}^{\nu_{n+1}\ldots\nu_s)}\,,
\end{align}
with the on-shell transverse projectors $\mathscr{P}_{\mu_1\mu_2\ldots\mu_s}^{\nu_1\nu_2\ldots\nu_s}=\Theta_{(\mu_1}^{\nu_1}\Theta_{\mu_2}^{\nu_2}\ldots\Theta_{\mu_s)}^{\nu_s}$, which is the $j=0$ term in the summation of \cref{eq:state_proj} and the symmetrization follows the definition in footnote~\ref{symmetrization_footnote}. 
For example, for the two-index and three-index-symmetric representations this becomes
\begin{align}
	\delta_{\mu_1}^{(\nu_1}\delta_{\mu_2}^{\nu_2)}&=\mathscr{P}_{\mu_1\mu_2}^{\nu_1\nu_2}+2 u_{(\mu_1}u^{(\nu_1}\mathscr{P}_{\mu_2)}^{\nu_2)}+u_{\mu_1}u_{\mu_2}u^{\nu_1}u^{\nu_2} \,,\\
	\delta_{\mu_1}^{(\nu_1}\delta_{\mu_2}^{\nu_2}\delta_{\mu_3}^{\nu_3)}&=\mathscr{P}_{\mu_1\mu_2\mu_3}^{\nu_1\nu_2\nu_3}+3u_{(\mu_1}u^{(\nu_1}\mathscr{P}_{\mu_2\mu_3)}^{\nu_2\nu_3)}+3u_{(\mu_1}u_{\mu_2}u^{(\nu_1}u^{\nu_2}\mathscr{P}_{\mu_3)}^{\nu_3)}+u_{\mu_1}u_{\mu_2}u_{\mu_3}u^{\nu_1}u^{\nu_2}u^{\nu_3} \,.\nonumber
\end{align}
The projectors used here single out the longitudinal components of fields but not traces. We ignore trace states; while they are propagating, in four-point Compton amplitudes they can be projected out from all diagrams that do not include loops of higher-spin states by choosing traceless external states. 

By inserting the resolution of the identity~\eqref{identityresolution} into the nonminimal interaction $\Lnonmin$ of \ref{fieldtheory1}, we can expose and identify the couplings of all the definite-spin components of $\phi_s$.\footnote{\baselineskip=14pt The projectors may be replaced with their off-shell-transverse version, constructed from $(\eta_{\mu\nu} - p^\mu p^\nu/p^2)$.  However, this yields a nonlocal Lagrangian. Moreover, transversality needs to be only an on-shell property, so using Eq.~\eqref{identityresolution} is sufficient.} For example, in the $\mathcal{O}(S^1)$ interaction $F_{\mu\nu}\phi_sM^{\mu\nu}\phib_s$, by using $u^\mu \rightarrow i \partial^\mu/m$, we get
\begin{align}\label{eq:lowspinfield}
\phi_sM^{\mu\nu}{\bar\phi}_s = 
\sum_{n=0}^{s} \frac{(-1)^n}{m^{2n}}\phi_s^{\rho_1\dots \rho_s} (M^{\mu\nu})_{\rho_1\dots \rho_s}^{\mu_1\dots \mu_s}
\partial_{(\mu_1}\ldots \partial_{\mu_n}\mathscr{P}_{\mu_{n+1}\ldots\mu_s)}^{\nu_{n+1}\ldots\nu_s}(\partial^n \phib_s){}_{\nu_{n+1}\ldots\nu_s} + \text{c.c}\,,
\end{align}
where $(\partial^n \phib_s){}_{\nu_{n+1}\ldots\nu_s} = \partial^{\nu_1}\dots  \partial^{\nu_n}\phib_s{}_{\nu_{1}\ldots\nu_s}$ is a field in the $(s-n, s-n)$ representation of the Lorentz group, and the projector $\mathscr{P}_{\mu_{n+1}\ldots\mu_s}^{\nu_{n+1}\ldots\nu_s}$ singles out its spin-$(s-n)$ component. 
In \cref{eq:lowspinfield} each term in the summation is given by partial derivatives and is thus not invariant under the photon gauge transformation. 
We only use this equation as a guide to construct an effective field theory in which an $s$-index tensor nonminimally couples to an $(s-n)$-index tensor.

Schematically, we identify $\mathscr{P}_{\mu_{n+1}\ldots\mu_s}^{\nu_{n+1}\ldots\nu_s}(\partial^n \phib_s){}_{\nu_{n+1}\ldots\nu_s}\equiv(\phi_{s-n})_{\mu_{n+1}\ldots\mu_s}$ as an off-shell spin-$(s-n)$ field and assign to it the kinetic term is given by $\mathcal{L}^{s-n}_{\text{min}}$ defined in \cref{eq:Lspins}. We further replace all the remaining partial derivatives by their covariant version. For the coupling $F_{\mu\nu}\phi_sM^{\mu\nu}\phib_s$ at the linear order in spin, this prescription leads to the following interaction between $\phi_s$ and $\phi_{s-1}$,
\begin{align}
	\frac{1}{m}F_{\mu\nu}\left[\phi_s^{\alpha_1\ldots\alpha_s}M^{\mu\nu}{}_{\alpha_1\ldots\alpha_s,\beta_1\ldots\beta_s}D_{\phantom{s-1}}^{(\beta_1}\phib_{s-1}^{\beta_2\ldots\beta_s)}+\text{c.c}\right].
\end{align} 
This interaction agrees with the one included in \ref{fieldtheory3} for $\widetilde{C}_1=\widetilde{C}_2$ (see \cref{eq:transition_L}). If we further relax the requirement that the interaction has to be mediated by the Lorentz generator, we  get one more gauge invariant structures and thus arrive exactly at \cref{eq:transition_L}.

Having identified the off-shell component fields that exist within the off-shell field $\phi_s$, we may explore how does the amplitude change if we restrict both the on-shell and the off-shell states to $(2s+1)$ states of a spin-$s$ particles. We study this by building the four-point Compton amplitude involving only massive spin-$s$ degrees of freedom with on-shell methods. On general grounds, we should find $\mathcal{A}^\text{FT2}$; to carry out this calculation, we need to find products of spin-$s$ polarization tensors and the projector, similarly to the products involving Lorentz generators we computed in \cref{sec:ss_rep}. We then subtract it from the corresponding amplitude $\mathcal{A}^{\text{FT1s}}$ to obtain the contribution from the lower-spin degrees of freedom, i.e. the difference between $\mathcal{A}^{\text{FT1s}}$ and $\mathcal{A}^{\text{FT2}}$. 
In \cref{sec:FourPointCompt}, the Compton amplitudes of \ref{fieldtheory2} are computed from fixed value of $s$ and then extrapolated to the generic case. {Here, we will keep $s$ arbitrary, but only consider the linear order of spin; this will be sufficient to illustrate the main points of our discussion .} 

We evaluate the products in question explicitly starting from low and fixed 
values of the spin, extrapolating to arbitrary $s$ and then taking the classical limit. 
We find,
\begin{align}
\label{Eq:ProductsProjector}
\polM_1^{(s)} \cdot \mathcal{P}^{(s)}(p_1+q_2) \cdot \polMb_4^{(s)} &= 
\polM_1^{(s)} \cdot \polMb_4^{(s)}\left(1  +
\frac{s  \varepsilon_1 \cdot q_2{\bar \varepsilon}_4 \cdot q_3}{ \varepsilon_1\cdot {\bar \varepsilon}_4m^2} + 
\ldots \right),  \\
\polM_1^{(s)} \cdot \mathcal{P}^{(s)}(p_1+q_2) \cdot M^{\mu\nu} \cdot \polMb_4^{(s)} &= 
\polM_1^{(s)}  \cdot M^{\mu\nu} \cdot \polMb_4^{(s)}  + 
\frac{i s ( p_1^{\mu} {\bar\varepsilon}_4^{\nu} - p_1^{\nu} {\bar\varepsilon}_4^{\mu} ) \varepsilon_1 \cdot q_2}{ \varepsilon_1\cdot {\bar \varepsilon}_4 \, m^2} \, \polM_1^{(s)} \cdot \polMb_4^{(s)} + \ldots \,,\nn
\end{align}
where we used the on-shell conditions and transversality and $q_2$ and $q_3$ are the momenta of the Compton amplitude photons.
%
%
We have omitted terms that do not contribute in the classical limit of the Compton amplitude at  $\mathcal{O}(S^1)$. 

Using Eqs.~\eqref{Eq:ProductsProjector} it is straightforward to compute the pole part of the Compton amplitude.  
To complete the amplitude we construct an ansatz for the missing contact term and fix it by demanding gauge invariance 
for the two photon external lines. 
We find that the difference $\mathcal{A}_{4, \text{cl}}^\delta$ between the amplitude without the spin-$s$ projector, ${\cal A}_{4, \text{cl}}^{\text{FT1s}}$, and the amplitude with the spin-$s$ projector, which is indeed $\mathcal{A}^\text{FT2}$, is given by
\begin{align}
\label{difference}
\mathcal{A}_{4, \text{cl}}^\delta = 
\mathcal{A}_{4, \text{cl}}^{\text{FT1s}}-{\cal A}_{4, \text{cl}}^{\text{FT2}} =
-(-1)^s \, \frac{\polM_1^{(s)} \cdot \polMb_4^{(s)}}{m^2} \frac{ 2i (1-C_1+D_1)^2}{p_1 \cdot q_2} 
p_{1} \cdot f_2 \cdot S(p_1) \cdot f_3 \cdot p_{1} 
\, .
\end{align}
This is exactly the difference between \cref{eq:s1_comp} and \cref{eq:A4FT2}.
The sign difference compared to Eq.~\eqref{ComptonFT4S1} reflects the negative norm of the spin-$(s-1)$ states 
that are part of $\phi_s$ compared to the positive norm of the analogous states in \ref{fieldtheory3}. 
Eq.~\eqref{difference} also manifests that choosing $D_1 = C_1-1$ for $\mathcal{A}^{\text{FT1s}}$ is equivalent to consistently inserting the spin-$s$ physical-state projector.

\subsection{Lower-Spin States and their Scaling in the Classical Limit}

Having identified the relevance of the lower-spin states for Compton amplitudes,  we now proceed to examine the processes whose classical limit receives contributions from such states.
While, as already noted, in \ref{fieldtheory1} such states have negative norm, we may either construct field theories such as \ref{fieldtheory3} in which their norm is positive so they are physical, or we may simply use maps such as \eqref{14map} or \eqref{14mapS2} to modify the amplitudes of \ref{fieldtheory1} to agree with amplitudes with physical  intermediate states.

\begin{figure}[tb]
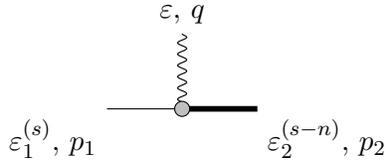

\FigLowerSpinThreePTsn
\caption{The three-point amplitude involving a massive spin-$s$ particle (thin line), a massive spin-$(s-n)$ 
particle (thick line) and a photon (wiggly line). 
}
\label{Fig:LowerSpin3PTsn}
\end{figure}

We wish to characterize the classical scaling of the transitions from 
the spin-$s$ to the spin-$(s-n)$ state via the emission of a photon. 
There are several distinct structures that can appear in such an amplitude, as illustrated for example in Eq.~\eqref{eq:A3min}. For illustrative purposes we focus on the first term in that equation which arises from the covariant derivative in the quadratic Lagrangian $\Lmin$ of \ref{fieldtheory1}; other interactions may be treated similarly with similar expected conclusions. We moreover interpret the lower-spin field as the longitudinal components of a higher-spin field, as discussed in Sec.~\ref{sec:ss_rep_non_trans}. Thus, the three-point amplitude we consider here and illustrated in \cref{Fig:LowerSpin3PTsn} is
\begin{equation}
\mathcal{A}^{s \rightarrow s-n}_{3, \text{min}}  = 
i \pol_3\cdot p_1 \, \polM^{(s)}_1 \cdot \left(u^n_2 \, \polMb^{(s-n)}_2\right)\,,
\end{equation}
where all momenta are outgoing, the matter momenta are $p_1$ and $p_2$, $u_i = p_i/m$ and the photon momentum is $q$.
Using the explicit form of the polarization tensors in Eq.~\eqref{eq:ExplicitLowSpinState}, this three-point amplitude becomes
\begin{equation}
\mathcal{A}^{s \rightarrow s-n}_{3, \text{min}}=
i \pol_3\cdot p_1 \, (\varepsilon_1 \cdot \bar\varepsilon_2)^{s-n}  
{s \choose n}^{1/2} \left( \frac{q \cdot \varepsilon_1}{m} \right)^n \,,
\label{threepts}
\end{equation}
where we used the on-shell conditions $p_2 = -p_1-q$ and $\varepsilon_1 \cdot p_1 = 0$. For $n\ll s$ and $1\ll s$ we may approximate ${s \choose n} \approx \frac{s^{n}}{n!}$. 
We may use the scaling of polarization tensors implied by their embedding in a nontransverse $(s,s)$ representation of their Lorentz group to obtain the scaling of the transition amplitude. Together with \cref{eq:eSeSm1,eq:esMes2,offdiagonalKdependence}, \cref{threepts} implies that the transition three-point amplitude $\mathcal{A}^{s \rightarrow s-n}_{3, \text{min}}$ depends on $q$ and $K$ as 
\begin{equation}
\mathcal{A}^{s \rightarrow s-n}_{3, \text{min}}
\sim 
q^n K^n \sim q^0 
    \,.
\label{Eq:3PTScaling}
\end{equation}
Thus, the transition three-point amplitudes scale as $q^0$ in the classical limit, so they are classical.

\begin{figure}[tb]
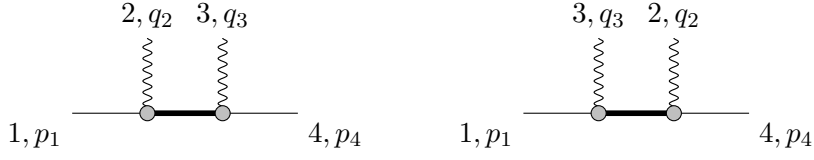

\FigLowerSpinExchangeInCompton
~~~~~~
\FigLowerSpinExchangeInComptonX
\caption{Representative diagram of the contribution of the spin-$(s-n)$ exchanges in the Compton amplitude. Legs 1 and 4 are massive spin-$s$ particles, legs 2 and 3 are photons, and the intermediate thick line corresponds to the spin-$(s-n)$ particle for some $n>0$.  }
\label{Fig:LowerSpinInCompton}
\end{figure}

We now discuss the contribution of three-point amplitudes to the residue of four-point amplitudes. Since in Eq.~\eqref{Eq:3PTScaling} the polarization tensors have already been used to generate the factors of $K$, the expression of the amplitudes that is useful for residue computation is
Eq.~\eqref{threepts} together with the fact that the sum of a product of spin-$(s-k)$ polarization tensors over all the physical states yields the projector onto the spin-$(s-k)$ states. It is then straightforward to see that the pole part of diagonal amplitudes, whose diagrams are illustrated in \cref{Fig:LowerSpinInCompton}, is 
\begin{align}
 &\mathcal{A}_4^{s\rightarrow s}\Big\vert_\text{spin-$(s{-}n)$}^{\text{exchange}} = \sum_{(s-n)\text{ states}}
\left[\frac{\mathcal{A}_{3: p_1, q_2, P}^{s\rightarrow s-n} 
\mathcal{A}_{3: -P, q_3, p_4}^{s-n\rightarrow s}}{2p_1 \cdot q_2} + \frac{\mathcal{A}_{3: p_1, q_3, P}^{s\rightarrow s-n} \mathcal{A}_{3: -P, q_2, p_4}^{s-n\rightarrow s}}{2p_1 \cdot q_3} \right]
\nonumber \\[3pt]
&\hskip 2.5 cm \sim  \frac{s^{n}}{2p_1 \cdot q_2}(\varepsilon_1\cdot{\bar\varepsilon}_4)^{s-n}
\left[
\left( \frac{q_2 \cdot \varepsilon_1}{m} \right)^n
\left( \frac{q_3 \cdot {\bar\varepsilon}_4}{m} \right)^n 
- 
\left( \frac{q_3 \cdot \varepsilon_1}{m} \right)^n
\left( \frac{q_2 \cdot {\bar\varepsilon}_4}{m} \right)^n 
\right]
,
\label{exchange}
\end{align}
where we assumed that the relevant higher-spin theory has standard factorization properties. The second term in \eqref{exchange} follows by interchanging $q_2$ and $q_3$.
In the large $s$ limit, \cref{eq:S1rep,eq:Srep2,eq:Srep1} imply that
\begin{align}
\left( \frac{q_2 \cdot \polm_1}{m}  \frac{q_3 \cdot \polmb_4}{m} \right)^n
&\rightarrow \frac{(\polm_1\cdot \polmb_4)^n}{(2 m s)^n}\left[ i q_2\cdot S(p_1)\cdot q_3 - \frac{1}{m s} q_2\cdot S(p_1)\cdot S(p_1) \cdot q_3\right]^n \nonumber\\[3pt]
&\rightarrow \frac{(\polm_1\cdot\polmb_4)^n}{(2ms)^n}\Big[iq_2\cdot S(p_1)\cdot q_3\Big]^n\,.
\label{scalingSsTOs}
\end{align}
We observe that for any $n$ the explicit factors of $s$ cancel in Eq.~\eqref{exchange}, i.e. the various factors combine such that the only spin dependence is through $(\varepsilon_1\cdot \bar\varepsilon_4)^s$ and $S^{\mu\nu}$.
However, since the square parenthesis in Eq.~\eqref{scalingSsTOs} scales as $q^n$ and the propagator in Eq.~\eqref{exchange} scales as $1/q$, only for $n=1$ the exchange term has a classical contribution. 
Heuristically, the existence of one matter propagator allows for transitions to spin states that differ from the external by one unit (i.e. $s \rightarrow s-1 \rightarrow s$).

A similar argument reveals the contribution of transition three-point amplitudes to off-diagonal $s\rightarrow s-m$ two-photon amplitudes. 
It is intuitive that intermediate spin-$(s-n)$ states can contribute if $0 \leq n \leq m$. For $n>m$, we find that the existence of one matter propagator in the four-point amplitude allows for the state $n=m+1$ to also contribute. 
Indeed, factorization together with Eq.~\eqref{threepts} imply that
\begin{align}
 &\mathcal{A}_4^{s\rightarrow s-m}\Big\vert_\text{spin-$(s{-}n)$}^{\text{exchange}} = \sum_{(s-n)\text{ states}}
\left[\frac{\mathcal{A}_{3: p_1, q_2, P}^{s\rightarrow s-n} 
\mathcal{A}_{3: -P, q_3, p_4}^{s-n\rightarrow s-m}}{2p_1 \cdot q_2} + \frac{\mathcal{A}_{3: p_1, q_3, P}^{s\rightarrow s-n} \mathcal{A}_{3: -P, q_2, p_4}^{s-n\rightarrow s-m}}{2p_1 \cdot q_3} \right]
\nonumber\\[3pt]
&\hskip 1.5 cm 
\sim  
\frac{s^{n-m/2}}{2p_1 \cdot q_2}(\varepsilon_1\cdot{\bar\varepsilon}_4)^{s-n}
\left[
\left( \frac{q_2 \cdot \varepsilon_1}{m} \right)^n
\left( \frac{q_3 \cdot {\bar\varepsilon}_4}{m} \right)^{n-m}
- 
\left( \frac{q_3 \cdot \varepsilon_1}{m} \right)^n
\left( \frac{q_2 \cdot {\bar\varepsilon}_4}{m} \right)^{n-m} 
\right] 
,
\label{exchangeTransition}
\end{align}
where we assumed that $n \ge m$
and that, as before, the relevant higher-spin theory has standard factorization properties.\footnote{\baselineskip=14pt
An expression analogous with Eq.~\eqref{exchangeTransition} can be written for $m>n$. Both in that expression and in Eq.~\eqref{exchangeTransition} the sum over intermediate states yields an on-shell transverse projector. Transversality of external states implies, however, that the momentum-dependent terms in that projector are subleading in the classical limit, which justifies why no projector is included in Eq.~\eqref{exchangeTransition}.
}
Eqs.~\eqref{eq:S1rep}, \eqref{eq:Srep2} and \eqref{eq:Srep1} imply that, as in the diagonal amplitude, the factors with an equal number of $\varepsilon_1$ and $\bar\varepsilon_4$
can be effectively written in the large $s$ limit as 
\begin{align}
\left( \frac{q_2 \cdot \varepsilon_1}{m} \frac{q_3 \cdot {\bar\varepsilon}_4}{m} \right)^{n-m}
\rightarrow \frac{(\varepsilon_1\cdot \bar\varepsilon_4)^{n-m}}{(2 m s)^{n-m}}\left[ i q_2\cdot S(p_1)\cdot q_3 - \frac{1}{m s} q_2\cdot S(p_1)\cdot S(p_1) \cdot q_3\right]^{n-m} \, ,
\end{align}
and similarly for $q_2\leftrightarrow q_3$. 
Eqs.~\eqref{eq:GeneralEpMtokuEpTOpolvectors} and \eqref{eq:qDotKToTheK} can be used to write the remaining unbalanced dependence on $\varepsilon_1$ and $\bar\varepsilon_4$ as a linear combination of $K$ and $S$ vectors of degree larger or equal to $m$, which contains at least one term with $m$ factors of $K$ and has an overall factor of $s^{-m/2}$ (see for e.g. Eq.~\eqref{eq:qDotKToTheK}).
The overall factors of $s$ cancel out, as in the case of diagonal amplitudes.
The terms with $m$ factors of $K$ and one power of $S$ if $n>m$ and the terms with $m$ factors of $K$ and no power of $S$ if $n=m$ exhibit classical scaling. 
In other words, suppressing factors of $q$ and $S$, we have $\mathcal{A}_4^{s\rightarrow s-m} \sim K^m$ as its three-point counterpart in Eq.~\eqref{Eq:3PTScaling}.
This dependence prompted us to restrict our analysis of \ref{fieldtheory3} to a single power of $K$ or an arbitrary power of $S$ and no $K$, i.e. $m \leq 1$, since these are the terms we may probe by considering a spin-$s$ and a spin-$(s-1)$ field.
It would be interesting to extend \ref{fieldtheory3} with further lower-spin fields and access nonlinear dependence on $S$ and $K$.
Consistency of the theory should lead to the cancellation of possible superclassical terms.

The arguments above can be repeated to analyze the possible intermediate states that can contribute to higher-point tree-level amplitudes. For example, starting with Eq.~\eqref{Eq:3PTScaling}, the two-pole part of a diagonal  $s\rightarrow s$ three-photon amplitude can receive contributions from suitable combinations of intermediate states of spin different from $s$.
Further contributions from single-pole terms depend on the scaling of four-point contact terms; for example, if it is the same as for the three-point amplitude,  $\mathcal{A}_4^{s\rightarrow s-n} \sim (q\cdot K/m)^n$, then such four-point amplitudes contribute to single-pole terms 
of the five-point amplitude. Such higher-point amplitudes are some of the ingredients of higher-PL spin-dependent calculations, so it would be interesting to investigate them further. 

\subsection{Lower-Spin States in the Compton Amplitude}

With the information we acquired from the analysis of the soft-region scaling of amplitudes with states of different large spin we may construct Compton amplitudes using a standard on-shell approach: We start with three-point amplitudes with the appropriate scaling and use them to construct the $O(S)$ exchange part of the Compton amplitude. We then fix the contact terms by demanding gauge invariance and that their dimension is the same as that of contact terms arising from the Lagrangians of \ref{fieldtheory1}, \ref{fieldtheory2} and \ref{fieldtheory3}.

The three-point amplitude is shown diagrammatically in Fig.~\ref{Fig:LowerSpin3PTsn}.  With $n=1$ and all-outgoing momenta, its expression that follows from a Lagrangian such as that of \ref{fieldtheory1} is
\begin{align}
\mathcal{A}_3^{s \rightarrow s-1} = 
(-1)^{s} \polM_1^{(s)} \cdot \mathbb{M}_3(p_1,p_2,\kk_3,\pol_3) \cdot \left(u_2 \polMb_2^{(s-1)}\right) \,,
\label{eq:3ptOperator}
\end{align}
where $\mathbb{M}_3$ is given in Eq.~(\ref{eq:A3min}). Using Eqs.~(\ref{eq:eSeSm1}) and~(\ref{eq:esMes2}), the linear-in-$S$ or $K$ part of the three-point amplitude to leading order in $s$ can be written as
\begin{equation}
\label{Eq:3ptSpinFlip}
\mathcal{A}_3^{s\rightarrow s-1} = (-1)^s \frac{2 \sqrt{s} (C_1-D_1-1) (p_1 \cdot \pol_3) (\varepsilon_1 \cdot q) (\varepsilon_1 \cdot {\bar\varepsilon}_2)^{s-1}}{m} + \ldots \,.
\end{equation}
where the ellipsis stands for terms of higher order in $s$ and $q$.

Next, we sew together two of these three-point amplitudes to obtain the residues of the two matter-exchange poles of the Compton amplitude corresponding to the two diagrams in Fig.~\ref{Fig:LowerSpinInCompton}. Focusing solely on the spin-$(s-1)$ exchange, we have
\begin{align}
\text{Res} \Big( \mathcal{A}_{4,\text{cl}}\Big\vert_\text{spin-$(s{-}1)$} \Big)\Big\vert_{2 p_1 \cdot q_2 = 0}  &=
(-1)^s
\frac{4 s (C_1-D_1-1)^2 
(p_1 \cdot \pol_2) (\varepsilon_1 \cdot q_2)
(p_4 \cdot \pol_3) ({\bar\varepsilon}_4 \cdot q_3) }{m^2}
 \nn \\
&\times\sum_{\text{phys. } \varepsilon_\ell \text{ states}} (\varepsilon_1 \cdot {\bar\varepsilon}_\ell)^{s-1}(\varepsilon_\ell \cdot {\bar\varepsilon}_4)^{s-1}\,,
\end{align}
where $\pol_2$ and $\pol_3$ are photon polarization vectors.
The physical state sum is evaluated using 
\begin{align}
\sum_{\text{phys. } \varepsilon_\ell \text{ states}} (\varepsilon_1 \cdot {\bar\varepsilon}_\ell)^{s-1}&(\varepsilon_\ell \cdot {\bar\varepsilon}_4)^{s-1} =
\varepsilon_{1\mu_1} \ldots \varepsilon_{1\mu_{s-1}} \,
\left(\mathcal{P}^{(s-1)}(\ell)\right){}^{\mu(s-1)}_{\nu(s-1)} \,
{\bar\varepsilon}_4^{\nu_1} \ldots {\bar\varepsilon}_4^{\nu_{s-1}}  \,,
\end{align}
and Eq.~(\ref{Eq:ProductsProjector}).
Given that the residue scales as $q$,  we may replace the projector with the identity, as all the other terms are subleading in small $q$. 
The residue becomes
\begin{align}
\text{Res} \Big( \mathcal{A}_{4, \text{cl}}\Big\vert_\text{spin-$(s{-}1)$} \Big)&\Big\vert_{2 p_1 \cdot q_2 = 0}  = \\
&(-1)^s
\frac{4 s (C_1-D_1-1)^2 
(p_1 \cdot \pol_2) (\varepsilon_1 \cdot q_2)
(p_4 \cdot \pol_3) ({\bar\varepsilon}_4 \cdot q_3) }{m^2}
(\varepsilon_1 \cdot {\bar\varepsilon}_4)^{s-1} \,. \nn
\end{align}

To complete the amplitude we need to add the other exchange channel, with a pole at $p_1\cdot q_3 = 0$, and to find the 
contact term so that the result is invariant under photon gauge transformations, $\varepsilon_{i} \rightarrow \varepsilon_{i}+\lambda q_i$ with $i=2$ and separately $i=3$. Allowing for at most two powers of momenta in the contact term, its effect
is only the replacements
\begin{align}
(p_1 \cdot \pol_2) (\varepsilon_1 \cdot q_2) 
\rightarrow
p_1^\mu f_{2,\mu \nu} \varepsilon_1^\nu\,,
\qquad
(p_4 \cdot \pol_3) ({\bar\varepsilon}_4 \cdot q_3) 
\rightarrow
p_4^\alpha f_{3,\alpha \beta} {\bar\varepsilon}_4^\beta\, ,
\end{align}
with $f_{i, \mu\nu}$ defined below Eq.~\eqref{spinlessCompton}. 
Thus, the classical Compton amplitude of two spin-$s$ particles due to an intermediate spin-$(s-1)$ exchange is
\begin{align}
\mathcal{A}^{s\rightarrow s}_{4, \text{cl}}\Big\vert_\text{spin-$(s{-}1)$} =
(-1)^s
\frac{(\varepsilon_1 \cdot \bar{\varepsilon}_4)^{s-1}}{m^2} \frac{4 s (C_1-D_1-1)^2}{2 p_1 \cdot q_2} 
p_1\cdot f_2\cdot \varepsilon_1 \,
 p_4\cdot f_3\cdot {\bar\varepsilon}_4 + 
 (2 \leftrightarrow 3) \,.
\end{align}
Finally, replacing the polarization vectors $\varepsilon_1$ and ${\bar\varepsilon}_4$ in terms of the spin tensor as in Eq.~(\ref{eq:S1rep}) and keeping only the classical terms leads to  
\begin{equation}
\mathcal{A}^{s\rightarrow s}_{4,\text{cl}}\Big\vert_\text{spin-$(s{-}1)$} = \mathcal{A}_{4, \text{cl}}^\delta 
= \mathcal{A}_{4, \text{cl}}^{\text{FT1s}}-\mathcal{A}_{4, \text{cl}}^{\text{FT2}} \, ,
\end{equation}
where for the second equality we used Eq.~\eqref{difference}.  Thus, we explicitly identify the difference between $ \mathcal{A}_{4, \text{cl}}^{\text{FT1s}}$ and $\mathcal{A}_{4, \text{cl}}^{\text{FT2}}$ as due to the propagation of an intermediate $(s-1)$-spin state.

\section{Discussion and Conclusion}
\label{sec:Conclusion}

In this paper we addressed a puzzle regarding the description and dynamical evolution of spinning bodies in Lorentz invariant theories, with an eye towards applications to the two-body problem in general relativity.  
Their gravitational or electromagnetic interactions are described via an effective field theory of point particles in terms of a set of higher-dimension operators each with a free Wilson coefficient. Ref.~\cite{Bern:2022kto} found that the amplitudes-based framework of Ref.~\cite{Bern:2020buy} leads to additional independent Wilson coefficients in observables compared to the usual worldline description.  
These additional Wilson coefficients appear to vanish identically for black holes, but seem to contribute to scattering observables for more general spinning objects starting at the second order in Newton's constant and at cubic order in the spin.

To identify the origin and the physics described by the extra Wilson coefficients we analyzed the simpler case of electromagnetic interactions of charged spinning bodies. This theory is inherently simpler than general relativity because it has no photon self-interactions and more importantly the analogous effects are already present at linear order in spin.
We constructed several such electromagnetic field theories: one with two physical propagating higher-spin fields, another with multiple physical and unphysical propagating higher-spin states packaged in a single higher-spin field, and finally one with a single quantum spin. When available, we also considered several possible classical asymptotic states. 
In the classical limit we found that simple maps connect the amplitudes of the various  cases and reached the conclusion that the presence of states beyond those of a spin-$s$ particle leads to additional Wilson coefficients. These Wilson coefficients govern transitions between states of different spin which in turn lead to changes in the magnitude of the classical spin vector even for conservative dynamics. While the magnitude of the spin vector can change in theories with additional propagating states, the magnitude of the spin tensor is conserved.

We found that these results have an interpretation in a more conventional worldline framework and exposed it by analyzing two distinct worldline theories. The first one corresponds to the standard construction~\cite{Porto:2006bt, JanSteinhoff:2015ist} where a spin supplementary condition is imposed. The second theory relaxes this constraint, introducing additional degrees of freedom.  As for field theories with transitions between states with different spin, the dynamics of this theory allows for changes in the magnitude of the spin vector along classical trajectories.

While the results of all of our field theories can be obtained as limits of results of these two worldline theories, we did not find a worldline theory that reproduces observables obtained from ${\cal A}^\text{FT1s}$ whose asymptotic states are limited to a single quantum spin. It would be interesting to pursue the construction of such a theory; to this end it may be profitable to interpret ${\cal A}^\text{FT1s}$ as a sequence of absorption amplitudes and match them with a worldline theory with additional non-asymptotic states, along the lines of Ref.~\cite{Goldberger:2020fot}.  Another interesting direction would be to generalize \ref{fieldtheory3}, which was constructed using spin $s$ and $(s-1)$ states, to include spin $(s-k)$ state with $k \ge 2$, in order to describe interactions beyond the spin-orbit case. 

We evaluated tree-level Compton amplitudes to provide a direct comparison between the various field and worldline theories. We carried out this comparison to second order in the spin tensor. Field theories restricted to propagate only the states of a spin-$s$ particle preserve the magnitude of the classical spin vector, and the results match those of the worldline with a spin supplementary condition imposed, compatible with Refs.~\cite{Kim:2023drc, Haddad:2023ylx}. 
In contrast, if states of different spin propagate and transitions are allowed between them, the field-theory Compton amplitudes contain additional Wilson coefficients and match those of the worldline with no spin supplementary condition.
The results of the theory with propagating states of a single spin-$s$ particle are reproduced for special values of the Wilson coefficients; thus, for these values, the SSC condition is effectively imposed (albeit not actively), and the spin gauge symmetry is restored. This holds true both for the field theory where some of the additional spin states were negative norm~\cite{Bern:2020buy} and for the alternative construction with all positive-norm states. 

To establish a closer connection between the extra degrees of freedom present in the various field-theory descriptions of spinning bodies and classical observables, we constructed a pair of two-body Hamiltonians where the obtained amplitudes match the field-theory amplitudes~\cite{Bern:2020buy}. The first of these Hamiltonians  is the standard two-body one including the standard spin-orbit terms.  The second incorporates the mass moment as a new (boost) degree of freedom, and is the one that can match both the field theories with transitions between states of different spins and the worldline with no spin supplementary condition imposed.
We carried out detailed comparisons of the impulse, and spin and mass-moment kicks through ${\mathcal O}(\alpha^2 S)$ between the predictions of these two-body Hamiltonians and the corresponding worldline approaches and found agreement to this order.
It would be interesting to generalize our field theory with two propagating fields to contain multiple propagating fields and in this way verify the connection to the worldline through ${\mathcal O}(\alpha^2 S^{k\ge 2})$. 

We also succeeded in finding a compact way to express scattering observables via an eikonal formula.  The spin eikonal formula of Ref.~\cite{Bern:2020buy} provides a direct connection between amplitudes and scattering observables and bypasses explicit use of the Hamiltonian. We found a generalization of this formula, which is valid through $\mathcal O(\alpha^2 S)$, compactly contains the intricate results of Hamilton's equations for scattering observables and includes extra degrees of freedom (in the form of the rest-frame boost vector) and all Wilson coefficients. It would be interesting to extent this comparison to higher powers of the spin and boost vectors.
While this eikonal formula was not derived from first principles, its existence strongly suggests that a first-principles derivation should exist.

Our primary conclusion is that, whether using a four-dimensional field-theory or a worldline description of spinning bodies,  the extra Wilson coefficients are directly associated with additional propagating degrees of freedom.  These extra coefficients induce a dynamical change in the magnitude of the rest-frame spin vector even for conservative dynamics. This change in spin magnitude is necessarily associated with a change in the mass moment, which in turn induces a change in the electric dipole moment.  It would be very interesting to identify physical systems 
where these additional degrees of freedom lead to observable effects whether in electrodynamics or general relativity.

We expect that carrying out similar field theory, worldline and effective two-body Hamiltonian constructions and comparisons for general relativity should be straightforward. We look forward to studying the phenomena described here in detail for the case of general relativity where they were originally observed.


\subsection*{Acknowledgments}                 

\vspace{-2truemm}

We thank Thibault Damour, Henrik Johansson, Callum Jones, Jung-Wook Kim, Maxim Pospelov, Francesco Serra, Mikhail Solon and Jan Steinhoff for useful discussions.  We also especially thank Juan Pablo Gatica for pointing out a typo in an earlier version.
Z.~B., and T.~S.  are supported by the U.S. Department of Energy
(DOE) under award number DE-SC0009937. 
D.~K. is supported by the Swiss National Science Foundation under grant no. 200021-205016.
A.~L. is supported by funds from the European Union’s Horizon 2020 research and innovation program under the Marie Sklodowska-Curie
grant agreement No.~847523 ‘INTERACTIONS’.  
R.~R. and F.~T.~are supported by
the U.S.  Department of Energy (DOE) under award number~DE-SC00019066.
We are also grateful to the Mani L. Bhaumik Institute for Theoretical Physics for support.


\newpage
\bibliography{ref}{}
\setlength{\bibsep}{0pt plus 0.1ex}
\bibliographystyle{utcaps}

\end{document}